\documentclass[camera,letterpaper,nomarginnotes,nonarrowgutter]{jpaper} 

\usepackage[sort,compress]{cite}
\usepackage{amsmath,amssymb,amsfonts}
\usepackage{algorithmic}
\usepackage{graphicx}
\usepackage{textcomp}
\usepackage{fancyhdr}
\usepackage{adjustbox}
\usepackage{siunitx}
\usepackage{pifont}

\def\BibTeX{{\rm B\kern-.05em{\sc i\kern-.025em b}\kern-.08em
    T\kern-.1667em\lower.7ex\hbox{E}\kern-.125emX}}

\usepackage{subcaption}
\usepackage{setspace}
\usepackage{dblfloatfix}
\usepackage{xspace}
\usepackage[binary-units=true]{siunitx}
\usepackage[us,12hr]{datetime}
\usepackage{makecell}
\usepackage[bookmarks=true,breaklinks=true,letterpaper=true,colorlinks,citecolor=blue,linkcolor=blue,urlcolor=blue]{hyperref}
\usepackage{marginnote} 
\usepackage{cleveref}
\crefformat{section}{\S#2#1#3} 
\crefformat{subsection}{\S#2#1#3}
\crefformat{subsubsection}{\S#2#1#3}

\newcommand{\ignore}[1]{}
\newcommand{\ignorerev}[1]{}

\newcommand{\trcd}{\texttt{{tRCD}}\xspace}
\newcommand{\tras}{\texttt{{tRAS}}\xspace}

\newcommand{\trp}{\texttt{{tRP}}\xspace}

\newcommand{\trfc}{\texttt{{tRFC}}\xspace}
\newcommand{\trefi}{\texttt{{tREFI}}\xspace}
\newcommand{\tcl}{\texttt{{tCL}}\xspace}

\newcommand{\trefw}{\texttt{{tREFW}}\xspace}

\newcommand{\cmdact}{\texttt{{ACT}}\xspace}
\newcommand{\cmdread}{\texttt{{RD}}\xspace}
\newcommand{\cmdwrite}{\texttt{{WR}}\xspace}
\newcommand{\cmdpre}{\texttt{{PRE}}\xspace}
\newcommand{\cmdrefresh}{\texttt{{REF}}\xspace}

\newcommand{\ARI}{\texttt{{ARI}}\xspace}

\newcommand{\actnack}{\texttt{ACT\_NACK}}
\newcommand{\tnackplain}{T_{\actnack{}}}
\newcommand{\tnack}{$\tnackplain$\xspace}

\newcommand{\mechunformatted}{SMD}
\newcommand{\mech}{\mechunformatted{}}
\newcommand{\mechlong}{Self-Managing DRAM}

\newcommand{\fr}{{SMD-FR}}
\newcommand{\vr}{{SMD-VR}}
\newcommand{\prp}{{SMD-PRP}}
\newcommand{\prpplus}{{SMD-PRP+}}
\newcommand{\drp}{{SMD-DRP}}
\newcommand{\sms}{{SMD-MS}}

\newcommand{\squishlist} {
    \begin{list}{$\bullet$} {
        \setlength{\itemsep}{-1pt}
        \setlength{\parsep}{2pt}
        \setlength{\topsep}{0pt}
        \setlength{\partopsep}{0pt}
        \setlength{\leftmargin}{1.0em}
        \setlength{\labelwidth}{1em}
        \setlength{\labelsep}{0.5em}
    }
}

\newcommand{\squishend} {
    \end{list}
}

\newif\ifcameraready
\camerareadytrue

\newif\ifiterations
\iterationsfalse

\newif\ifiscaadds
\iscaaddsfalse

\newif\ifhpcarevision
\hpcarevisionfalse

\newif\ifhpcalastminute
\hpcalastminutefalse

\newif\ifrevision
\revisionfalse

\newif\ifiscarevision
\iscarevisionfalse

\newif\ifmicroadds
\microaddsfalse

\newif\ifmicrorevision
\microrevisionfalse

\usepackage{soul}
\soulregister\cite7
\soulregister\ref7

\colorlet{LightYellow}{yellow!60!white}
\colorlet{LightBlue}{yellow!60!white}
\DeclareRobustCommand{\bgyellow}[1]{{#1}}

\definecolor{jazzberryjam}{rgb}{1.0, 0.65, 0.0}
\definecolor{mediumpersianblue}{rgb}{0.0, 0.4, 0.65}
\definecolor{coolblack}{rgb}{0.0, 0.18, 0.39}
\definecolor{bleudefrance}{rgb}{0.19, 0.55, 0.91}
\definecolor{ao}{rgb}{0.0, 0.0, 1.0}
\definecolor{babyblueeyes}{rgb}{0.63, 0.79, 0.95}

\usepackage[most]{tcolorbox}
\newtcolorbox{yellowbox}{
    colframe=yellow!60!white,
    colback =yellow!60!white,
        grow to left by=4mm,
        grow to right by=4mm,
        boxrule=0mm,
        before skip =0mm,
        after skip =0mm,
    top=0mm, bottom=0mm, right=3.25mm, left=3.25mm,
}

\newtcolorbox{bluebox}{
    colframe=LightBlue,
    colback =LightBlue,
        grow to left by=4mm,
        grow to right by=4mm,
        boxrule=0mm,
        before skip =0mm,
        after skip =0mm,
    top=0mm, bottom=0mm, right=3.25mm, left=3.25mm,
}

\newtcolorbox{yellowbox_break}{
    breakable,
    colframe=yellow!60!white,
    colback =yellow!60!white,
        grow to left by=4mm,
        grow to right by=4mm,
        boxrule=0mm,
        before skip =0mm,
        after skip =0mm,
    top=0mm, bottom=0mm, right=3.25mm, left=3.25mm,
}

\newlength\MyIndent
\ifcameraready
\newenvironment{yellowb}{}{}
\newenvironment{yellowb_break}{}{}
\else
\newenvironment{yellowb}
    {\begin{yellowbox}\setlength\parindent{\MyIndent}}
    {\end{yellowbox}}

\fi

\definecolor{dollarbill}{rgb}{0.52, 0.73, 0.4} 
\definecolor{amber}{rgb}{1.0, 0.49, 0.0}

\newcounter{version}
\ifiterations
\else
\fi

\ifiterations
    \newcommand{\atbcr}[2]{\ifnum#1=\value{version}\textcolor{red}{#2}\else{#2}\fi}
    \newcommand{\omcr}[2]{\ifnum#1=\value{version}\textcolor{blue}{#2}\else{#2}\fi}
    \newcommand{\agycr}[2]{\ifnum#1=\value{version}\textcolor{orange}{#2}\else{#2}\fi}
    \newcommand{\atbcrcomment}[2]{\ifnum#1=\value{version}\todo[size=\scriptsize, linecolor=orange, bordercolor=orange, backgroundcolor=white]{\textcolor{red}{Atb:~#2}}\else{}\fi}
    \newcommand{\omcrcomment}[2]{\ifnum#1=\value{version}\todo[size=\scriptsize, linecolor=orange, bordercolor=orange, backgroundcolor=white]{\textcolor{blue}{Onur:~#2}}\else{}\fi}
    \newcommand{\agycrcomment}[2]{\ifnum#1=\value{version}\todo[size=\scriptsize, linecolor=orange, bordercolor=orange, backgroundcolor=white]{\textcolor{orange}{Agy:~#2}}\else{}\fi}
    \paperwidth=\dimexpr \paperwidth + 2cm\relax
    \oddsidemargin=\dimexpr\oddsidemargin + 1cm\relax
    \evensidemargin=\dimexpr\evensidemargin + 1cm\relax
    \marginparwidth=\dimexpr \marginparwidth + 1cm\relax
\fi

\ifcameraready
    \newcommand{\atbcr}[2]{\ifnum#1=\value{version}\textcolor{red}{#2}\else{#2}\fi}
    \newcommand{\omcr}[2]{\ifnum#1=\value{version}\textcolor{blue}{#2}\else{#2}\fi}
    \newcommand{\agycr}[2]{\ifnum#1=\value{version}\textcolor{orange}{#2}\else{#2}\fi}
    \newcommand{\atbcrcomment}[2]{\ifnum#1=\value{version}\todo[size=\scriptsize, linecolor=orange, bordercolor=orange, backgroundcolor=white]{\textcolor{red}{Atb:~#2}}\else{}\fi}
    \newcommand{\omcrcomment}[2]{\ifnum#1=\value{version}\todo[size=\scriptsize, linecolor=orange, bordercolor=orange, backgroundcolor=white]{\textcolor{blue}{Onur:~#2}}\else{}\fi}
    \newcommand{\agycrcomment}[2]{\ifnum#1=\value{version}\todo[size=\scriptsize, linecolor=orange, bordercolor=orange, backgroundcolor=white]{\textcolor{orange}{Agy:~#2}}\else{}\fi}

    \newcommand{\hh}[1]{{#1}\xspace}
    \newcommand{\hht}[1]{{#1}\xspace}
    \newcommand{\hhf}[1]{{#1}\xspace}
    \newcommand{\hext}[1]{{{#1}}\xspace}
    \newcommand{\hhh}[1]{{{#1}}\xspace}
    \newcommand{\hhi}[1]{{{#1}}\xspace}
    \newcommand{\agy}[1]{{#1}\xspace}
    \newcommand{\atbcomment}[1]{}
    \newcommand{\atb}[1]{{#1}\xspace}
    \newcommand{\atbt}[1]{{#1}\xspace}
    \newcommand{\agycomment}[1]{{}}
    \newcommand{\param}[1]{{#1}}

    \newcommand{\reva}[1]{#1}
    \newcommand{\revb}[1]{#1}
    \newcommand{\revc}[1]{#1}
    \newcommand{\revd}[1]{#1}
    
    \newcommand{\revcommon}[1]{#1}

    \newcommand{\revlabel}[1]{}
    
    \newcommand{\hpcanew}[1]{#1}

    \newcommand{\hpcarevcommon}[1]{{#1}}
    \newcommand{\hpcareva}[1]{{#1}}
    \newcommand{\hpcarevb}[1]{{#1}}
    \newcommand{\hpcarevc}[1]{{#1}}
    \newcommand{\hpcarevd}[1]{{#1}}
    
    \newcommand{\hpcarevlabel}[1]{}

\fi

\ifhpcarevision
    \definecolor{goodgreen}{rgb}{0.0, 0.5, 0.0} 
    \definecolor{brightpink}{rgb}{1.0, 0.0, 0.5}

    \renewcommand{\hpcarevcommon}[1]{\textcolor{blue}{#1}}
    \renewcommand{\hpcareva}[1]{\textcolor{red}{#1}}
    \renewcommand{\hpcarevb}[1]{\textcolor{goodgreen}{#1}}
    \renewcommand{\hpcarevc}[1]{\textcolor{orange}{#1}}
    \renewcommand{\hpcarevd}[1]{\textcolor{purple}{#1}}

    \paperwidth=\dimexpr \paperwidth + 2cm\relax
    \oddsidemargin=\dimexpr\oddsidemargin + 1cm\relax
    \evensidemargin=\dimexpr\evensidemargin + 1cm\relax
    \marginparwidth=\dimexpr \marginparwidth + 1cm\relax
    
    \usepackage[colorinlistoftodos,prependcaption,textsize=small]{todonotes}
    \usepackage{xargs}
    \renewcommandx{\hpcarevlabel}[2][1=]{\todo[linecolor=blue,backgroundcolor=blue!25,bordercolor=blue,#1,size=\scriptsize]{#2}}
\fi

\ifrevision
    \definecolor{goodgreen}{rgb}{0.0, 0.5, 0.0}
    \definecolor{carrotorange}{rgb}{0.93, 0.57, 0.13}    
    \definecolor{brightpink}{rgb}{1.0, 0.0, 0.5}
    \newcommand{\reva}[1]{\textcolor{red}{#1}}
    \newcommand{\revb}[1]{\textcolor{goodgreen}{#1}}
    \newcommand{\revc}[1]{\textcolor{carrotorange}{#1}}
    \newcommand{\revd}[1]{\textcolor{purple}{#1}}
    
    \newcommand{\revcommon}[1]{\textcolor{blue}{#1}}
    
    \let\marginpar\marginnote
    \newcommand{\revlabel}[1]{\todo[size=\scriptsize, linecolor=blue, bordercolor=blue, backgroundcolor=white]{\textcolor{blue}{\textbf{#1}}}}
\fi

\ifiscarevision
    \paperwidth=\dimexpr \paperwidth + 2cm\relax
    \oddsidemargin=\dimexpr\oddsidemargin + 1cm\relax
    \evensidemargin=\dimexpr\evensidemargin + 1cm\relax
    \marginparwidth=\dimexpr \marginparwidth + 1cm\relax
  \newcommand{\cql}[2]{#2\todo[size=\small,color=orange]{\textbf{\textcolor{white}{#1}}}}
  \newcommand{\iql}[2]{#2\todo[size=\small,color=babyblueeyes]{\textbf{{#1}}}}
  
  \newcommand{\iqrev}[1]{\textcolor{ao}{{#1}}}
\else
  \newcommand{\cql}[2]{}
  \newcommand{\iql}[2]{}
  
  \newcommand{\iqrev}[1]{#1}
\fi

\ifmicrorevision
    \definecolor{goodgreen}{rgb}{0.0, 0.5, 0.0} 
    \definecolor{brightpink}{rgb}{1.0, 0.0, 0.5}

    \newcommand{\microrevcommon}[1]{\textcolor{blue}{#1}}
    \newcommand{\microreva}[1]{\textcolor{red}{#1}}
    \newcommand{\microrevb}[1]{\textcolor{goodgreen}{#1}}
    \newcommand{\microrevc}[1]{\textcolor{orange}{#1}}
    \newcommand{\microrevd}[1]{\textcolor{purple}{#1}}
    \newcommand{\microrevf}[1]{\textcolor{brightpink}{#1}}
    
    \paperwidth=\dimexpr \paperwidth + 2cm\relax
    \oddsidemargin=\dimexpr\oddsidemargin + 1cm\relax
    \evensidemargin=\dimexpr\evensidemargin + 1cm\relax
    \marginparwidth=\dimexpr \marginparwidth + 1cm\relax
    
    \usepackage[colorinlistoftodos,prependcaption,textsize=small]{todonotes}
    \usepackage{xargs}
    \newcommandx{\microrevlabel}[2][1=]{\todo[linecolor=blue,backgroundcolor=blue!25,bordercolor=blue,#1,size=\scriptsize]{#2}}
\else

    \newcommand{\microrevcommon}[1]{#1}
    \newcommand{\microreva}[1]{{#1}}
    \newcommand{\microrevb}[1]{{#1}}
    \newcommand{\microrevc}[1]{{#1}}
    \newcommand{\microrevd}[1]{{#1}}
    \newcommand{\microrevf}[1]{{#1}}
    
    \usepackage[colorinlistoftodos,prependcaption,textsize=small]{todonotes}
    \usepackage{xargs}
    \newcommandx{\microrevlabel}[2][1=]{}

\fi

\ifhpcalastminute
    \newcommand{\hpcanew}[1]{\textcolor{blue}{#1}}
    
    \newcommand{\hh}[1]{{#1}\xspace}
    \newcommand{\hht}[1]{{#1}\xspace}
    \newcommand{\hhf}[1]{{#1}\xspace}
    \newcommand{\hext}[1]{{{#1}}\xspace}
    \newcommand{\hhh}[1]{{{#1}}\xspace}
    \newcommand{\hhi}[1]{{{#1}}\xspace}
    \newcommand{\agy}[1]{{#1}\xspace}
    \newcommand{\atbcomment}[1]{}
    \newcommand{\atb}[1]{{#1}\xspace}
    \newcommand{\atbt}[1]{{#1}\xspace}
    \newcommand{\agycomment}[1]{{}}
    \newcommand{\param}[1]{{#1}\xspace}

    \newcommand{\reva}[1]{#1}
    \newcommand{\revb}[1]{#1}
    \newcommand{\revc}[1]{#1}
    \newcommand{\revd}[1]{#1}
    
    \newcommand{\revcommon}[1]{#1}

    \newcommand{\revlabel}[1]{}
    
\fi

\ifiscaadds
    \newcommand{\iscanew}[1]{\textcolor{blue}{#1}}
    \newcommand{\iscaonur}[1]{\textcolor{orange}{#1}}
    \renewcommand{\atbcomment}[1]{\textcolor{ao}{\textbf{[@atb:} #1\textbf{]}}}
    \renewcommand{\param}[1]{\textcolor{red}{\textbf{#1}}}
\else
    \newcommand{\iscanew}[1]{{#1}}
    \newcommand{\iscaonur}[1]{{#1}}
\fi

\ifmicroadds
    \newcommand{\micronew}[1]{\textcolor{orange}{#1}}
    
    \renewcommand{\atbcomment}[1]{\textcolor{ao}{\textbf{[@atb:} #1\textbf{]}}}
    \renewcommand{\param}[1]{\textcolor{red}{\textbf{#1}}}
\else
    \newcommand{\micronew}[1]{{#1}}
    
\fi

\newcommand*\circled[1]{\tikz[baseline=(char.base)]{
            \node[shape=circle,draw,inner sep=.5pt,fill=black,text=white] (char) {\scriptsize #1};}}

\newcommand{\addressdivergence}[1]{#1}
\newcommand{\addressdivergencetwo}[1]{#1}

\newcommand{\rowHammerGetsWorseCitations}[0]{\cite{kim2020revisiting, frigo2020trrespass, yaglikci2022understanding, orosa2021deeper, mutlu2017rowhammer, mutlu2018rowhammer, mutlu2019rowhammer, mutlu2023fundamentally}}

\newcommand{\mitigatingRowHammerAllCitations}[0]{\cite{AppleRefInc, rh-hp,rh-lenovo,greenfield2012throttling, kim2014flipping, kim2014architectural, bains14d, bains14c, aweke2016anvil, bains-merged, son2017making, seyedzadeh2018cbt,irazoqui2016mascat, you2019mrloc, lee2019twice, park2020graphene, yaglikci2021security, yaglikci2021blockhammer, frigo2020trrespass, kang2020cattwo, hassan2021utrr, qureshi2022hydra, saileshwar2022randomized, brasser2017can, konoth2018zebram, van2018guardion, vig2018rapid,  kim2022mithril, lee2021cryoguard, marazzi2022protrr, zhang2022softtrr, joardar2022learning, juffinger2023csi, yaglikci2022hira, saxena2022aqua, enomoto2022efficient, manzhosov2022revisiting, ajorpaz2022evax, naseredini2022alarm, joardar2022machine, zhang2020leveraging,loughlin2021stop, devaux2021method, han2021surround, fakhrzadehgan2022safeguard, saroiu2022price, saroiu2022configure, loughlin2022moesiprime, zhou2022lt, hong2023dsac, mutlu2023fundamentally, marazzi2023rega, di2023copy, sharma2022review, woo2023scalable, park2022row, wi2023shadow, kim2023ddr5, gude2023defending, guha2022criticality, france2022modeling, france2022reducing, bennett2021panopticon, arikan2022processor, tomita2022extracting, saxena2023pt, zhou2023dnndefender, woo2023rampart, kim2023how, olgun2024abacus, yaglikci2024spatial, bostanci2024comet, saroiu2024ddr5, saxena2024start, jedecddr5c, canpolat2024understanding,jaleel2024pride,saxena2024rubix}}
\newcommand{\valARI}{62.5} 

\newcommand{\affilETH}[0]{\textsuperscript{$\dagger$}}

\author{
{Hasan Hassan\affilETH}\qquad%
{Ataberk Olgun\affilETH}\qquad
{A. Giray Ya\u{g}l{\i}k\c{c}{\i}}\qquad
{Haocong Luo}\qquad%
{Onur Mutlu}\qquad\vspace{-3mm}\\\\
{\emph{ETH Z{\"u}rich}} %
}

\newcommand\extrafootertext[1]{%
    \bgroup
    \renewcommand\thefootnote{\affilETH}%
    \renewcommand\thempfootnote{\affilETH}%
    \footnotetext[0]{#1}%
    \egroup
}

\newcommand{\papertitle}[0]{\hhi{Self-Managing DRAM: A Low-Cost Framework for\\Enabling Autonomous and Efficient DRAM Maintenance Operations}}
\title{\Large{\papertitle{}}} 

\begin{document}

\maketitle
\extrafootertext{H. Hassan and A. Olgun are co-primary authors.}

\ifiterations
    \fancyhead{}
    \fancypagestyle{firstpage}
    {
        \fancyhead{}
        \fancyhead[C]{\textcolor{red}{CONFIDENTIAL DRAFT -- DO NOT DISTRIBUTE -- TO APPEAR IN MICRO'24} \\ \textcolor{blue}{\emph{Version 2.1~---~\today, \ampmtime}} }
    }
    \pagenumbering{arabic}
    \thispagestyle{firstpage}
    \fancyhead[C]{\textcolor{blue}{\emph{Version 2.1~---~\today, \ampmtime}}}
    \pagestyle{fancy}
\else 
    \pagenumbering{arabic}
    \renewcommand{\headrulewidth}{0pt}
    \fancyhf{} 
    \fancyfoot{} 
    \thispagestyle{plain}
    \pagestyle{plain}
\fi

\bstctlcite{IEEEexample:BSTcontrol}


\begin{abstract}

    \bgyellow{The memory controller is in charge of managing DRAM maintenance
    operations (e.g., refresh, RowHammer protection, memory scrubbing) \iscaonur{to reliably operate modern} DRAM chips. Implementing new maintenance operations often
    necessitates modifications in the DRAM interface, memory controller, and
    potentially other system components. Such modifications are only possible
    with a new DRAM standard, which takes a long time to develop, \iscaonur{likely} leading to
    slow progress in \iscaonur{the adoption of new architectural techniques in} DRAM \iscaonur{chips}.}

    %


    We propose a new low-cost DRAM architecture, \mechlong{} 
    (\mech{}), \iscaonur{that enables autonomous in-DRAM maintenance
    operations by transferring the responsibility for controlling maintenance operations from the memory controller to the \mech{} chip. 
    \iscaonur{To enable autonomous maintenance operations, we \omcr{2}{make} a single, simple modification 
    \omcr{2}{to the DRAM interface},
    such that an \mech{} chip rejects memory controller
    accesses to DRAM regions (e.g., a subarray or a bank) under maintenance, while
    allowing memory accesses to other DRAM regions.}
    Thus, \mech{} enables 1)} implementing new in-DRAM maintenance 
    mechanisms \hht{(or modifying existing ones)} with no \iscaonur{further} 
    changes in the DRAM 
    interface, memory controller, or other system components\iscaonur{, and 2) 
    overlapping the latency of a maintenance operation in one DRAM region with the latency of accessing data in another}.
    
    \iscaonur{We evaluate \mech{} and show that it 1)} 
    can be implemented \emph{without} \atbcr{2}{adding new pins}\atbcrcomment{2}{This is what we meant. Does it look better?} to the DDRx 
    interface \micronew{with low latency (0.4\% of row activation latency) and area 
    \atbcrcomment{3}{This does not assume new pins}
    (1.\atbcr{2}{1}\% of a \SI{45.5}{\square\milli\meter} DRAM chip) overhead}, 
    2) \iscaonur{achieves \param{4.1}\% average speedup across 20 four-core memory-intensive workloads over a \micronew{DDR4-based system/DRAM co-design technique} that intelligently parallelizes maintenance operations with memory accesses,} and 3) \iscaonur{\micronew{guarantees} forward progress for rejected memory accesses}. \omcr{2}{We believe and hope SMD can enable 
    innovations in DRAM architecture to rapidly come to fruition. We open source all SMD source code and 
    data at \url{https://github.com/CMU-SAFARI/SelfManagingDRAM}.}

\end{abstract}


\vspace{3mm}

\section{Introduction}
\label{sec:intro}

Advances in manufacturing technology 
enable
increasingly smaller DRAM cell \hht{sizes}, continuously reducing cost per bit
of a DRAM chip~\cite{lee2017design,mutlu2013memory, mutlu2014research, patel2024rethinking,
nair2013archshield,patel2022case}. However, as a DRAM cell becomes smaller and the distance
between adjacent cells shrinks, ensuring reliable \hh{and efficient} DRAM
operation \hht{becomes} an even more critical
challenge~\cite{awasthi2012efficient,cha2017defect,hong2010memory,kang2014co,kim2020revisiting,kim2014flipping,
lee2016technology, mandelman2002challenges, mutlu2013memory,
mutlu2019rowhammer,nair2013archshield, park2015technology,qureshi2015avatar,mutlu2023fundamentally}. A
modern DRAM chip requires three \omcr{2}{major} types of maintenance operations \revcommon{(described in detail in \cref*{sec:maintenance_mechanisms})} for
reliable and secure operation: \revcommon{1) \hhh{DRAM} refresh~\cite{chang2014improving, liu2012raidr,
qureshi2015avatar,nair2014refresh, baek2014refresh, bhati2013coordinated,
cui2014dtail, emma2008rethinking, ghosh2007smart, isen2009eskimo,
jung2015omitting, kim2000dynamic, luo2014characterizing, kim2003block,
liu2012flikker, mukundan2013understanding, nair2013case, patel2005energy,
stuecheli2010elastic, khan2014efficacy, khan2016parbor, khan2017detecting,
venkatesan2006retention, patel2017reaper, riho2014partial, hassan2019crow,
kim2020charge, nguyen2018nonblocking, kwon2021reducing}, 2) RowHammer protection~\mitigatingRowHammerAllCitations{}, and
3) memory scrubbing~\cite{jacob2010memory,
mukherjee2004cache,schroeder2009dram,meza2015revisiting, saleh1990reliability,
siddiqua2017lifetime, rooney2019micron, gong2018duo,han2014data,qureshi2015avatar,choi2020reducing,
sharifi2017online,alameldeen2011energy,naeimi2013sttram}.\footnote{Memory 
scrubbing is not \hht{always} required in
consumer systems but often used in cloud systems.}}
New DRAM chip \hht{generations
necessitate} making existing maintenance operations more \hhh{aggressive} (e.g.,
lowering the refresh period~\cite{jedecddr5c, lpddr5, rh-apple}) and
introducing new types of maintenance operations (e.g., targeted
refresh~\cite{hassan2021utrr,frigo2020trrespass,jattke2022blacksmith},
\hht{DDR5 RFM~\cite{jedecddr5c}, \omcr{2}{and PRAC~\cite{jedecddr5c}} as \atbt{{RowHammer}} defenses}).\atbcrcomment{3}{What about here?}\footnote{
\atbcr{2}{A very recent update to the DDR5 standard~\cite{jedecddr5c} introduces PRAC,
which is an on-DRAM-die read disturbance mitigation mechanism.
PRAC requires more changes to the DRAM interface and continues to use RFM. 
Note that PRAC is concurrent with this work, as the initial version of this 
paper~\cite{hassan2024smd} 
was placed on arXiv on 27 July 2022 and initial submission to \omcr{4}{the} MICRO \omcr{4}{2022} 
conference was made on \atbcr{3}{22 April} 2022.}
}

\iscanew{Two problems \iscaonur{likely} hinder \omcr{2}{the adoption of} 
effective and efficient maintenance mechanisms in
modern and future DRAM-based computing systems. \iql{IQE1}{}\iqrev{First, 
it is difficult to modify existing maintenance mechanisms and introduce 
new maintenance operations \iscaonur{because doing so often necessitates 
changes to the DRAM interface, which takes a long time (due to various 
issues related to standardization and agreement across many vendors
\omcr{2}{with conflicting interests~\cite{patel2024rethinking, patel2022case}})}.} 
Second, it is challenging to keep the overhead of DRAM maintenance mechanisms 
low as DRAM reliability characteristics worsen and DRAM chips require more 
aggressive maintenance operations. We expand on the two problems in the next two paragraphs.}




\bgyellow{Implementing new maintenance
operations (or changing existing ones) often necessitates modifications in the
DRAM interface, \microrevcommon{memory controller (MC)}, and potentially other system components. \iql{IQE1}{}\iqrev{Such modifications
are only possible with a new DRAM standard, which takes a long time to develop,
\iscaonur{likely} leading to slow progress in \iscaonur{the adoption of new architectural techniques in} DRAM \iscaonur{chips}.} 
For example, there was a five-year gap
between the DDR3~\cite{jedec2012ddr3} and DDR4~\cite{jedec2020ddr4} standards, and
an eight-year gap between the DDR4~\cite{jedec2020ddr4} and DDR5~\cite{jedecddr5c}
standards. While developing a new standard, DRAM vendors need to push their
proposal regarding the maintenance operations through the JEDEC committee \iql{IQE1}{}\iqrev{(which is the case even if the gap between subsequent DRAM standards were to substantially reduce)} \iscanew{such 
that} the new standard includes the desired changes to enable new maintenance 
operations. Thus, a flexible DRAM interface that
\iscaonur{decouples improvements to DRAM maintenance operations from interface changes (and hence the timeline needed to enable such changes) would}
allow DRAM \iscanew{designers} 
to quickly 
implement custom in-DRAM maintenance mechanisms and develop more \omcr{2}{robust,}
reliable\omcr{2}{,}
secure\omcr{2}{, safe, and more efficient} DRAM chips.}

\iscanew{A maintenance operation triggers more frequently and takes 
longer with worsening DRAM reliability characteristics, increasing density, \microreva{or increasing operating temperature with tighter DRAM-system integration (e.g., high-bandwidth memory DRAM temperature can reach \SI{97}{\celsius}~\cite{ramalingam2016hbm})}\microrevlabel{AQ3}.\atbcrcomment{2}{cannot find a bigger number}
While a maintenance operation executes, DRAM is unavailable to serve memory requests
(e.g., a periodic refresh operation in DDR4 prevents access to a DRAM rank
for \SI{410}{\nano\second}~\cite{jedec2020ddr4} and \atbcr{2}{in DDR5 prevents access
to a DRAM bank in all bank groups for \SI{190}{\nano\second}}~\cite{jedecddr5c}).
As a result, maintenance operations can significantly degrade system 
performance by delaying critical memory requests. \iscaonur{Enabling autonomous maintenance operations} without delaying critical memory
requests would improve system performance, energy efficiency, \atbcr{2}{and robustness}.
}

\omcr{2}{\agycr{2}{Enabling} autonomous operations with a flexible interface \agycr{2}{would provide two main benefits. First, it} would enable \agycr{2}{DRAM} manufacturers 
with \omcr{3}{\emph{breathing room}} to perform architectural optimizations that can greatly benefit 
from \agycr{2}{their DRAM-internal proprietary} knowledge
without 
having to expose 
\agycr{2}{such knowledge to off-chip components and third-party manufacturers, which may be impractical or undesirable~\cite{patel2024rethinking,patel2022case}.}
\agycr{2}{Second,}
\agycr{2}{doing so} divides the work nicely between the \agycr{2}{memory controller and the DRAM chip}\atbcr{2}{. DRAM}
manufacturers 
can focus on \agycr{2}{their critical challenges of designing better DRAM chips} 
\atbcr{2}{and}
memory controller manufacturers can focus on the critical tasks for 
memory controller without being burdened by other tasks that 
are not really easy to optimize at the \agycr{2}{memory controller level without 
proprietary DRAM-internal knowledge.}}

\textbf{Our goal} is to \hht{1) ease and accelerate the process of
implementing new in-DRAM maintenance operations and 2) enable more efficient maintenance operations.} \iscanew{To this end,} we propose \mechlong{} 
(\mech{}), a new \atbcr{2}{framework} that enables \iscanew{implementing} new \microrevb{in-}DRAM\microrevlabel{BQ1}
maintenance operations and modif\iscanew{ying} the existing ones \iscaonur{with a single, simple interface change that eliminates the need for future} changes in the DRAM interface, the MC, or other
system components.\atbcrcomment{2}{Cite in footnote}\footnote{\omcr{3}{The current
commodity DRAM interface (e.g., 
DDRx~\cite{jedec2012ddr3,jedec2020ddr4,jedecddr5c}, HBMx~\cite{jedec2015hbm,jedec2021hbm}) 
is completely processor centric (i.e., \omcr{4}{the memory controller}
on the processor chip 
dictates everything \omcr{4}{to the DRAM chips, giving} no breathing room for DRAM to do much \omcr{4}{on its own}). 
SMD, by slightly modifying the interface, enables the 
interface to move towards and more memory-centric approach 
where memory can do things autonomously. This is in line 
with the general memory-centric computing paradigm~\cite{ghose.ibmjrd19,mutlu2019processing,mutlu2020modern} where 
memory can also do computation.}}

\micronew{\mech{}'s \textbf{key idea}} is to allow a DRAM chip to \hht{autonomously}
\iscanew{and efficiently} perform maintenance operations \hht{by preventing memory} 
accesses \iscanew{\emph{only}} to a \iscanew{relatively small,} under-maintenance \emph{DRAM
region}\iscanew{, i.e., a designated section in a DRAM chip (e.g., a DRAM subarray or a bank)}, 
\iscanew{while allowing memory accesses to other DRAM regions}.\footnote{\hht{\iscanew{\mech{} has the key property that} the MC 
does \emph{not} even know which maintenance operations an \mech{} chip performs.} 
\hpcarevlabel{HPCA Reviewer D}\hpcarevd{This property of \mech{} could also be 
appealing to DRAM vendors~\cite{patel2024rethinking, patel2022case}
because it would allow vendor-, and chip-generation-specific DRAM resilience characteristics
(e.g., the degree and distribution of RowHammer vulnerability and retention failures)
to remain undisclosed (i.e., not exposed to other DRAM vendors or system designers).}}  
To prevent access to \iscanew{an under-maintenance} DRAM region, \hh{an \mech{} chip
\agy{\emph{rejects}}} a row activation command (i.e., \cmdact{}) issued by the
MC \iscanew{to the region.} 
To ensure that a DRAM region cannot be locked for too long, \mech{} enforces
a minimum delay between consecutive maintenance operations targeting the same region.
\ignorerev{The MC periodically re-issues a previously-rejected \cmdact{} command with a period of \ARI{} to 
serve a high-priority memory request. Thus, a memory request to a locked memory region is serviced
immediately after the the maintenance operation ends. The maximum stall time for such a memory request
is bounded by the the most time taking maintenance operation.}We comprehensively study \mech{} and show 
that it ensures forward progress for memory requests (\cref{subsec:forward-progress}).
\atb{While a DRAM region is under maintenance, the MC can access \atbt{\emph{other}}
regions. This \iscanew{way, a majority of memory accesses are \emph{not} delayed
by maintenance operations as \mech{} overlaps} the latency of a maintenance operation 
\iscanew{in a DRAM region} with memory accesses \iscanew{to other DRAM regions}.} 

%


\ignorerev{\mech{} requires simple changes
in the MC to enable re-issuing a previously-rejected \cmdact{} command, {but
at the same time it} simplifies MC design as {the MC} is no longer
responsible for managing DRAM maintenance operations.}
\hh{To enable practical adoption of \mech{}, we \omcr{1}{propose to} implement it with low-cost
modifications \omcr{1}{to} existing DRAM chips and MCs.} 
\hht{First, to inform the MC that a row activation is rejected, \iql{IQA3/\\IQD1}{}\iqrev{\mech{} \atb
{uses a} single uni-directional pin \atb{on the} DRAM chip \atbt{(which already exists in
DDR4/5, see~\cref{sec:hw_overhead})}}.
Second, \mech{} implements a \atbcr{1}{small (e.g., 16 bits per DRAM bank) memory structure called the} \emph{Lock Controller} \iscanew{in a DRAM chip} for managing regions under
maintenance and preventing access to them. 
\iscanew{Third, \mech{} adds a new row address latch for every lock region 
in a DRAM chip to enable accessing one lock region 
while another is under maintenance (building on the basic design proposed 
in prior works~\cite{kim2012case,zhang2014cream,chang2014improving}).}
\iql{IQD1}{}\iqrev{In \cref{sec:hw_overhead}, we show
that the\iscanew{se modifications} have low DRAM chip area \atbcrcomment{2}{I revised the area overhead analysis to use
44 bits instead of 64 bits (which yields 1.6\%) following SALP's analysis instead for per-region latches.}
(1.\atbcr{2}{1}\% of
a \SI{45.5}{\milli\meter\squared} DRAM chip) and \atbcr{1}{negligible} latency overhead (\emph{only} 0.4\%
\omcr{1}{increase in} row activation latency).}}

\vspace{1mm}
\atbcrcomment{3}{TODO: Double check all numbers.}
We demonstrate the \omcr{2}{usefulness} and versatility of \mech{} by implementing
\atbcr{1}{three} \iscanew{in-DRAM maintenance} mechanisms for DRAM refresh (\cref{subsec:smd_refresh}), RowHammer protection (\cref{subsec:smd_rowhammer_protection}), 
and memory scrubbing (\cref{subsec:smd_ecc_scrubbing}). 
\iscanew{We rigorously evaluate \mech{}'s impact on system performance and energy efficiency using
cycle-accurate memory system simulations (using Ramulator~\cite{ramulatorgithub,kim2015ramulator}), executing
a diverse set of 62 single-core and 60 four-core workloads.}
\iscanew{We compare \mech{}-based implementations of the evaluated maintenance mechanisms to 
their MC-based implementations. We make \param{\atbcr{3}{two}} key observations from our evaluation.}
\iscanew{First, an \mech{} chip that implements lock regions at \emph{subarray} granularity (i.e., a maintenance mechanism
locks a small designated section in a bank) and implements all three maintenance mechanisms 
(refresh, RowHammer protection, and scrubbing)
1)~improves average performance across all memory-intensive four-core workloads by \param{8.6}\% 
compared to a baseline system
with a DDR4 chip that supports bank-level refresh and a memory 
controller that intelligently schedules maintenance operations\atbcrcomment{2}{the 
first one is DARP for maintenance operations, second one is DSARP for maintenance operations.}
to avoid delaying main memory accesses (DARP~\cite{chang2014improving})
and
2)~outperforms a\omcr{2}{n optimized state-of-the-art} system in which the 
DDR4 chip can concurrently perform a maintenance operation in a subarray and a memory access in another subarray
(maintenance-access parallelization) and the memory controller can intelligently exploit maintenance-access parallelization (e.g., similar to DSARP's~\cite{chang2014improving} refresh-access parallelization) by \param{4.1}\%.} \microrevb{SMD provides 88.3\% of the speedup provided \omcr{2}{by an oracle (No-Refresh)} that completely mitigates the overhead of maintenance operations, on average across all evaluated \atbcr{2}{high memory intensity four core} workloads.}\microrevlabel{BQ2}
\iscanew{Second, \mech{} reduces DRAM energy 
consumption by \param{4.3}\% on average as it eliminates DRAM commands
issued by the memory controller to perform maintenance operations and reduces total execution time.}
\iscanew{We conclude that SMD practically provides better performance and energy benefits, \emph{without} having to modify the DRAM interface (e.g., by adding new timing parameters to the standard)
for every new type of maintenance mechanism, than \omcr{3}{prior} techniques 
that enable refresh-access parallelization (e.g.,~\cite{chang2014improving,zhang2014cream}).}
\atbcrcomment{2}{The comparison here is worded correctly. Does it still sound as if we have two baselines?}
\atbcrcomment{4}{Will highlight our sensitivity study results
for the extended version here}
\ignorerev{
\iscanew{Third, a \iscaonur{very low chip area overhead (\param{0.001\%} of a \SI{45.5}{\square\milli\meter} DRAM chip) 
implementation of} \mech{} that locks regions 
at bank granularity (i.e., a maintenance mechanism 
locks a\omcr{2}{n entire} DRAM bank) and performs only refresh operations for maintenance (similar to DDR4)
induces \param{4.5}\% average slowdown due to rejected row activations across all memory-intensive four-core workloads
compared to the same baseline system (DARP~\cite{chang2014improving}).}}

We \agy{expect and} hope that \mech{} will inspire researchers to develop new
DRAM maintenance mechanisms \iscanew{that more efficiently tackle
the \omcr{2}{scaling, robustness, and efficiency} problems of DRAM} and to 
enable practical adoption of innovative ideas by
\hht{providing freedom to both DRAM and MC designers}. \iql{IQB2}{}\iqrev{We 
\atbcr{1}{release} all \mech{} source code 
and data at \url{https://github.com/CMU-SAFARI/SelfManagingDRAM} so that 
others can replicate \omcr{2}{our work} and build on \omcr{2}{it}.}

We make the following \textbf{key contributions}:
\squishlist
    \item We propose \mech{}, \hh{a new \atbcr{1}{framework} to
    enable autonomous and efficient in-DRAM} maintenance operations
    \atbcr{1}{with} small changes \hht{to modern} DRAM chips and MCs.

    \item We use \mech{} to implement \hh{efficient} DRAM maintenance mechanisms
    for three use cases: \hhh{DRAM} refresh, RowHammer protection, and memory
    scrubbing. \omcr{3}{We also show that SMD can seamlessly
    enable optimized \omcr{4}{maintenance mechanisms, including} variable-rate refresh and probabilistic
    RowHammer protection techniques.}

    \item We rigorously evaluate the performance and energy of \micronew{\mech{}-based}
    maintenance mechanisms. \hht{\mech{} provides large performance and energy
    benefits while also improving \atbcr{1}{system} \omcr{2}{robustness} across a variety of 
    workloads.}

\squishend 
\section{Background}
\label{sec:background}
\subsection{DRAM Organization}
\label{subsec:dram_organization}

\begin{figure}[!h]
    \centering
    \includegraphics[width=\linewidth]{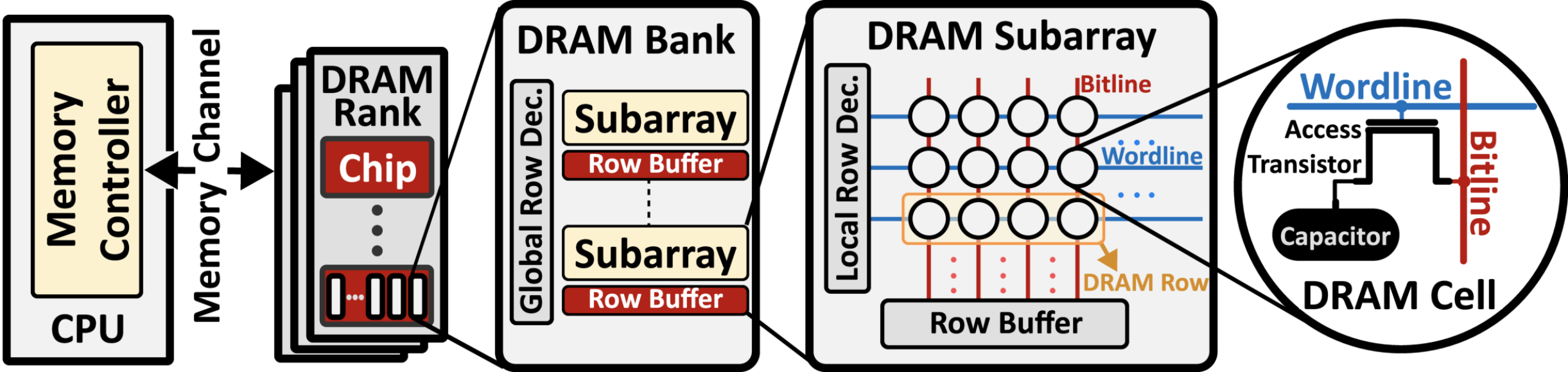}
    \caption{DRAM \atbcr{1}{rank, chip, bank, and subarray organization}}
    \label{fig:dram_organization}
\end{figure}

A computing system has one or more \emph{DRAM channels}, where each
channel has an independently operating I/O bus. As
Fig.~\ref{fig:dram_organization} illustrates, a \hht{Memory Controller (MC)}
interfaces with one or multiple \emph{DRAM ranks} via the channel's I/O bus. A
DRAM rank consists of a group of DRAM chips that operate in lockstep. Because
the I/O bus is shared across ranks, \atbcr{1}{memory} accesses 
to different ranks happen in serial
manner. A DRAM chip is divided into multiple \emph{DRAM banks}, each of which is
further divided into multiple two dimensional DRAM cell arrays, called
\emph{subarrays}. Within each subarray, DRAM cells are organized as \emph{rows}
and \emph{columns}. {A DRAM row consists of DRAM cells \atbcr{1}{that are} 
connected to the same \emph{wordline}.} A DRAM cell connects to a \emph{sense
amplifier} via \atbcr{1}{an \emph{access transistor} and} a \emph{bitline}, and all sense amplifiers in the subarray form a
\emph{row buffer}. Within each DRAM cell, the data is stored as \emph{electrical
charge} on the capacitor.

\subsection{Accessing DRAM}
\label{subsec:accessing_dram}

{\atbcr{1}{Reading or updating the value of a DRAM cell requires activating
(i.e., opening)
a DRAM row. To open a row, the memory controller issues a DRAM activate (\cmdact{}) 
command with a row address.}
The \emph{global row
decoder} and the \emph{local row
decoder} select a wordline based on the \atbcr{1}{row} address provided by
the \atbcr{1}{memory controller}.} 
\revcommon{Enabling the wordline copies the row's data to
the row buffer.}

\ignorerev{{Once a wordline is enabled, the row activation process (i.e., loading rows'
data into the row buffer) occurs in two steps. First, each cell in the
activating row share its charge with the corresponding bitline.} Due to the
charge sharing, the bitline voltage slightly deviates from its pre-activation
voltage level. Second, a sense amplifier detects the bitline voltage deviation
and gradually restores the cell capacitor back to its original charge level. The
data in the row buffer becomes accessible by a \cmdread{}/\cmdwrite{} command
when the bitline and cell voltage is restored to a certain level. Therefore, the
MC should follow an \cmdact{} with one or multiple \cmdread{}/\cmdwrite{}
\hht{commands} at least after the \emph{activation latency} (\trcd{}).}

\revcommon{\atbcr{1}{The memory controller can access data 
in the row buffer using DRAM read and write commands (}\cmdread{}/\cmdwrite{}). The
\atbcr{1}{delay between an \cmdact{} and the following \cmdread{} or \cmdwrite{} 
must be larger than \emph{activation latency} (\trcd{}).}}
\hhh{The \atbcr{1}{memory controller can issue} a \cmdpre{} 
command to close \atbcr{1}{the open} row, after which it can
activate a new row from the same bank.} \revcommon{The \atbcr{1}{memory controller
must wait for at least the \emph{restoration latency} (\tras{}) after
issuing an \cmdact{} before issuing a \cmdpre{}.} The memory controller 
\hht{follows} a \cmdpre{} with \hht{an} \cmdact{} to activate a new row at least
after \emph{precharge latency} (\trp{}).}
%
\ignorerev{The \hhh{time interval for issuing a \cmdpre{} after an \cmdact{} must be} at
least \emph{restoration latency} (\tras{}). A DRAM chip performs a precharge
operation in two steps. First, the DRAM chip disables the wordline of the
activated row to disconnect its cells from the bitlines. 
Second, the precharge circuitry, which is part of a sense amplifier,
precharges the bitline to prepare it for the next row activation. The MC
\hht{follows} a \cmdpre{} with \hht{an} \cmdact{} to activate a new row at least
after the \emph{precharge latency} (\trp{}).}

\ignorerev{
\subsection{Periodic DRAM Refresh}
\label{subsec:periodic_dram_refresh}

A DRAM cell capacitor loses charge over time~\cite{liu2013experimental,
liu2012raidr, saino2000impact}. Losing too much charge results in a bit flip. The
\emph{retention time} of a cell is the duration that its capacitor
stores the correct value. Cells in a DRAM chip have different retention times~\cite{hassan2017softmc, 
kang2014co, khan2014efficacy, kim2009new, lee2015adaptive, li2011dram, liu2013experimental,
liu2012raidr, patel2017reaper, qureshi2015avatar, venkatesan2006retention}. However, to
ensure data integrity, all \hht{DRAM cells} are uniformly refreshed at fixed
time intervals (typically 64, 32\hht{, or even \SI{16}{\milli\second}}), called
\emph{refresh window} (\trefw{}). To perform periodic DRAM refresh, the MC
issues a \cmdrefresh{} command at a fixed interval, denoted as \emph{refresh
interval} (\trefi{}) (typically 7.8 or \SI{3.9}{\micro\second}). A DRAM chip
refreshes several (e.g., $16$) rows upon receiving a single \cmdrefresh{}
command, and typically $8192$ \cmdrefresh{} commands refresh the entire DRAM
chip in a refresh window.
}
\section{Motivation \atbcr{1}{\& Goal}}
\label{sec:drawbacks_current_maintenance}

\hh{In current DRAM chips, the \hht{MC} is in charge of
managing DRAM maintenance operations such as periodic refresh, \atbcr{1}{RowHammer} 
protection, and memory scrubbing. When DRAM vendors \atbt{modify} a DRAM
maintenance mechanism \omcr{2}{to improve robustness, performance, or efficiency}, 
\atbcr{1}{often the DRAM interface needs to be modified as well,}  which makes such \omcr{2}{improvements}
\hht{very}
difficult \omcr{2}{to achieve}. As a result, implementing new or modifying existing maintenance
operations\iql{IQE1}{}\iqrev{, no matter how fast such implementations or modifications could be developed,} \hht{\iqrev{can only be realized after a}} multi-year effort by multiple parties that are part of
the JEDEC committee~\cite{patel2022case,patel2024rethinking,jedecjc45}.
A prime example to support our argument is the \hht{most
recent} DDR5 standard~\cite{jedecddr5c}\hht{, which took \omcr{2}{at least 8 years} to develop
after the initial release of DDR4\iscaonur{. Even though it might \emph{not} have taken DRAM designers
\omcr{2}{so long} to develop new maintenance mechanisms, the \atbcr{2}{DRAM chips implementing 
these} mechanisms are \atbcr{2}{released} only after
the new standard is released}}.
DDR5 also introduces changes to \hht{key issues
we study in this paper:} \hhh{DRAM} refresh, \atbt{RH} protection, and memory
scrubbing. We discuss these changes \hht{as motivating examples to show the
shortcomings of the status quo in DRAM}.\atbcrcomment{2}{Cannot find additional
examples in Minesh's arXiv papers.}}

\subsection{\hhh{DRAM} Refresh}

DDR5 introduces \emph{Same Bank Refresh (SBR)}, \hh{which refreshes one bank in
each bank group\atbcrcomment{2}{if you have 8 bank groups, you refresh 8 banks} at a time instead of
refreshing all banks \omcr{2}{concurrently} as in
DDR4~\cite{jedec2020ddr4,micronddr4}. \emph{SBR} improves \hh{bank} availability as the
MC can access the non-refreshing banks while \atbcr{1}{other} 
banks are being refreshed}.
\hh{DDR5 implements \emph{SBR} with a new \texttt{REFsb}
command~\cite{jedecddr5c}. \hhh{Implementing \texttt{REFsb}} necessitates
changes in the DRAM interface.} 

\subsection{\hhh{RowHammer Protection}}

\hh{In DDR4, DRAM vendors implement in-DRAM RowHammer protection mechanisms 
by
performing \atbcr{1}{Target} Row Refresh (TRR)~\cite{frigo2020trrespass,hassan2021utrr}
within the time \atbcr{1}{slack} available
when performing regular refresh. Prior works show that 
TRR is vulnerable to certain memory access
patterns~\cite{hassan2021utrr,frigo2020trrespass,jattke2022blacksmith}.
DDR5 specifies \emph{Refresh Management (RFM)}~\cite{canpolat2024understanding,jedecddr5c} that an MC implements
to aid \hht{in-DRAM RowHammer protection}. As part of RFM,
the MC \hht{uses} counters to keep track of row activations to each
DRAM bank. When a counter reaches a specified threshold value, the MC issues the
new \texttt{RFM} command to DRAM \hht{chips}. \hht{A DRAM chip then 
performs an undocumented operation \microrevlabel{CQ2}\microrevc{(\omcr{2}{one}
that may or may \emph{not} refresh one or \omcr{2}{more} 
victim rows based on an unknown victim row selection criteria)} with the time allowed for \texttt{RFM} to
mitigate RowHammer.}} 
\atbcr{1}{Implementing RFM} requires significant changes in the DRAM interface and\omcrcomment{2}{May want to elevate the footnote to the intro} the MC design.

\subsection{\hhh{Memory} Scrubbing}

DDR5 adds support for on-die ECC~\cite{kang2014co} and in-DRAM scrubbing~\cite{jedecddr5c}. 
\hh{A DDR5 chip
internally performs ECC encoding and decoding when the chip \hht{is accessed}.
To perform DRAM scrubbing, the MC must periodically issue the new
\emph{scrub command} (for manual scrubbing) or \emph{all-bank refresh command}
(for automatic scrubbing). Similar to \emph{Same Bank Refresh} and \emph{RFM},
enabling in-DRAM scrubbing \hht{necessitates} changes in the DRAM interface and
in the MC design.}

\atbcr{1}{\textbf{Our goal} is to accelerate the process of \omcr{2}{introducing} new in-DRAM
maintenance operations and enabl\omcr{2}{ing} more efficient maintenance operations.}\agycrcomment{2}{I think the last sentence belongs to the next section as the first sentence}

\section{Self-Managing DRAM \atbcr{1}{(\mech{})}}
\label{sec:sm_dram}

\atbcr{3}{We propose the Self-Managing DRAM (\mech{}) framework \omcr{2}{that decouples the responsibilities
of the DRAM chip and the memory controller with a single change
to the DRAM interface.}}
\atbcr{1}{Giving a DRAM chip the ability \omcr{2}{(i.e., the \omcr{3}{\emph{breathing room}})} to autonomously
perform maintenance operations renders DRAM interface modifications
for implementing future maintenance operations unnecessary.}
\mech{} is part of a continuum between 
fully master-slave \atbcr{2}{(completely controlled by one side)}
and fully request-reply \atbcr{2}{(equal partner)} interfaces.
\atbcr{2}{Today's \omcr{3}{commodity} DRAM interface \atbcr{3}{(e.g., DDRx~\cite{jedecddr5c,jedec2012ddr3,jedec2020ddr4}, HBMx~\cite{jedec2015hbm,jedec2021hbm})}
is completely on the fully master-slave end of the continuum
\omcr{4}{as the memory controller dictates everything
the DRAM chip does.}
SMD, in contrast,} tries to achieve a balance that enables faster innovation and easier adoption of new ideas \omcr{4}{by
giving the DRAM chip some freedom to autonomously perform
maintenance operations. We believe that doing so}
\omcr{3}{would ultimately more quickly enable better (more performant, efficient, robust) computing
systems.}

\subsection{Overview of \mech{}}
\label{sec:overview}


\mech{} has a flexible interface that enables efficient implementation of
multiple DRAM maintenance operations within the DRAM chip. 
The key idea of \mech{} is to provide a DRAM chip with the ability to
reject an \cmdact{} command via a \hht{single-bit}
\emph{negative-acknowledgment} (\actnack{}) signal. An \actnack{} informs the MC
that the DRAM row it tried to activate is under maintenance\hh{, and thus
temporarily unavailable}. Leveraging the ability to reject an \cmdact{}, a
maintenance operation can be implemented \emph{completely within} a DRAM chip.

\noindent
\textbf{\micronew{Organization}.} \mech{} preserves the general DRAM
interface and \atb{uses} a single
uni-directional pin \micronew{(which already exists in
DDR4/5, see~\cref{sec:hw_overhead})} in the physical interface of a DRAM chip. \atb{This} pin is
used for \atbcr{1}{sending} \hht{the} \actnack{} signal from the DRAM chip to the MC.

\begin{figure}[!h]
    \centering
    \includegraphics[width=\linewidth]{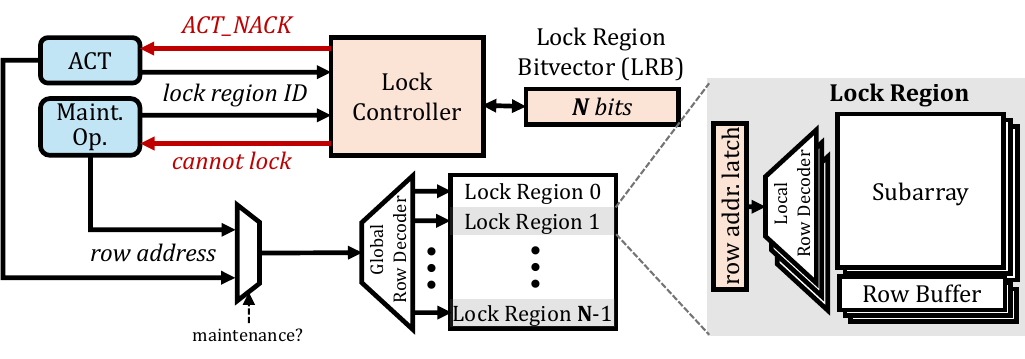}
    \caption{\mech{} bank organization \hht{in DRAM chip}}
    \label{fig:smd_bank_organization}
\end{figure}

Fig.~\ref{fig:smd_bank_organization} shows the organization of a bank in an
\mech{} chip. \mech{} divides \ignorerev{the rows in}a DRAM bank into multiple \emph{lock
regions} \micronew{of equal size}. \omcr{4}{\atbcr{3}{Each bank has at least one lock region.}}
\mech{} stores an entry in a \iql{IQC1}{}\iqrev{per-bank} \emph{Lock Region \omcr{3}{Bitvector} (LRB)} to indicate
whether or not a lock region is \hht{reserved} for performing a maintenance
operation.\atbcrcomment{2}{Address figure comments. Figures drawn in Visio so this will be a torture.}

\noindent
\textbf{\micronew{Operation}.} A maintenance operation takes place as the \emph{Lock Controller} sets
the \emph{lock bit} in the LRB entry that corresponds to the lock region \hh{on
which the maintenance operation is to be performed}. When the MC attempts to
open a row in a locked region, the Lock Controller generates an \actnack{}. \omcr{3}{When a maintenance operation
attempts to lock a region 1)~with an open row or 2)~that is already locked by another maintenance operation, the Lock
Controller responds with \emph{cannot lock}.}

A maintenance operation and an \cmdact{} can be performed concurrently \hht{in
the same bank} on different lock regions.\footnote{\hpcarevb{An ECC scrubbing operation (\cref{subsec:smd_ecc_scrubbing}) uses the data bus, 
which is a shared resource across all \atbcr{1}{lock} regions. Therefore, a scrubbing operation \microrevcommon{reserves} 
all lock regions inside a bank at once to prevent any other maintenance operation or memory access from happening at the same time.}} 
To enable this, \mech{} implements a
\emph{row address latch} in order to drive two local row decoders with two
independent row addresses.\footnote{\atb{One} row address latch per lock region
enables two or more maintenance operations to concurrently happen on different
lock regions within a bank. \revb{To} keep our design simple, we restrict a
maintenance operation to happen in one lock region while the MC
access\atb{es} another lock region. \hht{Our design can be easily extended to
increase concurrency of maintenance operations \omcr{2}{or accesses} across multiple lock regions,
\omcr{2}{similarly to SALP~\cite{kim2012case}}.}}
\hh{When a maintenance operation and an \cmdact{} arrive \hht{at} the same lock
region at the same time}, the \atbcr{2}{bank}
prioritizes the maintenance operation and
the \mech{} chip rejects the \cmdact{}.

\subsection{Region Locking Mechanism}
\label{subsec:locking_mechanism}




\microrevlabel{AQ7}\microreva{A maintenance operation can be performed \hhh{on a \emph{lock region}
only \atbcr{1}{while the region is locked}}. 
Therefore, a maintenance mechanism must lock the region that
includes the rows that should undergo maintenance. A maintenance mechanism can
\emph{only} lock a region that is \emph{not} locked, \atb{which} 
prevents different maintenance mechanisms from interfering with each other.}
\noindent
\textbf{Lock Region Size.}
A \emph{lock region} consists of a fixed number of \hht{consecutively-addressed}
DRAM rows. \hht{To simplify the design,} we set the lock region size \atb{to} 
one or multiple subarrays. This is because a
maintenance operation \hht{uses} the local row buffer in a subarray.
We design and evaluate \mech{} assuming \hht{a default} lock region
size of $16$ subarrays. \atbcr{2}{We evaluate SMD's sensitivity to 
lock region size (for a DRAM bank with 256 subarrays) and find that increasing 
the number of lock regions \atbcr{3}{in a bank} beyond 16 (i.e., having fewer than
16 subarrays in a lock region)
yields diminishing returns (\cref{sec:sensitivity-lock-region}).}



Modern DRAM chips typically use the density-optimized \emph{open-bitline}
architecture~\cite{keeth2007dram,itoh2013vlsi,chang2016low,luo2020clrdram}, which places sense amplifiers
on both sides of a subarray and \atbcr{1}{where} adjacent subarrays share sense
amplifiers. With the open-bitline architecture, the MC should \emph{not} access a row
in a subarray adjacent to one under maintenance. To achieve this, the Lock
Controller simply sends an \actnack{} when the MC attempts to activate a row in
a subarray adjacent to a locked region. Consequently, when SMD
locks a region that spans $16$ subarrays, it prevents the MC from
accessing $18$ subarrays. We evaluate SMD using
the density-optimized open-bitline architecture. In the
\emph{folded-bitline} architecture~\cite{keeth2007dram, chang2016low, itoh2013vlsi},
adjacent subarrays do \emph{not} share sense amplifiers with each other. 
Therefore, SMD
prevents access only to the subarrays in a locked region. 


\noindent
\textbf{Ensuring Timely Maintenance Operations.}
\microrevlabel{AQ7}\microreva{A maintenance mechanism \emph{cannot} lock a region \hhh{with} an active row.
To lock the region, the maintenance mechanism waits for the MC to precharge
the row. \micronew{DRAM standards specify the maximum allowed time for a row to remain active as
\micronew{9}x~\trefi{}~\cite{jedec2020ddr4,micronddr4,ddr4operationhynix,luo2023rowpress}). As such, a maintenance mechanism
is delayed at most by \micronew{9}x~\trefi{}.}}\atbcrcomment{1}{Incorporate numbers for DDR5?}


\subsection{\micronew{Controlling an \mech{} Chip}}
\label{subsec:control}
While introducing changes in the memory controller (MC) to handle \actnack{}, 
\mech{} also simplifies MC design and
operation as the MC no longer implements control logic to issue
DRAM maintenance commands. For example, the MC does \emph{not} 1) prepare a bank
for refresh by precharging it, 2) implement timing parameters relevant to
refresh, and 3) issue \cmdrefresh{} commands. The MC still maintains the bank
state information (e.g., which row is open in a bank) and respects
the DRAM timing parameters associated with \atbcr{1}{\emph{other} DRAM} commands 
(e.g.,
\cmdact{}).



\begin{figure}[!h]
    \centering
    \includegraphics[width=\linewidth]{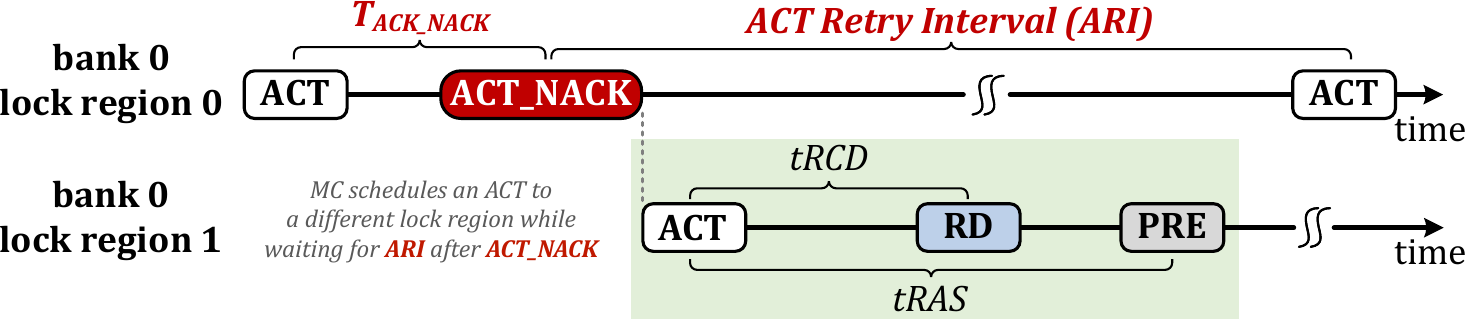}
    \caption{Handling \actnack{} \hht{in MC}}
    \label{fig:smd_act_nack_timeline}
\end{figure}

\noindent
\textbf{Handling an \actnack{} Signal.}
Fig.~\ref{fig:smd_act_nack_timeline}\atbcrcomment{2}{address figure comments} 
depicts a timeline that shows how the MC
handles an \actnack{} signal. \iql{IQB1}{}\iqrev{\hh{Upon receiving} an \actnack{}, the MC waits
for \emph{ACT Retry Interval} (\ARI) \hht{time} \hh{to re-issue} the same
\cmdact{}. \hh{It} \omcr{2}{can} keep re-issuing the \cmdact{} command once every \ARI{}
until the DRAM chip accepts the \cmdact{}.} \atb{To ensure that a DRAM chip 
\emph{cannot} lock a region for prolonged periods, \mech{} enforces a minimum
\atbcr{2}{(configurable)} delay of \ARI{} between the end of a maintenance operation and the start of the
next maintenance operation targeting the same region\atbcrcomment{2}{We set this value to \ARI{}}.
This allows the MC 
to serve a request that targets the locked region immediately after
the maintenance operation in this region ends.} While waiting for \ARI{}, the MC can
\hhh{activate} a row \hh{from a different lock region or bank}, to overlap the
\ARI{} latency with a useful operation.
\atbcr{2}{An enforced minimum delay of \ARI{} is sufficient to guarantee forward progress for memory
requests (\cref{subsec:forward-progress}).}
\atbcr{2}{The system designer can set the enforced minimum delay to a multiple of \ARI{}
to enforce a larger time \omcr{3}{period} between two maintenance operations targeting the same region.}

\noindent
\textbf{Setting \hht{the} \emph{ACT Retry Interval} (\ARI).}
Setting \ARI{} to a very low or high value relative to the expected duration of
the maintenance operations can have a negative impact on system performance and
energy efficiency. 
\iql{IQB1}{}\iqrev{We empirically find
that $\ARI=\SI{\valARI}{\nano\second}$\atbcrcomment{2}{Please see what I wrote above. Is it better?}
is a favorable configuration for the \param{three}
maintenance mechanisms that we evaluate in this work
(\cref{sec:maintenance_mechanisms}).}



\noindent
\textbf{\tnack{} Latency.}
An \mech{} chip sends an \actnack{} 
\tnack{} DRAM command bus cycles after receiving the \cmdact{}. 
%
%
\hht{\tnack{} \atb{should} \micronew{be low} so that the MC \atb{is} quickly notified when an \cmdact{} fails, and \micronew{the MC} can attempt
to activate a different row in the same bank while waiting for \ARI{}.}

The \tnack{} latency has three components: 1) the propagation delay (from the
MC to the DRAM chip) of the \cmdact{}, 2) the latency of
determining whether or not the row to be activated belongs to a locked region,
and 3) the propagation delay (from the DRAM chip to the MC) of
the \tnack{} signal. 

We estimate the overall \tnack{} latency based on the latency of \cmdread{} \omcr{2}{(read)}
in a conventional DRAM \iscanew{chip}. \cmdread{} has a latency breakdown that resembles the
latency breakdown of a \tnack{}. \cmdread{} command 1) propagates from the MC
to the DRAM chip, 2) accesses data in a portion of the row buffer in the
corresponding bank, and 3) sends the data back to the MC. In the DDR4
standard~\cite{jedec2020ddr4}, the latency between issuing an \cmdread{} and the first
data beat appearing on the data bus is defined as \tcl{} \hht{(typically
22~cycles for DDR4-3200)}. The latency components 1) and 3) of \cmdread{} are
similar to those of \tnack. Thus, the main difference between \tnack{} and
\tcl{} arises from the second component. \microrevlabel{FQ1}\microrevf{According to our evaluation, the
latency of accessing the Lock Region \omcr{3}{Bitvector} (LRB) is \SI{0.053}{\nano\second}
(\cref{sec:hw_overhead}). Given the relatively low complexity of the LRB
compared to the datapath that is involved during \cmdread{}, the
overall \tnack{}
latency can be designed to be much smaller than \tcl{}. \hht{We assume \tnack{}
$= 5$ cycles unless stated otherwise. In our evaluations \omcr{2}{(see~\cref{sec:evaluation})}, 
we find that small
\tnack{} latencies (e.g., $\leq \tcl{}$) have \omcr{2}{a} negligible effect on system
performance mainly because the number of rejected \cmdact{}s constitutes a small
portion of all \cmdact{}s.}}


\subsection{\actnack{} Divergence Across DRAM Chips}
\label{subsubsec:act_nack_divergence}

\addressdivergence{\mech{} maintenance operations \hht{take place} independently in each DRAM chip.
Therefore, when a DRAM rank, which can be composed of multiple DRAM chips
operating in lock step,
receives an \cmdact{} command, some of the chips may \hh{send an
\actnack{} while others do not.}}
\addressdivergence{Normally, \actnack{} divergence would \emph{not} happen for maintenance mechanisms
that perform the exact same operation at the exact same time in all DRAM chips 
\atb{(e.g., the Fixed Refresh mechanism, Section~\ref{subsec:smd_refresh})}.
However, a mechanism can also operate differently in each DRAM chip, e.g.,
\iscanew{a variable rate refresh mechanism~\cite{liu2012raidr,qureshi2015avatar,patel2017reaper}
\addressdivergencetwo{or a RowHammer protection mechanism (e.g.,~\cite{kim2014flipping,kim2022mithril,park2020graphene,jedecddr5c})}.}
As a result \hh{of this divergence}, the row becomes partially open: open in 
chips that do \emph{not} send \actnack{} and closed in chips that do. 

\addressdivergencetwo{The \actnack{} divergence problem does \emph{not} happen 
when each chip is individually controlled by the MC.}
\atb{As a solution \addressdivergencetwo{for systems where a rank of
chips operate in lock step,} the MC can employ one of the three 
strategies: Precharge, Wait, and Hybrid. We evaluate each technique under
common-case and worst-case scenarios with regard to when maintenance
operations happen across different SMD chips in the same rank in~\cref{subsec:divergence-evaluation}.}}

\noindent\textbf{Precharge.} 
\addressdivergence{With the \emph{Precharge} policy, the MC
issues a \cmdpre{} command to close the partially activated row when some DRAM
chips send \actnack{} but others do not. After closing the partially activated
row, the MC can attempt to activate a row from a different lock
region in the same bank.}

\noindent\textbf{Wait.} 
\addressdivergence{With the \emph{Wait} policy, the MC issues
multiple \cmdact{} commands until a partially activated row becomes fully
activated. When some chips send \actnack{} for a particular \cmdact{} but others
do not, the MC waits for \ARI{} and issues a new \cmdact{} to
attempt activating the same row in DRAM chips that previously sent \actnack{}.}

\noindent\textbf{Hybrid.} 
\addressdivergence{We also design a \emph{Hybrid} policy, where the memory
controller uses the \emph{Precharge} policy to close a partially activated row
if the request queue contains $N$ or more requests that need to access rows in
different lock regions in the same bank. If the requests queue has less than $N$
requests to different lock regions, the MC uses the \emph{Wait}
policy to retry activating the rest of the partially activated row.}


\subsection{Forward Progress for Memory Requests}
\label{subsec:forward-progress}

\micronew{The MC could fail to retry an \actnack{}'d memory request if an active DRAM row \atbcr{2}{(other
than the one accessed by \actnack{}'d memory request) prevents}\atbcrcomment{4}{you activate a row, it gets 
nack'd, 
you activate a row in another lock region, ARI passes, you cannot retry the NACK'd request 
because you are serving
reads to the active row.}\omcrcomment{4}{footnote?}\atbcrcomment{4}{I will add this if you think it is necessary. We are running out of space...} retrying the 
\actnack{}'d memory request every \ARI{}. Such retry failures could occur repeatedly 
and lead to \emph{temporary starvation} of a lock region (i.e., no request 
targeting that lock region is served for a prohibitively long time). 
Temporary starvation of a lock region is very unlikely to happen because
it requires a maintenance mechanism to repeatedly lock the same lock region for maintenance, 
which is \omcr{3}{exactly} \atbcr{2}{what} a \omcr{3}{\emph{carefully
designed}} maintenance mechanism \omcr{2}{should avoid}. Instead, a carefully designed maintenance mechanism
\omcrcomment{2}{alternates is likely not a good word}
alternates between different lock regions as it performs new maintenance operations (e.g., periodic 
refresh~\cref{subsec:smd_refresh}).}

\micronew{To prevent temporary starvation of lock regions irrespective of maintenance mechanism design, 
\omcr{2}{an implementation of}
\mech{} issues an \actnack{}'d memory request every \ARI{}. No 
memory request is \actnack{}'d by two separate maintenance operations
because an SMD chip does \emph{not} perform any maintenance operations 
for \omcr{2}{at least} \ARI{} after the end of a maintenance operation 
in a lock region (\cref{subsec:control}).} 
\micronew{Issuing a rejected memory request strictly every \ARI{} 
does \emph{not} prevent the MC from serving memory requests 
that access other lock regions.}\footnote{\microrevc{An \mech{} 
chip may reject a memory request’s activate command. This means that the memory scheduler, based 
on its scheduling algorithm, has deemed that request to be the most 
important request to schedule. Therefore, it should be in the best 
interest of the scheduling algorithm to serve this memory request 
as soon as possible. The ARI interval allows the memory controller 
to \omcr{3}{schedule} one or multiple other memory requests by leveraging lock-region-level 
parallelism.}} Within one \ARI{} (\SI{62.5}{\nano\second}) 
the MC can 1) activate a DRAM row other than the \actnack{}'d DRAM row, 
2) serve seven (based on the \omcr{2}{DDR4} standard values of $t_{RAS}$, $t_{RP}$, $t_{CCD\_L}$ 
timing parameters) READ/WRITE requests to this newly activated row, and 3) 
precharge the bank such that an activate command to the previously 
\actnack{}'d DRAM row can be issued every \ARI{}. 
\microrevlabel{CQ5}\microrevc{To ensure that the 
memory controller retries the rejected memory request in due time, we 
set the number of allowed column commands (RD/WR) to an open row 
before an older request (to another row) is serviced (i.e., the Cap parameter~\cite{mutlu2007stall}). 
Based on \ARI{}=\SI{62.5}{\nano\second}, we use a Cap of 7 in our evaluations.}

\noindent
\textbf{\micronew{Proof of Forward Progress.}} \micronew{We consider a memory request $R$ in the memory controller's request queue to have made forward progress if it is served by the memory controller.}

\noindent
\micronew{\textbf{Theorem.} If the MC retries $R$ every \ARI{} then $R$ makes forward progress.}\footnote{
\atbcr{3}{The MC is \emph{not} too busy to retry the request because we modify the page policy
by capping the number of column accesses in an \ARI{} to 7.}
}
\atbcrcomment{3}{The MC \omcr{4}{can never be} too busy to retry the request every \ARI{}. Buildup until the theorem was
supposed to explain this. Did it completely not do its job?}

\noindent
\micronew{\textbf{Proof.} We devise a \emph{contrapositive proof} for the above theorem. Suppose $R$ does \emph{not} make forward progress, i.e., $R$ is never dequeued from the request queue. $R$ must target a locked region because otherwise $R$ is dequeued when it is scheduled. The maintenance mechanism can only hold the lock for \emph{finite time}. When the maintenance mechanism releases the lock, no maintenance mechanism locks the same region for at least \ARI{}. If the MC retries $R$ within one \ARI{} after the lock is released, then $R$ is \emph{not} rejected by a maintenance mechanism, i.e., $R$ makes forward progress. However, if the MC retries $R$ every \ARI{}, then the MC also retries $R$ within one \ARI{} after the lock is released. Thus, if $R$ does \emph{not} make forward progress, then the MC does \emph{not} retry $R$ every \ARI{}. We infer from this that the theorem is true.} 

\noindent
\micronew{\textbf{Memory Request Stall Time.}}
{{\atbt{To limit the maximum potential memory request stall time, 
1) \microrevcommon{time-intensive} maintenance operations can be divided into many 
small, short\omcr{2}{-duration} maintenance operations, 2) the maximum time 
a maintenance operation takes can be specified
in the DRAM standard.}} In our design, the \omcr{2}{longest-duration} 
maintenance operation is \iscanew{ECC scrubbing} (Section~\ref{subsec:smd_ecc_scrubbing}),
which takes $\approx \SI{350}{\nano\second}$ to \iscanew{read a DRAM row \atbcr{2}{(assuming
no corrected ECC codewords need to be written back \omcr{3}{to the DRAM array})}.} \iscanew{This stall time
is comparable to what a memory request in a modern DDR4-based system \omcr{2}{could} experience.
In such a system, a request that arrives at the memory controller's request queues immediately after the memory 
controller schedules a periodic refresh operation (by issuing a $REF$ command) has to wait for
the refresh operation to complete (e.g., for \SI{350}{\nano\second}~\cite{jedec2020ddr4}).}}

\subsection{Impact to Request Scheduling}
\label{subsec:impact_to_scheduling}

\ignore{
\mech{} partitions DRAM into small regions to perform maintenance operations.
This allows \mech{} to pause accesses to small DRAM regions while the rest of
the DRAM remains available to access. In contrast, existing DRAM chips perform
maintenance operations at a larger granularity. For example, with existing DRAM
chips, MCs perform refresh at bank or rank granularity, pausing accesses to an
entire bank/rank. Although the MC can postpone a refresh operation for a small
time interval (e.g., 8x \trefi{}), it does so just based on information (e.g.,
queue occupancy) available to the MC, but without the knowledge of future
requests. As a result, processing units are stalled with existing DRAM chips
when requests enter the MC after a maintenance operation starts. \mech{}
significantly improves the overall availability of DRAM by pausing accesses to
small regions and improves performance as we show in
\cref{subsec:single_core_perf} and \cref{subsec:multi_core_perf} (e.g., for
memory intensive four-core workloads, a system with \mech{}-based refresh
achieves \param{8.57}\% average speedup compared to a system with conventional DDR4).
Therefore, \mech{} improves the overall availability of DRAM, and thus reduces
the performance overhead of DRAM maintenance operations.
}

\revc{\hpcarevlabel{Rev.A/C2}
\hpcareva{The key drawback of our \mech{} implementation is that it makes the 
synchronous DDRx interface \emph{less} predictable \omcr{2}{(i.e., slightly asynchronous)}
\omcr{3}{from the perspective of the MC}.
However, the practical\omcr{2}{ly-observed} performance overheads of this drawback are very low, as we demonstrate in~\cref{sec:evaluation}.}  

In this paper, we \hpcareva{comprehensively describe} 
only one of many different possible \reva{implementations} of SMD due to page limits. 
This \hpcareva{implementation} prioritizes \atbcr{3}{\emph{simplicity}}. However, the DRAM 
standard can more strictly specify exactly \emph{when} and for \emph{how long} 
SMD chips are allowed to perform maintenance operations. This stricter 
\hpcareva{implementation} would \omcr{2}{likely} make SMD chips as predictable as DDRx chips today. 
We hope that future work building on SMD \omcr{2}{can} comprehensively \omcr{2}{study such} issues.}
\hpcareva{To make it more evident that \mech{} can be implemented in a way that 
preserves the predictability of the DDRx interface from the perspective of the memory 
controller (i.e., in a way that keeps the DRAM interface synchronous), we \omcr{3}{briefly} describe a 
(more) predictable \mech{} implementation.}

\noindent
\hpcareva{\textbf{Designing a Predictable \mech{} Interface.} A simple way of preserving
the DDRx interface's synchronicity with \mech{} is to allow the DRAM chip
to perform a maintenance operation periodically \emph{only} at well-defined time intervals.
In other words, the DRAM chip can respond with \actnack{}s \emph{only} for the fraction of time 
during \omcr{2}{a mode in} which the chip is allowed to execute maintenance operations. We call this \omcr{2}{mode} 
the \emph{maintenance operation time} (MOT). Outside MOT, the DDRx operates as defined today 
(i.e., synchronously and predictably). During MOT, the memory controller is allowed to access DRAM cells
at the cost of potentially getting \actnack{}'d by the DRAM chip.
The new SMD standard would specify i) how long MOT is, ii) the period at which the interface enters MOT \microreva{(MOT\_period)}, 
iii) how entry to MOT is triggered (e.g., by a new MOT command), and iv) how many MOTs the memory controller
can postpone.}




\section{SMD Maintenance Mechanisms}
\label{sec:maintenance_mechanisms}

\hht{We propose \mech{}-based maintenance mechanisms for three use cases.}
\mech{} is not limited to these three use cases and it can be used to
support more operations in DRAM \omcr{2}{(\omcr{3}{as described in~}\cref{subsec:other_use_cases} \atbcr{4}{and \cref{sec:other_optimizations}})}. 

\subsection{Use Case 1: DRAM Refresh}
\label{subsec:smd_refresh}

In conventional DRAM, the MC periodically issues \cmdrefresh{} commands to
initiate a DRAM refresh operation. This approach is inefficient due to
\hht{\param{two} main} reasons. First, transmitting \omcr{2}{thousands of (e.g., $8192$)}
\cmdrefresh{}s over
the DRAM command bus within the refresh period \hht{(e.g., 64, 32, or even
\SI{16}{\milli\second} depending on the refresh rate~\cite{jedecddr5c})} consumes energy and
increases the command bus utilization. 
A \cmdrefresh{} \omcr{2}{command} may delay a command to another 
\atbcr{2}{DRAM component (e.g., bank, bank group, rank)} 
in the same channel,
which increases DRAM access latency~\cite{chang2014improving}. Second, an entire
bank becomes inaccessible while being refreshed although a \cmdrefresh{} command
refreshes only \omcr{2}{a small fraction of} rows in the bank. 


Leveraging \mech{}, we design \iscanew{two new efficient} maintenance mechanisms for \omcr{2}{periodic}
DRAM refresh:
\emph{Fixed-Rate Refresh} and \emph{Variable-Rate Refresh}. 
The Fixed-Rate Refresh mechanism aims to refresh all rows in a DRAM chip
uniformly at a fixed refresh period, similar to conventional DRAM refresh. The
Variable Refresh mechanism inspired by RAIDR~\cite{liu2012raidr} refreshes different DRAM rows at different
intervals based on their retention time characteristics.

\noindent
\textbf{Fixed-Rate Refresh (\fr{}).}
\fr{} refreshes DRAM rows in a fixed time interval (i.e., \trefi{}), similar to
conventional DRAM refresh. 
To limit the time that the MC waits for a
region to get unlocked, \fr{} refreshes $RG$ (Refresh Granularity) number of
rows from a lock region and \omcr{2}{then} switches to refreshing another region. 

\begin{figure}[!h]
    \centering
    \includegraphics[width=1\linewidth]{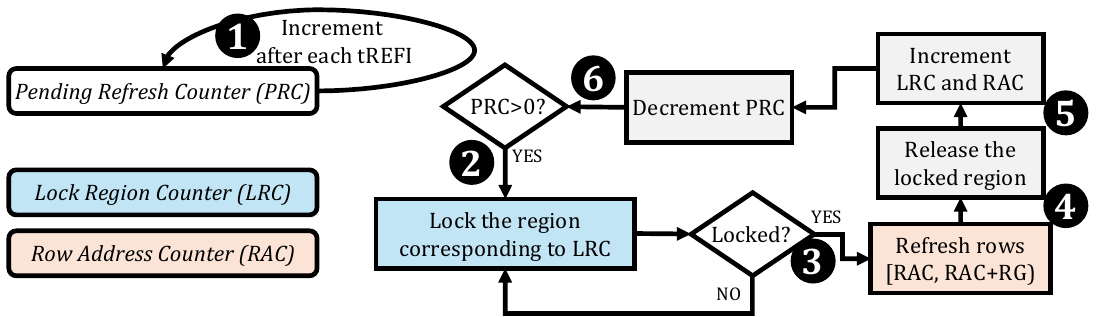}
    \caption{\omcr{3}{SMD-based} Fixed-Rate Refresh (\fr{}) Operation}
    \label{fig:smd_fixed_refresh}
    \vspace{-3pt}
\end{figure}

\fr{} (\atb{Fig.~\ref{fig:smd_fixed_refresh}})\atbcrcomment{2}{fix figure}
operates independently in each DRAM bank and \hhh{it} uses three counters
for managing the refresh operations: \emph{pending refresh counter}, \emph{lock
region counter}, and \emph{row address counter}.
The \emph{pending refresh counter} is
initially zero and \fr{} increments it by one at the end of \atbcr{1}{every} \trefi{}
interval~\circled{1}. \fr{} allows up to \omcr{2}{N} refresh operations to be
accumulated in the \emph{pending refresh counter}. \atbcrcomment{3}{this number is 4 for ddr5}\omcr{2}{We set N=8.}\footnote{DDR4~\cite{jedec2020ddr4,micronddr4,ddr4operationhynix} allows the MC to
postpone issuing up to $8$ \cmdrefresh{} commands in order to serve
\hht{pending} memory requests first~\cite{chang2014improving, stuecheli2010elastic}. \omcr{4}{DDR5 reduces this number to $4$~\cite{jedecddr5c}.}} 
Because the MC can keep a row open
for a limited time (\cref{subsec:control}), the \emph{pending refresh counter} never exceeds the value
\omcr{2}{N}.

The \emph{lock region} and \emph{row address} counters indicate the next row to
refresh. When \emph{pending refresh counter} is greater than zero, \fr{}
attempts to lock the region indicated by the \emph{lock region counter} every
clock cycle until it successfully locks the region~\circled{2}. \hht{In} some
cycles, \fr{} may fail to lock the region either because the lock region
contains an active row or the region is locked by another maintenance mechanism.
When \fr{} successfully locks the region, it initiates a refresh operation that
refreshes $RG$ number of 
rows in the lock region
starting from the row indicated by the \emph{row address counter}~\circled{3}.
{
We empirically find \iscanew{$RG = 8$}
to be a favorable design point.\footnote{\atbcr{2}{We evaluate RG = 1, 2, 4, 8, 16, 32, and 64. 
We find that RG~=~8 and RG~=~4 yield the two highest performance improvements
among all tested RG values. Making $RG$ larger increases the latency of a single refresh
operation, which prolongs the time that a lock region remains unavailable for
access. In contrast, \omcr{3}{a too} small $RG$ causes switching from one lock region to another
very often, increasing interference with accesses.}}}

After the refresh operation completes, \fr{} releases the locked
region~\circled{4} and  increments \emph{only} the \emph{lock region counter}.
\hht{When} the \emph{lock region counter} \hht{rolls back to zero}, \fr{} also
increments the \emph{row address counter}~\circled{5}. Finally, \fr{}
decrements the \emph{\omcr{4}{pending} refresh counter}~\circled{6}.




\subsubsection{Variable Refresh (\vr{})}
\label{sec:variablerefresh}

In conventional DRAM, all rows are uniformly refreshed with the same
refresh period. However, the \hht{actual data retention times of different rows}
in the same DRAM chip \hht{greatly vary} mainly due to manufacturing
\hht{process variation} and design-induced
variation~\cite{liu2012raidr,qureshi2015avatar,liu2013experimental,patel2017reaper,lee2017design,nair2014refresh}.
In fact, only hundreds of ``weak'' \hht{rows} across an entire
\SI{32}{\giga\bit} DRAM chip require to be refreshed at the default rate, and
\hht{a vast} majority of the \hht{rows} can correctly operate when the refresh
period is doubled or quadrupled~\cite{liu2012raidr}. Eliminating unnecessary
refreshes to DRAM rows that do not contain weak cells can significantly mitigate
the performance and energy consumption overhead of DRAM
refresh~\cite{liu2012raidr}. 

We \hht{develop} \emph{Variable Refresh} (\vr{}), a mechanism \hht{that}
refreshes different rows at different refresh rates depending on the retention
time characteristics of the weakest cell in each row. 
Our \vr{} design demonstrates the versatility of \mech{} in supporting different
DRAM refresh mechanisms. 

The \atb{key} idea of \vr{} is to group DRAM rows into multiple retention time
bins and refresh a row based on the bin that it belongs to. To achieve low
design complexity, \vr{} uses only two bins: 1) \emph{retention-weak} rows that
have retention time less than $RT_{weak\_row}$ and 2) rows that have retention
time more than $RT_{weak\_row}$. \hht{Inspired by RAIDR~\cite{liu2012raidr},
\vr{} stores the addresses of retention-weak rows using a per-bank Bloom
Filter~\cite{bloom1970space}, which is a space-efficient probabilistic data
structure for representing set membership. We assume retention-weak rows are
already inserted into the Bloom Filters by the DRAM vendors during
post-manufacturing tests.\footnote{\hht{Alternatively, \mech{} can be used to
develop a maintenance mechanism that performs retention profiling. 
We leave the development and analysis of it to future work.}}}

\begin{figure}[!h]
    \centering
    \includegraphics[width=.95\linewidth]{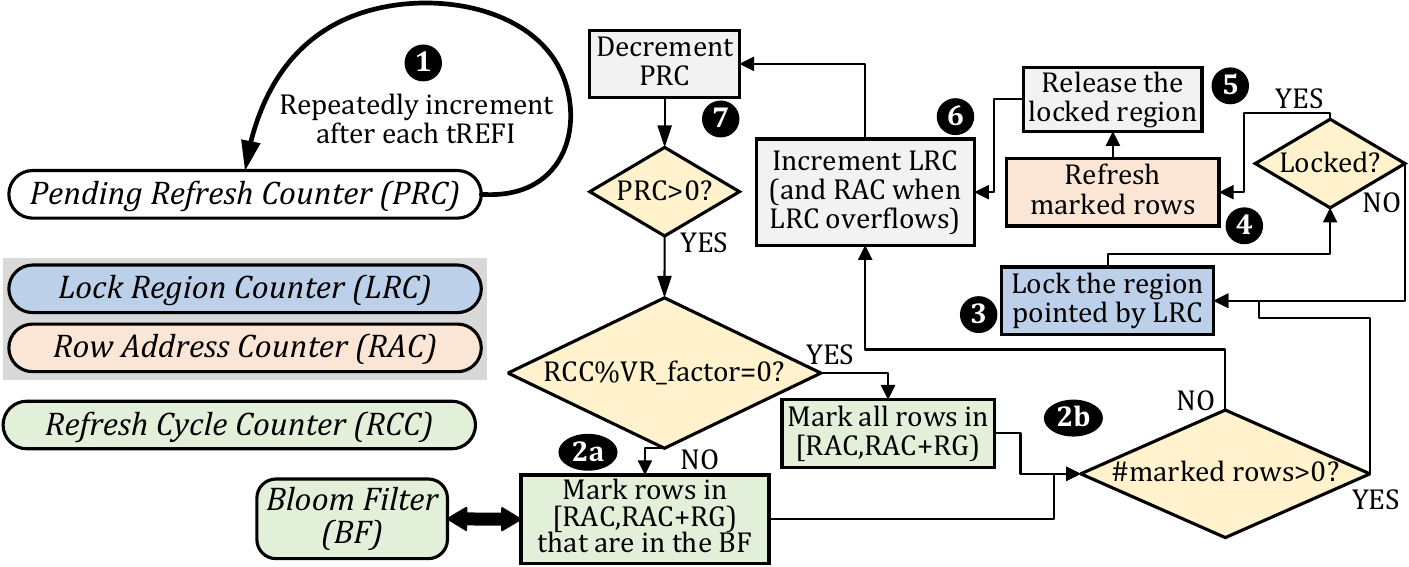}
    \caption{Variable Refresh (\vr{}) Operation.}
    \label{fig:smd_variable_refresh}
\end{figure}

The operation of \vr{} resembles the operation of \fr{} with the \hht{key}
difference that \vr{} \hht{sometimes} skips refreshes to a row that is not in
the Bloom Filter, i.e., a row with high retention time.
Fig.~\ref{fig:smd_variable_refresh} illustrates how \vr{} operates. \vr{} uses
the same three counters \hht{as} \fr{}. 
\vr{} \hht{also} uses a \emph{refresh cycle counter}, which is \hht{used to
indicate the refresh period when all rows (including retention-strong rows) must
be refreshed.} The \emph{refresh cycle counter} is initially zero and gets
incremented at the end of every refresh period, i.e., when the entire DRAM is
refreshed. 

\vr{} increments the \emph{pending refresh counter} by one at the end of a
\trefi{} interval~\circled{1}. When the \emph{pending refresh counter} is
greater than zero, \vr{} determines whether or not the $RG$ number of rows,
starting from the address indicated by the \emph{lock region} and \emph{row
address} counters, are retention-weak rows by testing their row addresses using
the bank's Bloom Filter~\circled{2a}. \vr{} refreshes the rows that are present
in the Bloom Filter every time when it is their turn to be refreshed, as
indicated by the \emph{lock region} and \emph{row address} counters. In
contrast, \vr{} refreshes the rows that are \emph{not} present in the Bloom
Filter \emph{only} when the \emph{refresh cycle counter} has a value that is
multiple of $VR\_factor$~\circled{2b}, specified \atbt{as $VR\_factor = RT_{weak\_row}/RefreshPeriod$.} 



Unless stated otherwise, we assume $RT_{weak\_row} =
\SI{128}{\milli\second}$ and $RefreshPeriod=\SI{32}{\milli\second}$. Therefore,
\vr{} refreshes the DRAM rows that are not in the Bloom Filter once every four
consecutive refresh \hht{periods} as these rows can retain their data correctly
for at least four refresh periods.
After determining which DRAM rows need refresh, \vr{} \hht{operates in a way
similar to \fr{}.}


\subsection{Use Case 2: RowHammer Protection}
\label{subsec:smd_rowhammer_protection}


Repeatedly
activating and precharging (i.e., hammering) a DRAM row \hht{(aggressor)}
causes \micronew{RowHammer}
\omcr{2}{bitflips}~\cite{kim2014flipping} in the cells of a nearby DRAM
row (victim)~\cite{kim2014flipping, mutlu2019rowhammer, mutlu2023fundamentally, yang2019trap,gautam2019row,jiang2021quantifying,park2016experiments,
park2016statistical, ryu2017overcoming, walker2021dram,yang2016suppression}.
DRAM manufacturers equip their \hht{existing} DRAM chips with
in-DRAM RowHammer protection mechanisms, generally referred to as Target Row
Refresh (TRR)~\cite{hassan2021utrr,frigo2020trrespass}. At a high level, TRR protects against RowHammer by
detecting an aggressor row and refreshing \hht{its} victim rows. Because a
conventional DRAM chip \emph{cannot} initiate a refresh operation by itself, TRR
refreshes victim rows by taking advantage of the slack time available in the
\hht{refresh} latency (i.e., \trfc{}), originally used to perform \emph{only}
periodic DRAM
refresh~\cite{hassan2021utrr,frigo2020trrespass,jattke2022blacksmith}.
Recent
works (e.g., \cite{hassan2021utrr,frigo2020trrespass,jattke2022blacksmith,deridder2021smash})
demonstrate a variety of \hht{new} RowHammer access patterns that circumvent the
TRR protection in chips of all \hht{three} major DRAM vendors, \omcr{2}{demonstrating} that the
existing DRAM interface is not well suited to enable strong RowHammer protection
\hht{especially as RowHammer becomes a bigger problem with DRAM technology
scaling}~\rowHammerGetsWorseCitations{}.
\atbcr{2}{We adapt two existing RowHammer protection mechanisms (PARA~\cite{kim2014flipping} and Graphene~\cite{park2020graphene})
to be implemented freely in DRAM. These adaptations}
overcome the
limitations of the existing TRR mechanisms by initiating victim row refresh
within the DRAM chip.

\begin{figure}[!h]
    \centering
    \includegraphics[width=.9\linewidth]{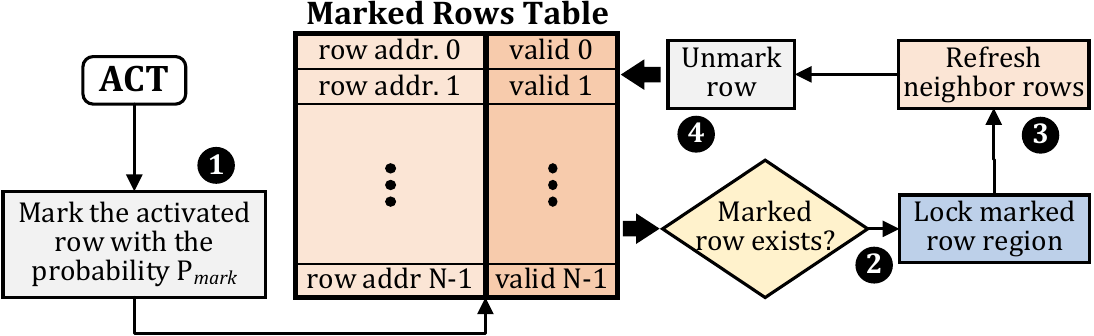}
    \caption{Probabilistic RowHammer Protection (\prp{}).}
    \label{fig:smd_para}
\end{figure}

\noindent
\textbf{\hht{Probabilistic} RowHammer Protection (PRP).}
Inspired by PARA~\cite{kim2014flipping}, we implement an in-DRAM maintenance
mechanism called \hht{Probabilistic RowHammer Protection (\prp{})}. \hht{The
high level idea is to refresh the nearby rows of an activated row with a small
probability. PARA is proposed as a mechanisms in the MC, which makes it
difficult to adopt since victim rows are not always known to the MC. \mech{}
enables us to overcome this issue by implementing the PARA-inspired \prp{}
mechanism completely within the DRAM chip. In addition, \prp{} avoids explicit
\cmdact{} and \cmdpre{} commands to be sent over the DRAM bus.} 
Fig.~\ref{fig:smd_para} illustrates the operation of \prp{}. 

On a DRAM row activation, \prp{} marks the activated row as an aggressor with
\hht{a small} probability of $P_{mark}$~\circled{1}. \hht{It} marks \hht{an
aggressor} row using a \hht{per-bank} \emph{Marked Rows Table (MRT)}, that
contains an entry for each lock region in the bank. \hht{An} entry consists of
the \hht{marked row address} and a \emph{valid} bit. The bit length of the
address depend on the size of a lock region. For example, an MRT entry has a
13-bit address field when the lock region size is $8192$ rows. When MRT contains
a marked row, \prp{} locks the corresponding region~\circled{2} and refreshes
the neighbor rows of the marked row~\circled{3}. \hht{This step can easily
accommodate blast radius~\cite{yaglikci2021blockhammer} and any address
scrambling.} Once the neighbor rows are refreshed, \prp{} \hht{unlocks the}
region and unmarks the row in MRT~\circled{4}.

\ignorerev{
\noindent
\textbf{\prp{} with Aggressor Row Detection (\prpplus{}).}
\hht{\prp{} refreshes victim rows with a small probability on every row
activation, even does so for a row that has been activated only a few times.
This \hht{results} in unnecessary victim refreshes, especially for high
$P_{mark}$ values that strengthen the RowHammer protection as DRAM cells become
more vulnerable to RowHammer with technology scaling.}

\hht{We propose \prpplus{}, which detects potential aggressor rows and
probabilistically refreshes only their victim rows.} The key idea of \prpplus{}
is to track frequently-activated rows in a DRAM bank and refresh the neighbor
rows of these rows using the region locking mechanism of \mech{}. 

\prpplus{} tracks frequently-activated rows, within a rolling time window of
length $L_{RTW}$, using two Counting Bloom Filters (CBF)~\cite{fan1998summary}
that operate in time-interleaved manner. CBF is a Bloom Filter variant that
\hht{represents the upperbound for} the number of times an element is inserted
into the CBF. \hht{We refer the reader to prior work for background on
CBFs~\cite{yaglikci2021blockhammer,pontarelli2016improving}.}
Depicted in Fig.~\ref{fig:smd_blockhammer}, \prpplus{} operation is based on 1)
detecting a row that has been activated more than $ACT_{max}$ times within the
most recent $L_{RTW}$ interval and 2) refreshing the neighbor rows of this row.
\hht{$ACT_{max}$ must be set according to the minimum hammer count needed to
cause a RowHammer bit flip in a given DRAM chip. A recent work shows that $4.8K$
activations can cause bit flips in LPDDR4 chips~\cite{kim2020revisiting}. We
conservatively set $ACT_{max}$ to $1K$. $L_{RTW}$ must be equal to or larger
than the refresh period in order to track all activations that happen until rows
get refreshed by regular refresh operations. We set $L_{RTW}$ equal to the
refresh period.}

\begin{figure}[!h]
    \centering
    \includegraphics[width=\linewidth]{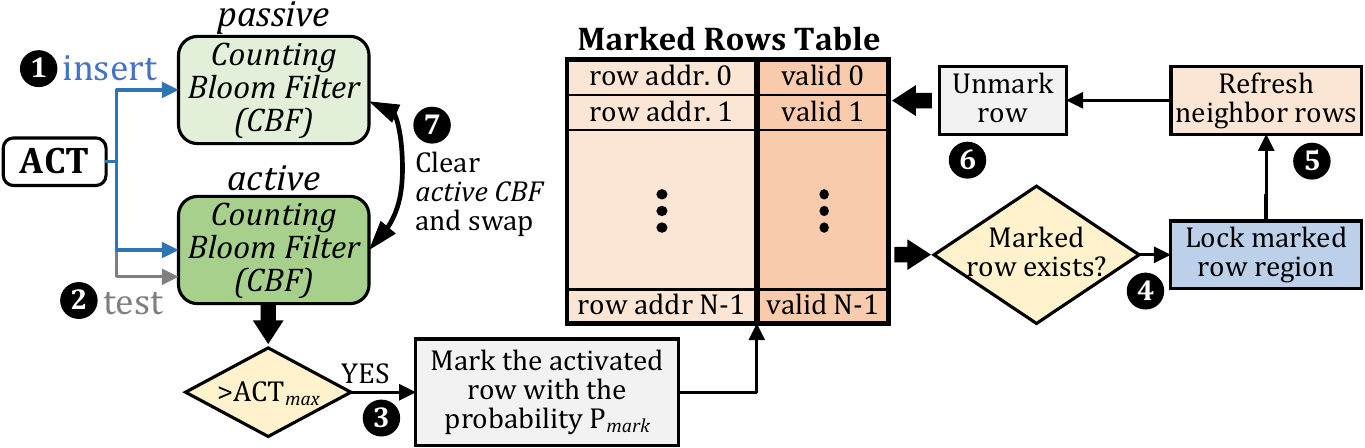}
    \caption{\prpplus{} Operation.}
    \label{fig:smd_blockhammer}
\end{figure}

\hht{Initially, one of the CBFs is in \emph{active} mode while the other is in
\emph{passive} mode.} When activating a row, \prpplus{} inserts the address of
the activated row to \hht{both} CBFs~\circled{1}. \hht{Then}, \prpplus{} tests
the active CBF to check if the accumulated insertion count exceeds the
$ACT_{max}$ threshold~\circled{2}. If \hht{so}, \prpplus{} marks the row in the
\emph{Marked Rows Table} to refresh its neighbor rows with \hht{the}
probability of $P_{mark}$~\circled{3}. \prpplus{} does not always mark the row
because it is impossible to reset \emph{only} the counters that correspond to
the marked row address in the CBFs. \hht{If always marked, a} subsequent
activation of the same row \hht{would again cause} the row to be marked,
\hht{leading to unnecessary neighbor row refresh} until all CBF counters are
reset\hht{, which happens} at the end of the $\frac{L_{RTW}}{2}$ interval.
\hht{After a row is marked, steps~\circled{4}-~\circled{6} are same as
steps~\circled{2}-~\circled{4} in \prp{}. Finally, at the end of an
$\frac{L_{RTW}}{2}$ interval,}
\prpplus{} clears the active CBF and swaps the two CBFs to continuously track
row activations within the most recent $L_{RTW}$ window~\circled{7}.
}

\noindent
\atbcrcomment{3}{How to make this figure nicer?}\iscanew{\textbf{Deterministic RowHammer Protection.}} We use 
\mech{} to implement \drp{}, a deterministic RowHammer 
protection \omcr{2}{technique} based
on the Graphene mechanism~\cite{park2020graphene} \omcr{2}{that}
keep\micronew{s} track of frequently activated DRAM rows. 
Different from Graphene, we
implement \drp{} completely within a DRAM chip, whereas Graphene requires the MC
to issue neighbor row refresh operations.\footnote{
We refer the reader to~\cite{park2020graphene} for more details on the Graphene
mechanism and its security proof. The operation of \drp{} is similar to the
operation of Graphene with the difference \omcr{2}{that}
\drp{} \omcr{2}{is} implemented completely
within a DRAM chip, which does not affect the underlying operation. Thus, the
security proof of Graphene applies to \drp{}.}

\begin{figure}[!h]
    \centering
    \includegraphics[width=\linewidth]{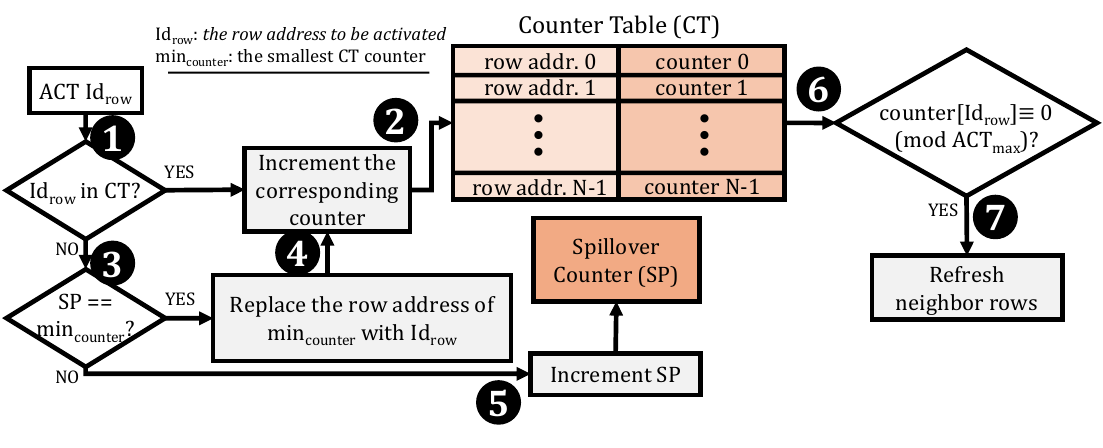}
    \vspace{-7mm}
    \caption{\omcr{3}{SMD-based} Deterministic RowHammer Protection (\drp{})}
    \label{fig:smd_graphene}
\end{figure}

The key idea of \drp{} is to maintain a per-bank \emph{Counter Table (CT)} to
track the $N$ most-frequently activated DRAM rows within a certain time interval
(e.g., refresh period of \trefw{}). Fig.~\ref{fig:smd_graphene} illustrates the
operation of \drp{}.\atbcrcomment{2}{Fix figure}\omcrcomment{4}{
try to address onur's comments on the figure
}


When \drp{} receives an \cmdact{}, it checks if the activated row address
($Id_{row}$) exists in CT~\circled{1}. If so, \drp{} increments the
corresponding CT counter by one~\circled{2}. Otherwise,
\drp{} finds the smallest counter value ($min_{counter}$) in CT and compares it
to the value of the \emph{spillover counter (SP)}~\circled{3}, which is
initially zero. \microrevlabel{FQ3}\microrevf{The key idea of SP is to track the activation count
of all DRAM rows that are \emph{not} actively tracked by any of the CT entries.
SP's value is an upper bound for the maximum number of activations any (not-actively-tracked)
DRAM row received.} 
If SP is equal to $min_{counter}$, \drp{} replaces the row
address corresponding to $min_{counter}$ in CT with $Id_{row}$~\circled{4} and
increments the corresponding CT counter by one~\circled{2}. If SP is smaller
than $min_{counter}$, \drp{} increments SP by one~\circled{5}. When a CT counter
is incremented in step~\circled{2}, \drp{} checks if the counter value is a
multiple of $ACT_{max}$~\circled{6}, which is the maximum number of times a row
can be activated without refreshing its neighbors \omcr{2}{(determined based on the RowHammer threshold~\cite{park2020graphene})}. If so, \drp{} refreshes the
neighbor \omcr{2}{rows} of $Id_{row}$~\circled{7}. To prevent the counters from overflowing,
\drp{} resets the CT counters and SP on every \trefw{} interval. 

To ensure that no row is activated more than $ACT_{max}$ without refreshing its
neighbor rows, the number of CT counters ($N$) must be configured \hhh{as follows}:

\vspace{-10pt}
\begin{equation}
    N > ({ACT_{\trefw}}/{ACT_{max}}) - 1
\end{equation}

\noindent where $ACT_{\trefw}$ is the maximum number of activations that the
MC can perform within a \trefw{} interval in a single bank. {In our evaluations, 
we set $ACT_{max}=512$ to show that the overheads of \drp{} are small even for 
DRAM chips that are \atbcr{1}{significantly} vulnerable to RowHammer.\microrevlabel{AQ4}\footnote{\microreva{When the 
\omcr{2}{maintenance operation time (MOT) mode} 
(\cref{subsec:impact_to_scheduling}) is enabled, 
\drp{} is configured with a smaller $ACT_{max}$ to provide the 
same security guarantee as when the MOT \omcr{2}{mode} 
is \emph{not} 
enabled. The reduction in $ACT_{max}$ is calculated as the 
number of ACT commands that the memory controller can issue to 
an aggressor row outside MOT. For example, if the DRAM chip 
switches into MOT mode every 3.9 µs (i.e., the MOT\_period is 
3.9µs), $ACT_{max}$ should be 3900/45 (MOT\_period/tRC) smaller 
than when the MOT \omcr{2}{mode} is \emph{not} enabled.}}}


\subsection{Use Case 3: Memory Scrubbing}
\label{subsec:smd_ecc_scrubbing}

\sloppypar{
To mitigate the increasing bit errors mainly caused by the continued DRAM
technology scaling, DRAM vendors equip their DRAM chips with on-die Error
Correction \hht{Codes}
(ECC)~\cite{patel2019understanding,patel2020bit,nair2016xed,kang2014co,patel2021harp,micron2017whitepaper,oh2014a,kwak2017a,kwon2017an}.
%
On-die ECC is designed to correct a single bit error assuming that a failure
mechanism is unlikely to incur more than one bit error in a
codeword~\cite{patel2020bit}. However, even \atbcr{1}{if} the assumption always
holds, a failure mechanism can gradually incur two or more bit \hht{errors} over
time.\atbcrcomment{3}{We are trying to motivate scrubbing here rather than make a statement about multi-bit errors (maybe I misunderstood your comment). I am not sure how to clarify the text further for now. I will revise based on your next feedback.} A widely-used technique for preventing the accumulation of an
uncorrectable number of bit errors is \emph{memory scrubbing}. Memory scrubbing
describes the process of periodically scanning the memory for bit errors in
order to correct \hht{them before more errors occur}.
}

\hht{We propose \mech{}-based Memory Scrubbing (\sms{}), which is an in-DRAM
maintenance mechanism that periodically performs scrubbing on DRAM chips with
on-die ECC.\iql{IQD2}{}\footnote{\iqrev{To preserve ECC transparency and enable system-level policies that depend on ECC scrubbing information (e.g., detecting and replacing faulty DRAM chips), \sms{} exposes critical ECC scrubbing information (e.g., the address of the DRAM row with the most corrected errors) to the memory controller via DRAM mode-status registers, similar to \omcr{2}{what} ECC Error Check and Scrub does in DDR5~\cite{jedecddr5c}.}}
Compared to conventional MC-based scrubbing, \sms{}
\omcr{2}{avoids} moving data to the MC by \omcr{2}{performing}
scrubbing \omcr{3}{completely} within the DRAM chip, \omcr{2}{thereby} 
reduc\omcr{2}{ing} 
the performance and energy overheads of \omcr{2}{MC-based} memory scrubbing.}

\sms{} operation resembles the operation of \fr{}. Similar to \fr{}, \sms{}
maintains \emph{pending scrub counter}, \emph{lock region counter}, and
\emph{row address counter}. \sms{} \hht{increments} the pending scrub counter at
fixed intervals of $tScrub$. When the pending scrub counter is greater than
zero, \sms{} attempts to lock the region indicated by the \emph{lock region
counter}. After locking the region, \sms{} performs a scrubbing
operation on the
row indicated by the \emph{row address counter}. 
The scrubbing operation takes
more time than a refresh operation as performing scrubbing on a row consists of
three time consuming steps for each codeword in the row: 1) reading the
codeword, 2) performing ECC decoding and checking for bit errors, and 3)
encoding and writing back the new codeword into the row only when the decoded
codeword contains a bit error. Refreshing a row takes $tRAS+tRP \approx 50ns$,
whereas scrubbing a row takes $tRCD+128*tBL+tRP \approx \SI{350}{\nano\second}$
when no bit errors are detected.\footnote{\hht{We assume ECC decoding/encoding
hardware is fully pipelined. In case of a bit error, writing a corrected
codeword incurs $4*tBL = \SI{2.5}{\nano\second}$ latency.}}
Therefore, \sms{} keeps a region locked for much longer than \fr{}.
When the
scrubbing operation is complete, \sms{} releases the lock region and increments
the \emph{lock region} and \emph{row address} counters, and decrements the
\emph{pending scrub counter}. 

\noindent
\iql{IQA4}{}\iqrev{\textbf{Rank-Level Memory Scrubbing.} Server memory modules typically use rank-level
ECC~\cite{patel2021harp, kim2015bamboo, synopsysRASreport,gong2018duo,kim2015bamboo,nair2016xed}.
\microrevlabel{DQ7}\microrevd{We discuss why rank-level ECC and in-DRAM ECC
scrubbing should be done separately but in a combined way in server modules. 
During a conventional DRAM \cmdread{} operation, a DRAM chip with in-DRAM ECC (e.g., a DDR5 chip~\cite{jedecddr5c}) 
performs error correction for the in-DRAM ECC codeword \emph{without} 1) writing 
the corrected codeword back and 2) keeping a record of the correction in mode-status registers~\cite{jedecddr5c}.
\atbcrcomment{2}{Yes. Which is the case for DDR5 at least.}
Thus, in a rank-level ECC scrubbing operation, the memory
controller \emph{cannot} distinguish between an in-DRAM-ECC-corrected codeword and a codeword with no errors.}
\microrevd{There\omcr{2}{fore}, the memory controller writes}
{back every codeword that it reads from DRAM for \emph{every} rank-level ECC scrubbing operation. Doing so} could incur performance overheads 
due to the long data bus turn-around latency in DRAM (e.g., $tCCD_{L}\__{WTR}$ in DDR5 is approximately \SI{30}{\nano\second}~\cite{luo2024ramulator2,jedecddr5c})
and energy overheads due to increased data movement between the MC and the DRAM chips. Instead, using \mech{}, the computing system could use 
both rank-level ECC scrubbing and in-DRAM ECC scrubbing to 1) leverage the (typically) stronger error correction 
capability of rank-level ECC and 2) leverage the performance and energy efficiency of in-DRAM ECC while preventing the accumulation of errors.}



\subsection{\revd{Other Use Cases}}
\label{subsec:other_use_cases}



\revd{To demonstrate\revlabel{RDC1} that SMD \omcr{3}{can enable many
other use cases}\agycr{2}{,} 
we \omcr{2}{describe \omcr{3}{three}} other use cases.}


\noindent
\revd{\textbf{Online DRAM Error Profiling.}}
\revd{\mech{} enables implementing a maintenance mechanism for online
profiling of DRAM errors that may occur when operating with increased refresh
period and reduced DRAM timing parameters~\cite{khan2014efficacy,liu2012raidr,kim2018solar,hassan2016chargecache,chang2016understanding,lee2017design,son2013reducing,hamamoto1998retention,liu2013experimental,patel2017reaper,qureshi2015avatar,chang2016low,patel2022case,patel2024rethinking,kim2018dram,kim2019d,yaglikci2024spatial}. Such a maintenance mechanism can be
used to \omcr{2}{exploit the variation in} DRAM cell characteristics \omcr{2}{within a} DRAM chip.}
\microrevc{To profile a DRAM region, the maintenance mechanism can use \mech{} to lock \omcr{2}{the} 
region and prevent the MC from accessing it while the region is
being profiled.}\microrevlabel{CQ3}
\microrevc{Because the profiling operation can cause bit flips, the
maintenance mechanism should temporarily buffer the original data that is stored
in the profiled region~\cite{kim2018solar,patel2017reaper}. For example, for profiling errors that occur at reduced
\trcd{}, the maintenance mechanism should first safely copy the data stored in a
row to another storage space (e.g., an unused row or an SRAM buffer), before
attempting to activate the row with reduced \trcd{}. The profil\omcr{2}{ing} mechanism can
internally share the profiling results with other maintenance mechanisms (e.g., \omcr{3}{maintenance 
mechanisms 1)~}for
skipping refreshes to rows with high retention 
times~\cite{patel2017reaper,qureshi2015avatar,liu2012raidr,khan2014efficacy,liu2013experimental}
\omcr{2}{\omcr{3}{and 2)~}changing the aggressiveness of RowHammer protection mechanisms based on per-row vulnerability~\cite{yaglikci2024spatial}}) and mechanisms for
improving DRAM access latency and 
energy efficiency~\cite{hassan2016chargecache,kim2018solar,lee2017design,ghose2018your}.}

\noindent
\revb{\textbf{\revlabel{RBQ1}Processing in/near Memory.} In the presence of an in-DRAM 
processing engine \microrevc{(as in e.g.,~\cite{devaux2019true,kim2022aquabolt,seshadri2013rowclone,seshadri2017ambit,ghose.ibmjrd19,mutlu2019processing,mutlu2020modern})}
\atbcrcomment{2}{Cite overview papers and more here},
\mech{} can help resolve access conflicts between the in-DRAM processing engine
and the MC. To do so, \mech{} can treat the in-DRAM processing
engine as a maintenance mechanism. The in-DRAM processing engine can use \mech{}
to lock a DRAM region that it will operate on.} 
\microrevlabel{CQ3}\microrevc{Because \mech{} does not allow
the MC to activate a row in a locked region, only the in-DRAM
processing engine will have access to the locked region until it completes the
processing and releases the region.}

\noindent
\atbcr{2}{\textbf{Power Management.}}\atbcrcomment{2}{One could argue the
majority of energy savings come from powering down the whole chip, so powering
down small regions would not help so much. There is one paper called GreenDIMM (?)
that proposes to control power delivery to subarrays and save energy. Should
we integrate that into here somehow??}
\agycr{2}{Modern DRAM chips implement self-refresh and power-down 
modes to save energy~\cite{jedecddr5c,jedec2020ddr4}. 
The whole DRAM chip is inaccessible in power-down mode, 
such that the DRAM chip does \emph{not} execute DRAM 
commands (e.g., ACT, PRE, RD, WR), but internally 
refreshes DRAM cells (in self-refresh mode) to maintain data integrity. 
Although a DRAM chip consumes significantly lower energy in 
self-refresh and power-down modes, a memory request 
accessing a small piece of data (e.g., 64B of a cache block out of 16GB of a DRAM module) 
causes the DRAM chip to \omcr{3}{completely} get out of self-refresh or power-down modes and consume 
significantly larger energy. To make matters 
worse, such mode switching
introduces additional delays for memory requests~\cite{jedec2020ddr4,jedecddr5c,thomas2012predictor}. 
To overcome these challenges, SMD can be leveraged to 
\omcr{3}{partially} power down DRAM chips by powering down rarely 
accessed memory regions, such that DRAM energy consumption 
can be reduced for a large portion of the DRAM chip while 
memory requests targeting frequently accessed memory 
regions do \emph{not} experience additional delays. 
\atbcr{3}{Because SMD does not allow the MC to activate a row in a locked (e.g., powered down) region,
only activate commands targeting powered-up regions would proceed without getting \actnack{}'d.}
SMD
would reject activate commands targeting powered-down memory regions
until th\atbcr{3}{ose} regions are powered on.}

\subsection{Other Maintenance Operation Improvements}
\label{sec:other_optimizations}
\atbcr{3}{SMD can seamlessly enable optimizations for the three maintenance
operations we extensively describe (SMD-FR~\cref{subsec:smd_refresh}, 
SMD-DRP~\cref{subsec:smd_rowhammer_protection}, and SMD-MS~\ref{subsec:smd_ecc_scrubbing}).}
At a high-level, any optimization idea or technique that 
is independent from the organization of multiple DRAM chips into a
module or a rank (i.e., any idea that can be leveraged by some 
technique in a single DRAM chip) can be implemented in an \mech{} chip. 
Implementing such an idea or technique in an \mech{} chip does \emph{not} 
require further changes to the DRAM interface.

\noindent
\textbf{\microrevb{\atbcr{3}{Refresh Overhead Mitigation.}}}
\microrevlabel{BQ1}\microrevb{A large set of prior work proposes various ideas and techniques that mitigate DRAM refresh overhead (e.g.,~\cite{chang2014improving, liu2012raidr,
qureshi2015avatar, nair2014refresh, baek2014refresh, bhati2013coordinated,
cui2014dtail, emma2008rethinking, ghosh2007smart, isen2009eskimo,
jung2015omitting, kim2000dynamic, luo2014characterizing, kim2003block,
liu2012flikker, mukundan2013understanding, nair2013case, patel2005energy,
stuecheli2010elastic, khan2014efficacy, khan2016parbor, khan2017detecting,
venkatesan2006retention, patel2017reaper, riho2014partial, hassan2019crow,
kim2020charge, nguyen2018nonblocking, kwon2021reducing}). 
Many of these ideas and techniques could improve in-DRAM autonomous
maintenance operations. \microrevb{One prior technique mitigates DRAM refresh overhead~\cite{nguyen2018nonblocking} by cleverly leveraging DRAM module organization and rank-level ECC to alleviate the DRAM bandwidth overhead of refresh operations. While \mech{} does \emph{not} preclude this technique from being implemented, implementing it as mainly described in~\cite{nguyen2018nonblocking} would require modifications to the DRAM interface. However, the key idea of~\cite{nguyen2018nonblocking} could still be leveraged inside an \mech{} chip by combining \atbcr{2}{1)~on-die ECC~\cite{patel2019understanding,patel2020bit,nair2016xed,kang2014co,patel2021harp,micron2017whitepaper,oh2014a,kwak2017a,kwon2017an}, which already exists (e.g., in DDR5~\cite{jedecddr5c})
and 2)~finer-granularity (\atbcr{2}{DRAM-mat}-level) refresh operations (as described in~\cite{nguyen2018nonblocking}) in a future DRAM design.}}}



\noindent
\textbf{\atbcr{3}{Other RowHammer Mitigations.}}
Many prior works \omcr{4}{provide} RowHammer mitigation \omcr{4}{techniques} (e.g.,~\cite{olgun2024abacus,yaglikci2024spatial, bostanci2024comet,
seyedzadeh2017counter,son2017making,yaglikci2021blockhammer, park2020graphene,
you2019mrloc, seyedzadeh2018cbt, aweke2016anvil, cojocar2019exploiting,
kim2014flipping, frigo2020trrespass, herath2015these, van2018guardion, konoth2018zebram, qureshi2022hydra, woo2023scalable, wi2023shadow, kim2023ddr5, bennett2021panopticon}). \atbcr{3}{The key ideas of these works can be implemented
using SMD to enable more performance-, area-, and energy-efficient and robust in-DRAM RowHammer protection.}

\section{\atbcr{2}{Hardware \omcr{3}{Implementation and} Overhead}}
\label{sec:hw_overhead}

\subsection{\atbcr{2}{\omcr{3}{DRAM} Interface Modifications}}

\hht{An \mech{} chip \atb{needs to} send \actnack{} signals to \atb{the}
memory controller (MC).} \atb{\mech{} can \iql{IQA3}{}\iqrev{1) repurpose the existing 
\texttt{alert\_n} pin in DDR4 chips~\cite{jedec2020ddr4, kwon2014understanding}} or 2)} introduce an extra physical pin 
to transmit \actnack{} signals. 

\noindent
\atb{\textbf{1) \texttt{alert\_n}.}} \texttt{alert\_n} is currently
used to inform the MC that the DRAM chip detected a CRC or parity check failure
on the issued command, and thus the MC must issue the command again. The
\texttt{alert\_n} signal can simply be asserted not only on CRC or parity check
failure, but also when the MC attempts accessing a row in a locked region. This 
approach does \atbt{\emph{not}} require \atbt{an} additional physical pin. \atb{However, \iqrev{1)}~assigning 
multiple \atbt{meanings} to an already ambiguous\omcr{2}{ly-defined} \texttt{alert\_n} 
signal~\cite{jedec2020ddr4} could complicate MC design and \iql{IQC2}{}\iqrev{2) \texttt{alert\_n}, as currently defined in the standard~\cite{jedec2020ddr4}, is an open drain signal that is \atbcr{3}{set and reset slowly}}\atbcrcomment{2}{I assume this means that the signal
has some variable delay in the range of tens of nanoseconds. We placed
the statement here to address a comment. It uses the terminology
the reviewer uses.}\atbcrcomment{3}{removed the precise qualifier}.}

\noindent
\atb{\textbf{2) Introducing a new pin.} To potentially simplify MC design}, 
\mech{} can introduce a new pin to transmit
\hpcarevlabel{Rev.D/C2}\actnack{} signals. \hht{\emph{Only} one new pin between the MC and \hpcarevd{all} memory
device\hpcarevd{s} (e.g., \hpcarevd{all DRAM chips in a memory channel or a rank}) is enough for systems that use rank-based 
DRAM chip organizations \microrevlabel{DQ4}(e.g., DDR\microrevd{5} DIMMs). \hpcarevd{For example, \microrevd{96} new pins are needed for 
a high-end system with \microrevd{12} memory \microrevd{module channels} equipped with 4-rank \microrevd{DDR5 modules} in each channel \microrevd{(each DDR5 rank has two memory channels)}.
However, a system of this scale already has \microrevd{\omcr{2}{>6K} total} processor pins \microrevd{(e.g., \omcr{2}{6,096} in AMD SP5 Socket~\cite{sp5socket})}, and \microrevd{96} new pins would amount to
a relatively small \microrevd{1.6\%} increase in pins.} \atb{Similar to how \omcr{2}{the} \texttt
{alert\_n} signal works, the module can} send the MC a single \actnack{} when 
any of the per-chip \actnack{} signals \omcr{2}{is} asserted.}

\subsection{\atbcr{2}{DRAM Chip Modifications}}


%

\noindent
\textbf{DRAM Chip Modifications for SMD.} We use CACTI \atbcr{3}{6.0}~\cite{muralimanohar2009cacti} 
to evaluate the hardware overhead of
the changes that \mech{} introduces \omcr{2}{in} a conventional DRAM bank
\hht{(highlighted in Fig.~\ref{fig:smd_bank_organization})} assuming
\SI{22}{\nano\meter} technology.\hpcanew{\footnote{\hpcanew{Capitalizing on the latest DRAM technology libraries to implement the changes that SMD introduces would likely provide more accurate area overhead estimations. However, such libraries and tools are proprietary.}}}
\hht{Lock Region \omcr{3}{Bitvector} (LRB)} is a small \omcr{3}{bitvector}
that stores a single bit for each lock region \iql{IQC1}{}\iqrev{in a DRAM bank} to indicate whether or not the
lock region is under maintenance. In our evaluation, we assume that a bank is
divided into $16$ lock regions. Therefore, LRB consists of only $16$ bits, which
are indexed using a 4-bit lock region address. According to our evaluation, an
LRB incurs \emph{only} \SI{32}{\micro\meter\squared} area overhead per bank. The
area overhead of all LRBs in a DRAM chip is \emph{only} 0.001\% of a
\SI{45.5}{\milli\meter\squared} DRAM chip. The access time of an LRB is
\SI{0.053}{\nano\second}, which is \emph{only} 0.4\% of typical row activation
latency (\trcd{}) of \SI{13.5}{\nano\second}.\atbcrcomment{2}{What is wrong? I do not understand. These are what CACTI reports.}
 
\revc{\revlabel{RCC2}\mech{} adds a \hht{per-region} RA-latch to enable accessing one lock region
while another is under maintenance. An RA-latch stores a pre-decoded row
address, which is provided by the global row address decoder, and drives the
local row decoders where the row address is fully decoded. \reva{This design builds 
on the basic design proposed in SALP~\cite{kim2012case} and refresh-access 
parallelization introduced in~\cite{chang2014improving,zhang2014cream}.} According to our
evaluation, \hht{all} RA-latches incur \hht{a} total area overhead of \atbcr{2}{\param{1.1\%}} \omcr{2}{in}
a \omcr{2}{typical} \SI{45.5}{\milli\meter\squared} DRAM chip. An RA-latch has only
\SI{0.028}{\nano\second} latency\atbcrcomment{2}{This is not CACTI-based. We follow SALP's methodology (multiply area numbers reported by a paper on latches) for RA-latches.}, which is negligible compared to \trcd{}.}

\noindent
\textbf{DRAM Chip Modifications for Maintenance Mechanisms.}
Besides these changes that are the core of the \mech{} substrate, a
particular maintenance mechanism may incur additional area overhead.
We evaluate the DRAM chip area overhead of the \omcr{3}{three} maintenance mechanisms presented
in \cref{subsec:smd_refresh}\atbcr{3}{,~\cref{subsec:smd_rowhammer_protection}, 
and \cref{subsec:smd_ecc_scrubbing}.} 
The simple \omcr{4}{fixed-rate} refresh mechanism, \fr{},
requires only \SI{77.1}{\micro\meter\squared} additional area in a DRAM chip.
The DRAM chip area overhead of \iscanew{the} refresh
mechanism is less than 0.1\% of a typical \SI{45.5}{\milli\meter\squared} DRAM
chip. 
\drp{} requires a large
\emph{Counter Table} with 1224 counters per bank for the RowHammer threshold
value $ACT_{max}=512$ that we use in our performance evaluation.
\drp{} requires \SI{3.2}{\milli\meter\squared} area, which is 7.0\% of a
typical \SI{45.5}{\milli\meter\squared} DRAM chip.\footnote{\omcr{3}{Note that the area overhead comes
especially from Graphene's expensive CAM structures~\cite{park2020graphene}. Cheaper RowHammer
mitigation mechanisms can be implemented with SMD, as we describe in~\cref{sec:other_optimizations}.}}
The control logic of \sms{} is similar to the control
logic of \fr{} and it requires only \SI{77.1}{\micro\meter\squared} additional
area excluding the area of the ECC engine, which is already implemented by DRAM
chips that support in-DRAM ECC.\atbcrcomment{2}{We did not open source these. Can be part of our TODOs for the extended version.}

\subsection{\atbcr{2}{Memory Controller Modifications}}
\iql{IQB1/\\IQD1}{}\iqrev{We slightly modify the MC's scheduling mechanism to retry a rejected
\cmdact{} command as we explain in \cref{subsec:control}. \hht{Upon
receiving an \actnack{}, the MC marks the bank as precharged.}
\hht{An existing} MC already implements control circuitry to pick an appropriate
request from the request queue and issue the necessary DRAM command based on the
DRAM bank state (e.g., \cmdact{} to a precharged bank or \cmdread{} \hht{if the
corresponding row is already open}) by respecting the DRAM timing parameters.
The \emph{ACT Retry Interval (ARI)} is simply a new timing parameter that
specifies the minimum time interval for issuing an \cmdact{} to a lock region
after receiving an \actnack{} \hht{from} the same region. Therefore, \mech{} can
be implemented in existing MCs with \hht{only} slight modifications by
leveraging the existing request scheduler \hht{(e.g.,
FR-FCFS~\cite{mutlu2007stall})}.} 

\noindent
\textbf{\iql{IQB1}{}\iqrev{Tracking Locked Regions.}}
\iqrev{To apply the ARI timing parameter (e.g., to an ACT command targeting a \atbcr{2}{locked} region), the 
memory controller tracks which regions are locked. The memory controller could store the address of the locked 
region in every DRAM bank (since only one region can be under maintenance at a time in a bank). 1 rank address 
bit, 4 bank address bits, and 4 lock region address bits (9 bits) are sufficient to track every 
under-maintenance lock region in 2 ranks and 16 banks (\atbcr{2}{32} unique banks) \atbcr{2}{across all chips on 
a DRAM module}. Therefore, the storage cost for tracking locked regions is only 288 bytes for the evaluated 
memory channel with a dual-rank x8 memory module.} 

\noindent
\textbf{\iql{IQB1}{}\iqrev{Address Mapping Schemes.}} \iqrev{\mech{} does \emph{not} require modifications to how a physical address is mapped to a memory-controller-visible DRAM address. In~\cref{sec:evaluation}, we evaluate \mech{} using a mapping scheme that interleaves consecutive cache blocks across channels, \atbcr{2}{columns, ranks, banks, and rows, in that order.}\atbcrcomment{2}{Put cache block (CB) 0 in channel 0 column 0, CB 1 in channel 1 column 0, CB 2 in channel 0 column 1, ...} 
We also evaluate \mech{} using the mapping scheme described in~\cite{kaseridis2011minimalist} that aims to exploit bank-level parallelism. For this mapping scheme, SMD-FR provides 8.7\% speedup over the baseline system, on average across 4c-high workloads. We leave detailed evaluation of \mech{} with different address mapping schemes for future work.}


\vspace{0.5em}
\section{Experimental Methodology}
\label{sec:methodology}
\vspace{0.5em}


We extend Ramulator~\cite{kim2015ramulator,ramulatorgithub} to implement
\hht{and evaluate} the \param{three} \mech{} maintenance mechanisms \hht{(\fr{}, \drp{}, and \sms{})} that we describe in
\cref{subsec:smd_refresh}, \cref{subsec:smd_rowhammer_protection}, and \cref{subsec:smd_ecc_scrubbing}. We use DRAMPower~\cite{drampowergithub,
chandrasekar2011improved} to evaluate DRAM energy consumption. We use Ramulator
in CPU-trace driven mode \hht{executing} traces of representative sections of
\hht{our workloads collected with} a custom Pintool~\cite{luk2005pin}. \hht{We}
warm-up the caches by fast-forwarding 100 million (\atb{M}) instructions. We simulate each
\hht{representative trace} for 500\atb{M} instructions (for \hht{multi}-core
simulations, until each core executes at least 500\atb{M}
instructions).

We use the system configuration provided in Table~\ref{table:system_config}\atbcrcomment{3}{Fix spacing} in
our evaluations. Although our evaluation is based on DDR4 DRAM, the
modifications required to enable \mech{} can be adopted in other DRAM standards,
as we explain in \cref{sec:hw_overhead}.

\vspace{1em}

\begin{table}[h!] \caption{Simulated system
    configuration}
    \centering \renewcommand{\arraystretch}{1.4}
    \resizebox{\linewidth}{!}{
    \begin{scriptsize}
    \begin{tabular}{m{2.1cm} m{4.9cm}}
        \hline
        \textbf{Processor} & \SI{4}{\giga\hertz} \& 4-wide issue CPU core, 1-4
        cores, 8~MSHRs/core, 128-entry instruction window\\
        \hline
        \textbf{Last-Level Cache} & \SI{64}{\byte} cache-line, 8-way associative, \SI{4}{\mebi\byte}/core \\
        \hline
    \textbf{Memory Controller} & \makecell[l]{64-entry read/write request
        queue,\\ FR-FCFS-Cap~\cite{mutlu2007stall} \omcr{2}{with Cap=7}}\\
        \hline
        \textbf{DRAM} & DDR4-3200~\cite{jedec2020ddr4}, \SI{32}{\milli\second} refresh period,
            4~channels, 2~ranks, 4/4~bank groups/banks, 128K-row bank, 512-row
            subarray, \SI{8}{\kibi\byte} row size \\
        \hline
    \end{tabular}
    \end{scriptsize}
    } 
\label{table:system_config}
\end{table}
\vspace{1em}

\noindent
\reva{\textbf{\revlabel{RAC4}DDR4 Baseline.} The baseline system 1)~uses per-rank refresh\footnote{\reva{The DDR4 standard does \emph{not} support per-bank refresh.}}
and 2)~does \emph{not} perform ECC scrubbing.}


\noindent
\textbf{Workloads.} We evaluate \param{\hpcanew{62}} single-core applications from four benchmark
suites: SPEC CPU2006~\cite{spec2006}, \hpcanew{SPEC CPU2017~\cite{spec2017},} TPC~\cite{tpc}, STREAM~\cite{stream}, and
MediaBench~\cite{fritts2005mediabench}. We classify the workloads in three
\iscanew{memory intensity groups measured using}
\hht{misses-per-kilo-instructions} (MPKI) in the last-level cache (LLC): low ($MPKI\!<\!1$), medium ($1\!\leq\!MPKI\!\leq\!10$), and high ($MPKI\!\geq\!10$). 
We randomly combine
single-core workloads to create multi-programmed workloads. \hht{Each
multi-programmed workload group}, \texttt{4c-low}, \texttt{4c-medium}, and
\texttt{4c-high}, contains $20$ four-core workloads.
\atbcrcomment{2}{20 LLLL, 20 MMMM, and 20 HHHH workload mixes.}

\noindent
\textbf{Metrics.} We use Instructions Per Cycle (IPC) to evaluate the
performance of single-core workloads. For multi-core workloads, we evaluate the
system throughput using the weighted speedup
metric~\cite{eyerman2008system,snavely2000symbiotic, michaud2012demystifying}.\atbcrcomment{2}{We did not look at other metrics.}

\noindent
\iscanew{\textbf{Comparison Points.} We compare \mech{} to memory controller/DRAM co-design techniques. \microrevlabel{DQ1}\microrevd{First,
DARP~\cite{chang2014improving} intelligently schedules per-bank refresh commands to idle banks while the
memory controller is in write mode\atbcr{1}{. Doing so reduces} delays imposed by refresh operations on memory demand requests 
(e.g., load instructions executed by the processor).}
Second, DSARP~\cite{chang2014improving} implements DARP and \omcr{2}{also} modifies DRAM 
chip \atbcr{1}{design} and the \atbcr{1}{DRAM} interface to \atbcr{2}{concurrently} perform a refresh operation in one subarray 
and a memory access in another subarray, thereby hid\atbcr{1}{ing} the latency of refresh operations and reduc\atbcr{1}{ing} delays they impose on memory requests.
We modify DARP and DSARP such that they \omcr{2}{also} \atbcr{1}{reduce delays imposed on memory 
requests by}
Deterministic RowHammer 
Protection's (\cref{subsec:smd_rowhammer_protection}) victim row refresh
operations.} 

\noindent
\iscanew{\textbf{\omcr{3}{Parameter Values of} Maintenance Mechanisms \omcr{3}{and SMD}.}}
\iscanew{\fr{} (\cref{subsec:smd_refresh}) refreshes a DRAM row
every \SI{32}{\milli\second}. \atbcr{3}{\fr{} refreshes
8 DRAM rows when it locks a DRAM region (RG~=~8) with each refresh operation
and postpones up to 8 such
refresh operations (N~=~8).}} 
Based on~\cite{liu2012raidr}, for SMD-VR, we conservatively assume 0.1\%
of rows in each bank need to be refreshed every \SI{32}{\milli\second} while
the rest retain their data correctly for \SI{128}{\milli\second} and more. SMD-VR 
uses a 8K-bit Bloom 
Filter with 6 hash functions. 
SMD-PRP refreshes the victims of an activated row
with a high probability, i.e., $P_{mark}$= 1\%.
\atb{\drp{} \atbcr{3}{(\cref{subsec:smd_rowhammer_protection})} refreshes the victims before the 
\atb{aggressor} is activated $ACT_{max}$ (512) times during a \trefw{} (\SI{32}{\milli\second}).}
\reva{\sms{} \atbcr{3}{(\cref{subsec:smd_ecc_scrubbing})} operates with an aggressive 5-minute scrubbing
period.}\footnote{\hht{We analyze \drp{}, and \sms{} at various
configurations. Their \hhh{performance} overheads are relatively
low even at more aggressive settings.}}
\atbcr{2}{We configure SMD with 16 lock regions (\cref{subsec:locking_mechanism})
in a DRAM bank unless stated otherwise.
\atbcr{3}{SMD uses an ACT Retry Interval ($ARI$) of \SI{62.5}{\nano\second} (\cref{subsec:control}).}}

\begin{figure*}[!t]
    \centering
    \includegraphics[width=\linewidth]{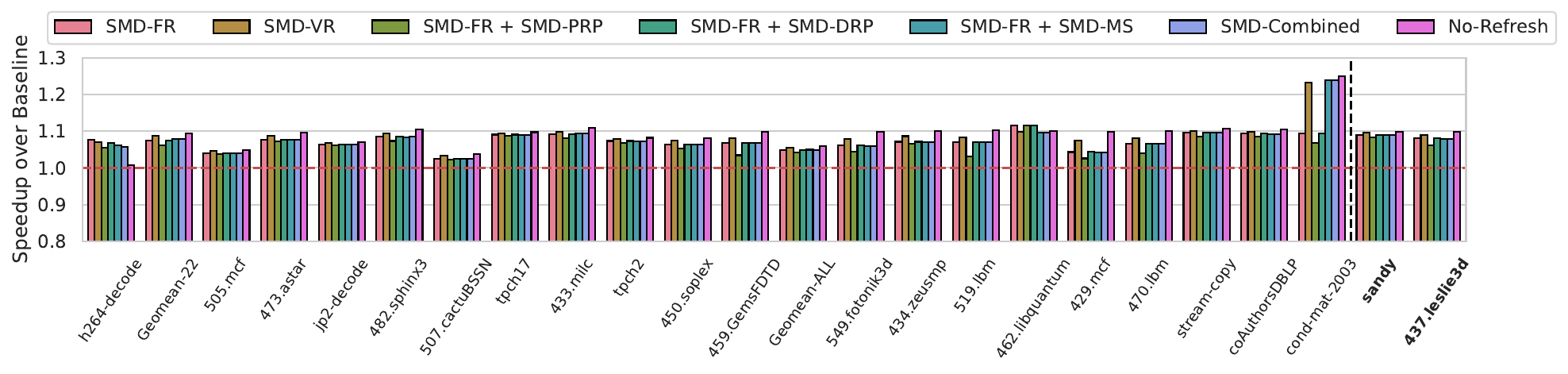}
    \caption{Single-core speedup \hht{over \revcommon{the baseline} DDR4 system \revcommon{(horizontal \microrevcommon{dashed} \atbcr{2}{red} line).}} \microrevcommon{The baseline system is the same DDR4 baseline for each configuration.} \atbcr{3}{{Note that y-axis starts at $y=0.80$.}}}
    \label{fig:single_core_perf}
\end{figure*}

\noindent
\iscanew{\textbf{Evaluated System Configurations.} \atbcr{3}{We evaluate 10 different systems:
1)~the DDR4 baseline system as described in Table~\ref{table:system_config},
2)~SMD-FR that can concurrently perform a refresh in a lock region and a memory access in another,
3)~SMD-VR inherits SMD-FR's abilities
and refreshes different rows at different refresh rates,
4)~SMD-FR + SMD-PRP inherits SMD-FR's abilities
and can concurrently perform a probabilistic victim
row refresh operation in a lock region and a memory
access in another,
5)~SMD-FR + SMD-DRP\atbcrcomment{4}{Fixed instances of
DPD} inherits SMD-FR's abilities and can concurrently
perform a deterministic (based on Graphene's~\cite{park2020graphene} tracking
algorithm)
victim row refresh operation in a lock region and a memory access in another,
6)~SMD-FR + SMD-MS\footnote{\atbcr{3}{We evaluate SMD-DRP and SMD-MS with
SMD-FR as the refresh mechanism since SMD-FR is very simple and
easy to implement.}} inherits SMD-FR's abilities and performs in-DRAM memory scrubbing,
7)~SMD-Combined implements all three SMD-based maintenance mechanisms (SMD-FR + SMD-DRP + SMD-MS),
8)~DARP-Combined inherits the abilities of DARP~\cite{chang2014improving} and implements
the memory-controller-based versions of fixed rate refresh (FR), 
deterministic RowHammer protection (DRP), and memory scrubbing (MS),
9)~DSARP-Combined inherits the abilities of DSARP~\cite{chang2014improving} and implements
the memory-controller-based (MC-based) versions of all three maintenance mechanisms, and
10)~No-Refresh is a hypothetical system configuration that
does \emph{not} perform any maintenance operations.}
\atb{MC-based implementations
of \mech{} maintenance mechanisms are functionally equivalent 
to \mech{}-based ones (e.g., \omcr{4}{SMD-Combined}
and \omcr{4}{MC-Combined} are configured
to provide the same RowHammer prevention guarantees).}}


\section{\omcr{2}{Performance and Energy Results}}
\label{sec:evaluation}

We evaluate the performance and energy efficiency of \mech{}-based maintenance
mechanisms. 
\atbcr{2}{Our results show that 1) substantial system performance and DRAM energy benefits
accompany the system robustness improvements provided by SMD that performs
periodic refresh, RowHammer protection, and memory scrubbing in DRAM,
2) \atbcr{3}{performance and energy of SMD incorporating these techniques is very similar to}
those of a hypothetical system that does \emph{not} perform any DRAM maintenance operations, 
and 3) SMD outperforms a \omcr{3}{state-of-the-art} work (DSARP~\cite{chang2014improving})
that can concurrently perform a maintenance operation
in one DRAM subarray and a memory access in another.}\atbcrcomment{2}{Justify why we
do not have DRP or MS alone.}\atbcrcomment{3}{These are the most important findings imo. We can add more
based on your feedback.}

\subsection{Single-core Performance}
\label{subsec:single_core_perf}

\iscanew{Fig.~\ref{fig:single_core_perf} shows the speedup of the 22 single-core workloads 
\atbcr{2}{that have at least 10 Last-Level Cache Misses per Kilo Instruction (LLC MPKI)}\atbcrcomment{3}{LLC MPKI can be said here as we 
do not use it anywhere else} 
\omcr{3}{with} 
\atbcr{2}{SMD-FR, SMD-VR, SMD-FR + SMD-PRP, \atbcr{3}{SMD-FR +} SMD-DRP, \atbcr{3}{SMD-FR +} SMD-MS}, \omcr{3}{SMD-Combined,} \atbcr{3}{and} No-Refresh, over the baseline system \atbcr{2}{that only performs memory-controller-based periodic refresh operations}.} \omcr{3}{We make five major observations.}

\atbcr{2}{First, SMD-FR provides 4.8\% average speedup across all
workloads. Although SMD-FR and DDR4 have similar average latency
to refresh a single DRAM row, SMD-FR outperforms \omcr{3}{baseline}
DDR4 refresh 
due to two main
reasons: \omcr{3}{SMD-FR} 1)~enabl\omcr{3}{es} 
concurrent access and refresh of two different
\emph{lock regions} within a DRAM bank and 
2)~completely eliminat\omcr{3}{es} the
\cmdrefresh{} commands on the DRAM command 
bus\omcr{3}{, thereby reducing} the command bus utilization.
Second, SMD-VR provides a higher 5.5\% average speedup
across all workloads as it mitigates the overhead of periodic 
refresh operations by refreshing different rows at
different refresh rates.
Third, SMD-FR + SMD-DRP (SMD-FR + SMD-PRP) provides 4.8\%
(4.2\%)
average 
speedup across all workloads. 
We observe that
for a RowHammer threshold of 512 row activations (\cref{subsec:smd_rowhammer_protection}),
SMD-DRP and SMD-PRP can almost completely \emph{overlap} the latency of 
victim row refresh operations in one lock region with 
the latency of a memory access
to another lock region.
Fourth, SMD-FR + SMD-MS provides 5.0\% average speedup across
all workloads. SMD enables low-overhead
in-DRAM memory scrubbing
for a relatively high memory scrubbing rate (each DRAM 
cell gets scrubbed every 5 minutes).
The scrubbing operations induce negligible performance overhead: The average speedup 
provided by SMD-FR + SMD-MS is \omcr{3}{within <0.1\%} of
SMD-FR \omcr{3}{if} we exclude cond-mat-2003.\footnote{\atbcr{2}{\atbcr{3}{SMD-FR + SMD-MS 
provides a relatively high 23.9\%
speedup for cond-mat-2003. \microrevf{SMD-FR provides a smaller speedup than SMD-Combined \emph{only} 
for cond-mat-2003. For the same workload, SMD-FR has 5.1\% higher average memory access latency than 
SMD-Combined. Even though SMD-Combined rejects (\actnack{}s) more activate commands (15622) than SMD-FR (14286) on average across all four DRAM channels, in SMD-Combined, load instructions at the head of the processor core's reorder buffer commit earlier 
on average than in SMD-FR. Consequently, SMD-Combined is less frequently bottlenecked by reorder buffer stalls (15.5\% of execution time) 
than SMD-FR (40.2\% of execution time). We attribute SMD-FR's relatively poor performance for cond-mat-2003 to reorder buffer stalls.
We hypothesize that the duration and the frequency of \actnack{}s in SMD-FR affect main memory request scheduling
in a way that older requests in the reorder buffer experience higher memory latency.
}}}}
Fifth, SMD-Combined provides 5.0\% average speedup across all workloads. 
SMD-Combined provides 84.7\% of the speedup provided 
by No-Refresh on average across all evaluated single-core workloads,
while \omcr{3}{also} improving system robustness with 
RowHammer protection and memory scrubbing \omcr{3}{(which No-Refresh
does \emph{not} do)}.}

\ignorerev{
\iscanew{We make \param{\microrevd{four}} key observations. 
First, SMD-Combined provides a comparable average speedup of \param{5.\atbcr{1}{0}}\% to
No-Refresh's \param{5.\atbcr{1}{9}}\% as SMD-Combined eliminates a large fraction of the maintenance overhead by allowing 
the memory controller to access lock regions that are \emph{not} under maintenance.
\iql{IQE5}{}\iqrev{Second, SMD-Combined outperforms MC-Combined-DSARP and MC-Combined-DARP by \param{0.6}\% and \param{3.9}\%, respectively, \microrevd{on average across all workloads}\microrevlabel{DQ1}. 
While MC-Combined-DSARP can concurrently access main memory and perform a maintenance operation, 
it needs to issue a DRAM command (e.g., a per bank refresh command for periodic refresh) for each maintenance operation
and delay other DRAM commands that the memory controller issues to serve main memory requests.
In contrast, SMD-Combined autonomously performs each maintenance operation inside the DRAM chip and does \emph{not} incur delays
for DRAM commands (except those incurred by \actnack{}s).} MC-Combined-DARP 
\emph{cannot} concurrently perform a maintenance operation and access main memory, thereby \atbcr{1}{MC-Combined-DARP} performs worse than both SMD-Combined and MC-Combined-DSARP. We make similar observations on SMD-FR's, DSARP's, and DARP's performance. These observations are in line with prior work~\cite{chang2014improving}.
Third, intelligent refresh scheduling (DARP) provides
\param{1.1}\% average speedup across all workloads by reducing the impact of refresh operations on 
the latency of memory requests. SMD-FR-1LR incurs \param{2.9}\% slowdown over DARP. Although SMD-FR-1LR
locks a bank \emph{only} when there is no active row in a bank, \microrevlabel{DQ1}\microrevd{it does \emph{not} have a global view over 
memory requests in memory controller read/write queues, unlike DARP. \atbcr{1}{Therefore, SMD-FR-1LR} \emph{cannot} opportunistically perform 
refresh operations while the memory controller is in write mode.}}
\microrevlabel{DQ2}\microrevd{Fourth, we observe that SMD-Combined and No-Refresh provide 
substantially higher speedups for cond-mat-2003 than other workloads. We find 
that cond-mat-2003's average memory latency is 15.8\% and 16.7\% smaller than the DDR4 baseline, for SMD-Combined 
and No-Refresh, repectively. We attribute 1) the decrease in average memory latency to 
the absence of long-latency (\SI{350}{\nano\second}) per-rank refresh operations, and 2) the speedups provided by
SMD-Combined and No-Refresh to cond-mat-2003's sensitivity to memory latency.}\microrevlabel{DQ3 \& FQ4}
\microrevf{SMD-FR provides a smaller speedup than SMD-Combined \emph{only} 
for cond-mat-2003. For the same workload, SMD-FR has 5.1\% higher average memory access latency than 
SMD-Combined. Even though SMD-Combined rejects (\actnack{}s) more activate commands (15622) than SMD-FR (14286) on average across all four DRAM channels, in SMD-Combined, load instructions at the head of the processor core's reorder buffer commit earlier 
on average than in SMD-FR. Consequently, SMD-Combined is less frequently bottlenecked by reorder buffer stalls (15.5\% of execution time) 
than SMD-FR (40.2\% of execution time). We attribute SMD-FR's pathological performance for cond-mat-2003 to reorder buffer stalls.
We hypothesize that the duration and the frequency of \actnack{}s in SMD-FR affect main memory request scheduling
in a way that older requests in the reorder buffer experience higher memory latency.
}
}

\subsection{Multi-core Performance}
\label{subsec:multi_core_perf}
\begin{figure}[!h]
    \centering
    \includegraphics[width=\linewidth]{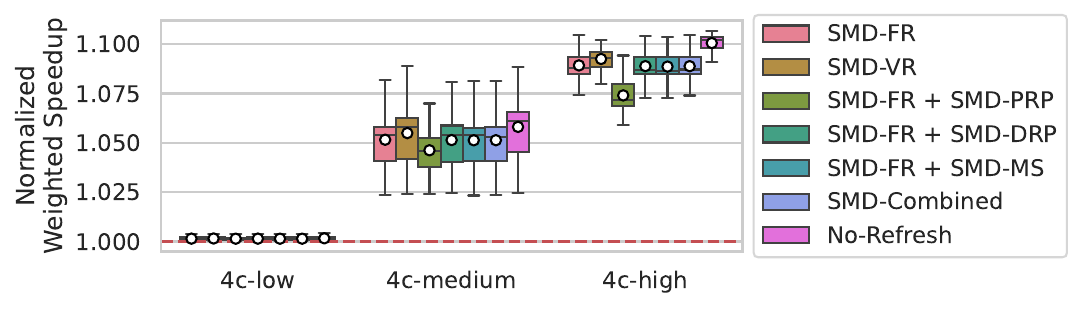}
    \caption{\atbcr{3}{Normalized weighted speedup 
    (\omcr{4}{to} the baseline DDR4 system represented by the horizontal red line) for four-core
    workloads}} 
    \label{fig:multi_core_perf}
\end{figure}

\atbcrcomment{4}{WARNING: Figure 7 appears before 6}
Fig.~\ref{fig:multi_core_perf} shows weighted speedup (normalized to the
weighted speedup of the DDR4 baseline) \omcr{3}{of} \hht{60 four-core workloads (20 per
memory intensity level)} \atbcr{3}{with SMD-FR, SMD-VR, SMD-FR + SMD-PRP, SMD-FR + SMD-DRP,
SMD-FR + SMD-MS, SMD-Combined, and No-Refresh}. \hht{The \iscanew{white circles inside each box} (error lines) represent the average
(minimum and maximum) \omcr{3}{normalized} weighted speedup across the 20 workloads in the
corresponding group.} 
\omcr{3}{We make \param{three} major observations.}

\atbcr{2}{First, SMD provides substantial performance improvements for 
memory intensive workloads with relatively 
high LLC MPKI (i.e., 4c-medium and 4c-high workloads).
For example, SMD-Combined
provides 5.1\% (8.2\%) and 8.9\% (10.5\%) average (maximum) speedup
across all evaluated 4c-medium and 4c-high workloads, respectively.
SMD does so by overlapping the latency of a majority of the maintenance operations
with the latency of memory demand requests (e.g., load instructions). 
For 4c-high workloads, SMD-Combined provides 88.3\% of
the average speedup provided by No-Refresh
\omcr{3}{even though SMD-Combined performs periodic refresh (\cref{subsec:smd_refresh}),
RowHammer protection (\cref{subsec:smd_rowhammer_protection}),
and frequent memory scrubbing (\cref{subsec:smd_ecc_scrubbing})
that No-Refresh does \emph{not} perform.}
Second, all evaluated SMD configurations provide similar speedups.
For example, SMD-FR, SMD-FR + SMD-DRP, SMD-FR + SMD-MS, and SMD-Combined
all provide \omcr{4}{approximately} 8.9\%; SMD-VR provides 9.2\%
and SMD-FR + SMD-PRP provides 7.4\% average speedup across
the evaluated 4c-high workloads.
\omcr{3}{This indicates that SMD is able to effectively hide
the latencies of RowHammer protection and memory scrubbing
operations.}
Third, SMD provides small performance improvements for non-memory-intensive
workloads with relatively low LLC MPKI (i.e., 4c-low workloads). For such workloads,
completely eliminating maintenance operations (i.e., No-Refresh) 
also provides small performance improvements. SMD-Combined provides
88.8\% of the speedup provided by No-Refresh for 4c-low workloads.}

\ignorerev{
\iscanew{from Fig.~\ref{fig:multi_core_perf}}. 
\iql{IQE5}{}\iqrev{\iscanew{First, SMD-Combined provides \param{8.6}\% and \param{4.1}\% speedups on average across 
4c-high workloads over MC-Combined-DARP and MC-Combined-DSARP, respectively.
We attribute these speedups to i) \mech{}'s maintenance-access parallelization and ii) \mech{}'s ability to perform
maintenance operations autonomously inside the DRAM chip \emph{without} the memory controller having to issue DRAM commands.}} 
Second, SMD-FR-1LR incurs an average \param{4.5}\% slowdown on average across 4c-high workloads compared to DARP. While DARP can reduce
the overheads of periodic refresh by intelligently scheduling refresh operations, SMD-FR-1LR \emph{cannot}. \iql{IQA2}{}\iqrev{We attribute SMD-FR-1LR's overheads to the relatively high rate of \actnack{} commands (not shown in the figure) it issues: SMD-FR-1LR issues an \actnack{} command for every 11.7 activate commands, on average across all 4c-high workloads.}
{Third, we observe that i) the speedup and slowdown trends for the tested mechanisms are similar for the average 4c-medium workload, and ii) the tested mechanisms provide little performance improvement for 4c-low workloads.}
}




\iscanew{We conclude that SMD provides substantial system performance benefits by
concurrently performing a maintenance operation and \omcr{3}{one or more}
memory access in different lock regions in a bank.}
\ignorerev{
and 
ii) does \emph{not} significantly hurt system performance 
when SMD performs maintenance at bank granularity (i.e., implements one lock region per DRAM bank).
}


\ignorerev{
\hht{\textbf{Performance Summary.}} The new DRAM refresh mechanisms \fr{} and
\vr{} significantly improve system performance, achieving \hht{speedup} close to
that provided by the hypothetical ``NoRefresh'' DRAM. \hht{\mech{} enables
efficient and robust RowHammer protection (\prp{} and \drp{}) and memory
scrubbing (\sms{}) that still provide significant speedup when integrated with
\fr{}. Overall, \mech{} enables DRAM that is both faster and more robust than
DDR4.} 
}
\subsection{Energy Consumption}
\label{subsec:energy_results}

\begin{figure}[!h]
    \centering
    \includegraphics[width=\linewidth]{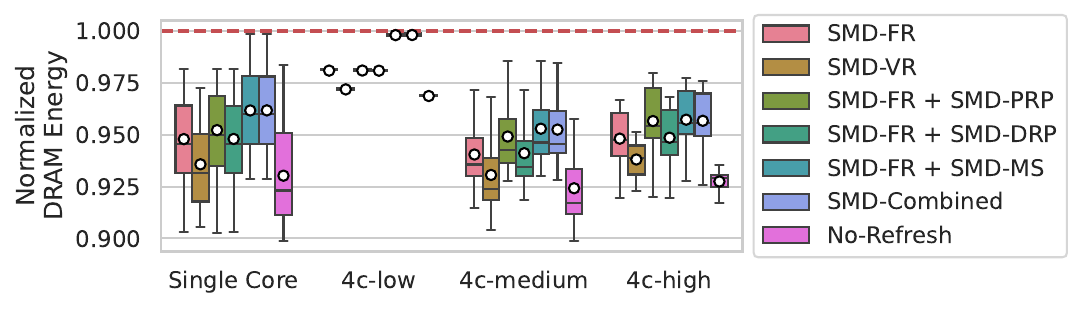}
    \caption{\atbcr{3}{Normalized DRAM energy consumption 
    (\omcr{4}{to} the baseline DDR4 system represented by the horizontal red line) for four-core
    workloads}}
    \label{fig:dram_energy}
\end{figure}

Fig.~\ref{fig:dram_energy} shows the DRAM energy consumption
\atbcr{3}{(normalized to the DDR4 baseline) across all 
tested single-core and four-core workloads 
with SMD-FR, SMD-VR, SMD-FR + SMD-PRP, SMD-FR + SMD-DRP, SMD-FR + SMD-MS, 
SMD-Combined, and No-Refresh. The white circles (error lines) 
represent the average (minimum and maximum) 
normalized DRAM energy consumption.
A lower value on the y-axis
indicates smaller DRAM energy consumption.} We make three major observations. 

\atbcr{2}{First, all SMD configurations reduce DRAM energy consumption of 
all evaluated workloads compared to the baseline.} We attribute the 
reduction in DRAM energy to i) the reduced DRAM background energy 
consumption because SMD shortens the execution time for these 
workloads and ii) \atbcr{1}{the elimination of} DRAM commands 
for maintenance operations (e.g., $REF$ for periodic refresh) 
on the power-hungry DDRx bus.
\atbcr{2}{Second, SMD provides substantial DRAM energy savings
for memory intensive workloads with relatively high LLC MPKI. 
For example, for 4c-medium and 4c-high workloads, on average,
SMD-Combined reduces DRAM energy by 4.8\% and 4.3\%.
For 4c-high workloads, SMD-Combined provides 59.6\% of the average DRAM
energy savings provided by No-Refresh, while \omcr{3}{also} improving system robustness 
by
conducting periodic refresh,
RowHammer protection, and periodic scrubbing operations inside the DRAM chip.}
Third, SMD-VR's energy savings (6.9\% on average
across all 4c-high workloads) are higher than other
SMD configurations because SMD-VR mitigates DRAM refresh overhead.

\atbcr{2}{We conclude that SMD, by autonomously performing maintenance operations inside the 
DRAM chip, significantly reduces DRAM energy consumption
compared to the baseline system that issues DRAM commands over
the power-hungry DDRx bus to perform maintenance operations.}
 
\ignorerev{
\iscanew{First, SMD-Combined and SMD-FR reduce DRAM energy by \param{3.9}\% (\param{4.3}\%) and \param{5.2}\% (\param{5.2}\%) across all single-core (4c-high) workloads compared to the baseline, respectively. SMD-Combined performs close to the hypothetical No-Refresh configuration, providing \param{59.7}\% of the energy reduction benefits of No-Refresh on average across 4c-high workloads. We attribute the reduction in DRAM energy to i) the reduced DRAM background energy consumption because SMD shortens the execution time for these workloads and ii) \atbcr{1}{the elimination of} DRAM commands for maintenance operations (e.g., $REF$ for periodic refresh) on the power-hungry DDRx bus. Second, SMD-FR-1LR increases average single-core (4c-high) workload DRAM energy consumption by $<\!$ \param{0.1}\% (\param{3.0}\%). While SMD-FR-1LR also eliminates DRAM commands for maintenance operations, it induces energy overheads mainly because of the increased execution time for the evaluated workloads.}
}

\subsection{\atbcr{2}{Performance Comparison \omcr{3}{to DARP and DSARP}}}

Fig.~\ref{fig:multi_core_perf} shows weighted speedup (normalized to the
weighted speedup of the DDR4 baseline) for \hht{60 four-core workloads (20 per
memory intensity level)}
\atbcr{3}{with DARP-Combined, DSARP-Combined, and SMD-Combined.}
\hht{The \iscanew{white circles inside each box} (error lines) represent the average
(minimum and maximum) \omcr{3}{normalized} weighted speedup
across the 20 workloads in the
corresponding group.} 

\begin{figure}[!ht]
    \centering
    \includegraphics[width=1\linewidth]{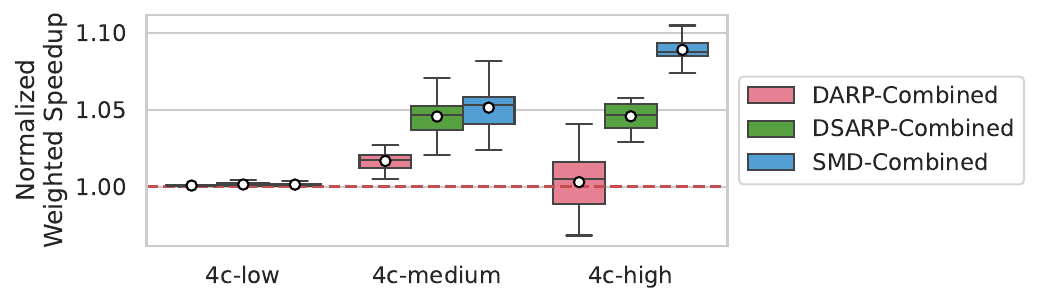}
    \caption{\atbcr{3}{Normalized weighted speedup (\omcr{4}{to} the baseline DDR4 system represented by the horizontal red line) of four-core workloads with DARP-Combined, DSARP-Combined, and SMD-Combined.}}
    \label{fig:performance-comparison}
\end{figure}

\ignorerev{
\iql{IQE5}{}Second, SMD-Combined outperforms MC-Combined-DSARP and MC-Combined-DARP by \param{0.6}\% and \param{3.9}\%, respectively, \microrevd{on average across all workloads}\microrevlabel{DQ1}. 
While MC-Combined-DSARP can concurrently access main memory and perform a maintenance operation, 
it needs to issue a DRAM command (e.g., a per bank refresh command for periodic refresh) for each maintenance operation
and delay other DRAM commands that the memory controller issues to serve main memory requests.
In contrast, SMD-Combined autonomously performs each maintenance operation inside the DRAM chip and does \emph{not} incur delays
for DRAM commands (except those incurred by \actnack{}s). MC-Combined-DARP 
\emph{cannot} concurrently perform a maintenance operation and access main memory, thereby \atbcr{1}{MC-Combined-DARP} performs worse than both SMD-Combined and MC-Combined-DSARP. We make similar observations on SMD-FR's, DSARP's, and DARP's performance. These observations are in line with prior work~\cite{chang2014improving}.
}

\atbcr{2}{
We observe that SMD-Combined outperforms both DARP- and DSARP-Combined on average
across all workloads in each memory intensity category.}
\iql{IQE5}{}\iqrev{\iscanew{For example, SMD-Combined provides 
\param{8.6}\% and \param{4.1}\% speedups on average across 
4c-high workloads over \omcr{3}{DARP-Combined} 
and \omcr{3}{DSARP-Combined}, respectively.
We attribute these speedups to i) \mech{}'s maintenance-access parallelization and ii) \mech{}'s ability to perform
maintenance operations autonomously inside the DRAM chip \emph{without} the memory controller having to issue DRAM commands.}}
While \omcr{3}{DSARP-Combined} can concurrently access main memory and perform a maintenance operation, 
it needs to issue a DRAM command (e.g., a per bank refresh command for periodic refresh) for each maintenance operation
and delay other DRAM commands that the memory controller issues to serve main memory requests.
In contrast, SMD-Combined autonomously performs each maintenance operation inside the DRAM chip and does \emph{not} incur delays
for DRAM commands (except those incurred by \actnack{}s). 
\omcr{3}{DARP-Combined}
\emph{cannot} concurrently perform a maintenance operation and 
access main memory, thereby \omcr{3}{DARP-Combined} performs 
worse than both SMD-Combined and \omcr{3}{DSARP-Combined}. 

\atbcr{2}{We conclude that SMD outperforms a system that 
allows the memory controller to concurrently perform a maintenance
operation and a memory access.}

\ignorerev{Second, SMD-FR-1LR incurs an average \param{4.5}\% slowdown on average across 4c-high workloads compared to DARP. While DARP can reduce
the overheads of periodic refresh by intelligently scheduling refresh operations, SMD-FR-1LR \emph{cannot}. \iql{IQA2}{}\iqrev{We attribute SMD-FR-1LR's overheads to the relatively high rate of \actnack{} commands (not shown in the figure) it issues: SMD-FR-1LR issues an \actnack{} command for every 11.7 activate commands, on average across all 4c-high workloads.}
{Third, we observe that i) the speedup and slowdown trends for the tested mechanisms are similar for the average 4c-medium workload, and ii) the tested mechanisms provide little performance improvement for 4c-low workloads.}
}





\ignore{
\subsection{\hpcarevcommon{Isolating \mech{}'s Performance Benefits}}

\hpcarevlabel{CC1}\hpcarevcommon{We evaluate \mech{}'s performance without the added benefits of memory accesses and maintenance operation parallelization (i.e., \mech{} with a single lock region) using all four-core heavily memory-intensive (4c-high) workloads. This configuration of \mech{}, called \mech{}-Combined-1LR, implements the same maintenance mechanisms as \mech{}-Combined does. We compare \mech{}-Combined-LR to \mech{}-Combined. We also evaluate a DRAM chip that leverages refresh-access parallelization~\cite{chang2014improving,zhang2014cream}, called DSARP. A system that implements DSARP can parallelize refresh and memory access operations to different subarrays. We compare DSARP to \mech{}-FR. Figure~\ref{fig:performance-isolation} shows the speedup of 4-core workloads using these three mechanisms (\mech{}-FR, DSARP, \mech{}-Combined-1LR) and NoRefresh.}

\begin{figure}[!h]
    \centering
    \includegraphics[width=1.\linewidth]{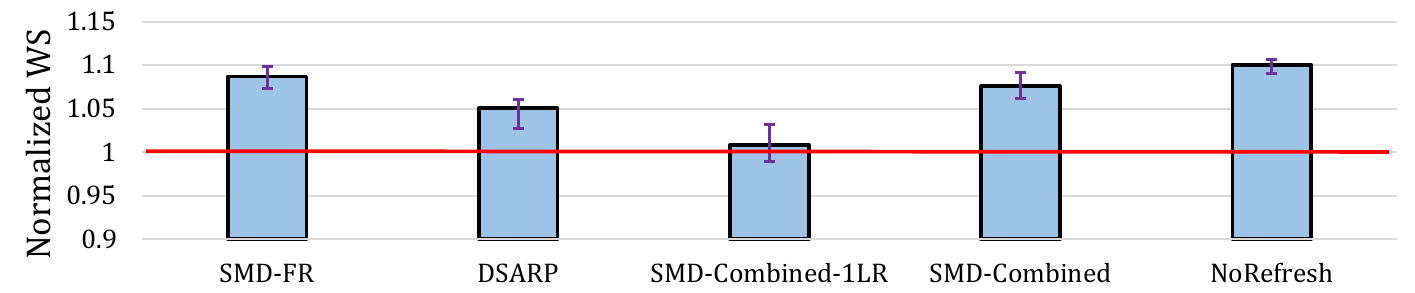}
    \caption{\hpcarevcommon{Four-core speedup for SMD with a \emph{single lock region} and other mechanisms \hht{normalized to \revcommon{the baseline} DDR4 system \revcommon{(horizontal black line)}}.}}
    \label{fig:performance-isolation}
\end{figure}

\hpcarevcommon{We make two major observations from Figure~\ref{fig:performance-isolation}. First, \mech{}-Combined-1LR improves performance by 0.9\% (3.2\%) on average (at maximum) across all workloads. \mech{}-Combined-1LR improves performance less than (and sometimes degrades performance) \mech{}-Combined because it \emph{cannot} concurrently perform maintenance operations and memory accesses. Second, DSARP improves performance by 5.0\% (6.0\%) on average (at maximum) across all workloads. We conclude that being able to concurrently execute maintenance and memory access operations at the same time increases \mech{}'s performance benefits, and \mech{}-FR provides comparable performance benefits to a DRAM chip that can leverage refresh-access parallelization (DSARP).}

\subsection{\hpcarevc{\mech{}'s Performance Sensitivity to Number of Cores}}

\hpcarevlabel{Rev.C/C3}\hpcarevc{The four-core workloads with high memory intensity (4c-high) largely saturate the DRAM bandwidth offered by four channels. Therefore, four cores are sufficient to capture the performance impact of \mech{} for cases where DRAM bandwidth utilization is high. To demonstrate this, we evaluate the performance of \mech{}-FR for a \SI{32}{\milli\second} refresh period using highly memory-intensive 8-core (i.e., 8c-high) workloads. The speedup is comparable to the speedup of 4c-high workloads. At 32ms refresh period, SMD-FR provides 8.1\% average speedup for 8c-high workloads while it provides 8.7\% speedup for 4c-high workloads. SMD provides a slightly lower speedup for 8c-high because of the increased memory request queueing delay, as eight-core workloads produce more memory requests than four-core workloads.}
}
\section{\iqrev{\atbcr{2}{Design Choices and Sensitivity Analyses}}}
\label{sec:discussion}

\atbcr{3}{We 1) describe and evaluate an SMD policy
that can prioritize activate commands targeting a locked region
over maintenance operations, 2) analyze SMD's performance
and energy sensitivities to the number of lock regions, refresh
period, number of cores in the system, memory scrubbing rate, 
RowHammer thresold, and RowHammer blast radius, 3) compare
SMD-based probabilistic RowHammer protection's (SMD-PRP) performance and energy to 
MC-based probabilistic RowHammer protection's (PARA~\cite{kim2014flipping}).}

\subsection{\iqrev{Prioritizing Memory Requests}}
\label{sec:smd_optimized}

\iql{IQA1}{}\iqrev{To minimize the impact of rejected activate commands on system performance, an \mech{} chip could apply a different activate command rejection policy that prioritizes activate commands over maintenance operations \omcr{2}{either selectively or globally}. We describe and estimate the performance impact of one such policy.}

\noindent
\iqrev{\textbf{``Pause \microrevcommon{Maintenance}'' Policy (SMD-PMP).} The key idea of SMD-PMP is to pause an ongoing maintenance operation for a lock region when the memory controller issues an activate command to the lock\omcr{2}{ed} region. The \mech{} chip could resume the maintenance operation when the memory controller precharges the bank (i.e., finishes accessing the lock region). For example, the \mech{} chip sequentially refreshes eight rows in a lock region (\cref{subsec:smd_refresh}) during a periodic refresh. If the chip receives an activate command to this lock region while only four out of eight rows have been refreshed, the chip does \emph{not} continue refreshing the fifth row, but yields control of the lock region to the memory controller.}

\noindent
\iqrev{\textbf{Latency of Pausing Maintenance Operations.} Even if the \mech{} chip decides to pause the ongoing maintenance operation, the activate command \emph{cannot} immediately proceed because the lock region might \emph{not} be in the precharged state (i.e., the DRAM bank might \emph{not} be ready to activate a row). The DRAM chip requires up to $tRAS + tRP$ (\cref{subsec:dram_organization}) to bring the under-maintenance lock region's state to the precharged state (in case an activate command was just issued by the maintenance mechanism), depending on the lock region's state when the activate command is received.}

\begin{figure}[!h]
    \centering
    \includegraphics[width=\linewidth]{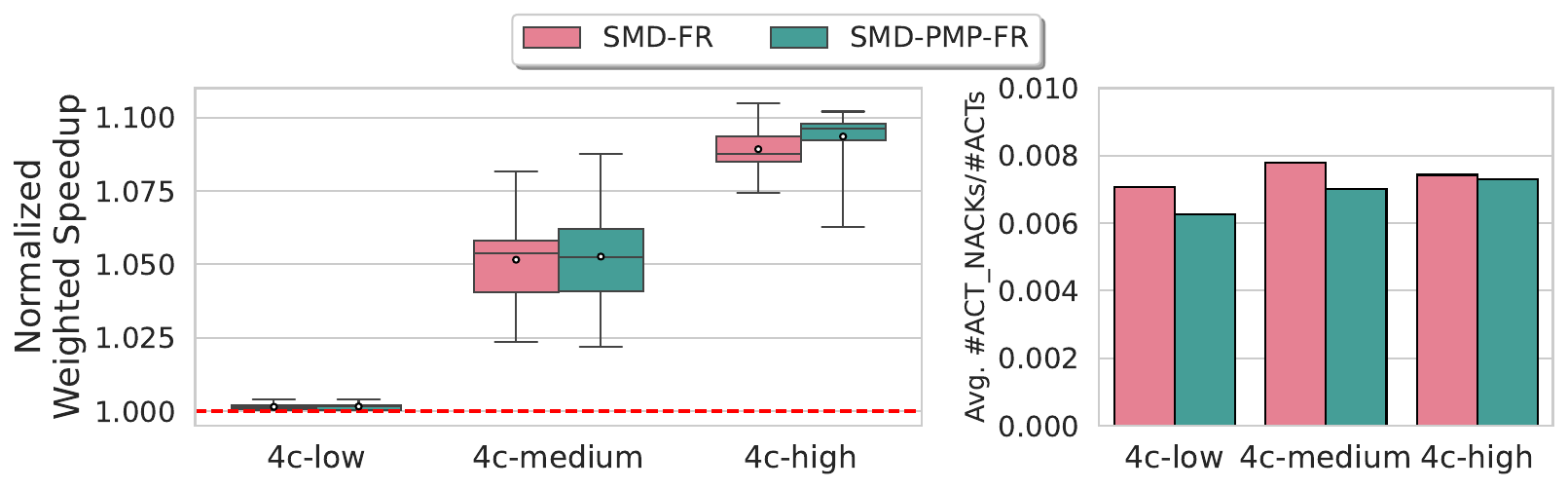}
    \caption{\iqrev{Four-core \omcr{3}{normalized} weighted speedup (left) with \omcr{3}{SMD-FR and SMD-PMP-FR}, \omcr{4}{and} rate of rejected activation commands (right)}}
    \label{fig:smd_prp_fr}
\end{figure}

\noindent
\iqrev{\textbf{Estimating the Performance of SMD-PMP.} We model SMD-PMP-FR \omcr{2}{(building on SMD-FR~\cref{subsec:smd_refresh})} that refreshes \atbcr{3}{\emph{only} one} DRAM row in a lock region before releasing the lock (i.e., SMD-PMP-FR locks a region for at most $tRAS + tRP$). Fig.~\ref{fig:smd_prp_fr} shows \omcr{4}{normalized} weighted speedup \omcr{3}{of} 60 four-core workloads (20 per memory intensity level) \atbcr{3}{with SMD-FR and SMD-PMP-FR} on the left and the number of \actnack{}s over the number of issued activate commands (which we call ``\actnack{} rate''), averaged across all workloads in every intensity level on the right. \atbcr{3}{SMD-FR is configured as described in~\cref{sec:methodology} and it refreshes 8 DRAM rows
in a lock region before releasing the lock.}}

\iqrev{\atbcr{2}{We observe that SMD-PMP-FR provides a larger average speedup than SMD-FR for 4c-high workloads. SMD-PMP-FR's \actnack{} rate is 2\% smaller than SMD-FR's, which explains the 0.4\% larger average speedup it provides over SMD-FR. 
We conclude that an SMD-PMP design that pauses maintenance operations to serve activate commands \atbcr{1}{could} improve performance.}
}

\subsection{\atbcr{2}{Sensitivity to Number of Lock Regions}}
\label{sec:sensitivity-lock-region}
\begin{figure}[!ht]
    \centering
    \includegraphics[width=1.0\linewidth]{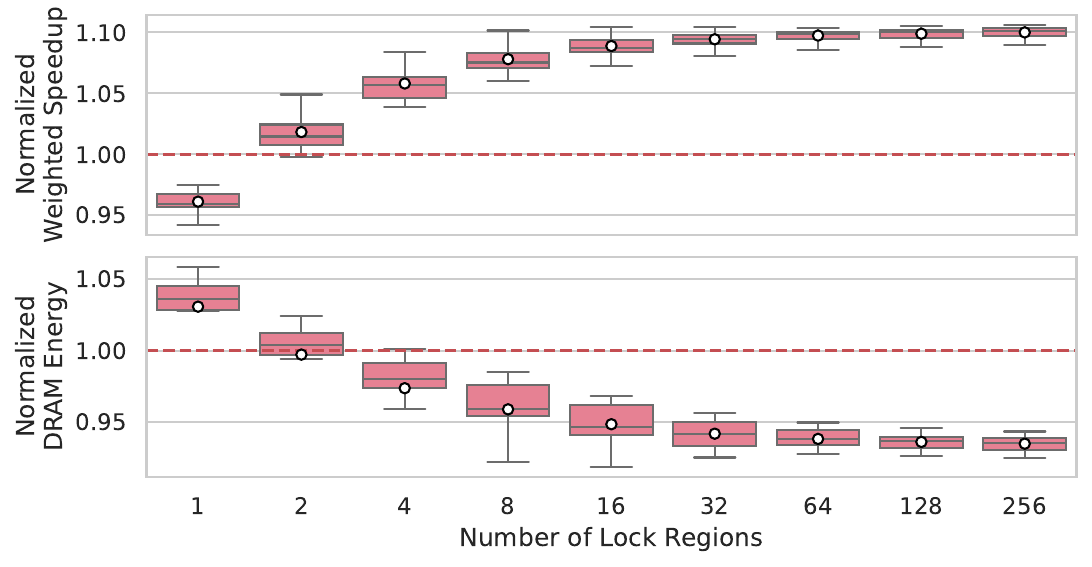}
    \caption{\atbcr{3}{Normalized weighted speedup of 4c-high workloads with SMD-FR (top). Normalized DRAM energy
    of 4c-high workloads with SMD-FR (bottom). Dashed red lines indicate the performance
    and DRAM energy of the baseline DDR4 system}}
    \label{fig:lock_region_sweep}
\end{figure}

\atbcr{2}{To understand SMD\omcr{3}{'s} sensitivity to the number of 
lock regions in a DRAM bank, we evaluate \omcr{3}{the performance and 
DRAM energy of} 
SMD-FR \omcr{3}{(that refreshes 8 DRAM rows
in a lock region before releasing the lock)
as we vary the number of} lock regions in a bank.
Fig.~\ref{fig:lock_region_sweep} (top) shows \omcr{3}{normalized} weighted speedup
across 20 four-core high memory intensity workloads (y-axis)
\omcr{3}{for different} number of lock regions (x-axis) \omcr{3}{with SMD-FR} and Fig.~\ref{fig:lock_region_sweep} (bottom) shows
\omcr{3}{normalized DRAM energy across the same workloads
and the number of lock regions}.}

\atbcr{2}{We make two key observations from Fig.~\ref{fig:lock_region_sweep}. First,
SMD-FR provides \omcr{3}{higher} speedup from 1.8\% to 10.0\% \omcr{3}{(averaged across all 20 workloads)}
as the number of lock regions increases from 2 to 256. This is because more lock
regions allow SMD to concurrently perform more maintenance operations and memory accesses in a bank. 
Increasing the number of lock
regions beyond 16 provides diminishing returns as
16 lock regions per bank
provides 88.8\% of the speedup provided by 256 lock regions.
Second, SMD-FR with \omcr{3}{\emph{only}} one lock region per bank causes 3.9\% average slowdown across all workloads.
SMD-FR with one lock region \emph{cannot} concurrently perform maintenance operations and memory accesses
in a bank because every maintenance operation locks the whole bank.
We attribute the overheads of SMD-FR with one lock region per bank to the relatively high rate of \actnack{} 
commands (not shown in the figure) it issues: This configuration of SMD issues an 
\actnack{} command for every 11.7 activate commands, on 
average across all 4c-high workloads. \atbcr{3}{SMD-FR with
16 lock regions per bank, in contrast, issues an \actnack{} command
\emph{only} every 134.6 activate commands, on average across all 4c-high workloads.}} 
\atbcr{3}{We make
similar observations for DRAM energy.}

\atbcr{2}{We conclude that SMD with 16 lock regions in a bank yields
robust performance \omcr{3}{and DRAM energy} improvements at relatively low
DRAM chip area cost (see~\cref{sec:hw_overhead}).}

\subsection{\iqrev{Sensitivity to DRAM Refresh Period}}
\label{sec:smd_scaling}
\iql{IQA2}{}\iqrev{We evaluate \mech{} for refresh periods of \omcr{4}{\SI{32}{\milli\second},} \SI{16}{\milli\second}, \SI{8}{\milli\second}, and \SI{4}{\milli\second}. 
\ignorerev{
\atbcr{3}{Figure~\ref{fig:smd_scaling} shows the weighted speedup across 20 four-core high memory intensity workloads.}
}
We observe that the performance benefit of \fr{} increases as the refresh period reduces, achieving 3.6$\times{}$ speedup at \SI{4}{\milli\second} refresh period \atbcr{2}{averaged}
across \texttt{4c-high}
workloads. \omcr{2}{As the refresh period becomes smaller,} the baseline system more frequently performs time-intensive periodic refresh operations (each of which takes $tRFC$=\SI{350}{\nano\second}). At a refresh period of \SI{4}{\milli\second}, the memory controller issues a $REF$ command every $tREFI$ of \SI{493.75}{\nano\second}. Thus, in the baseline system, a DRAM rank is busy 70.9\% of the time doing refresh. \mech{} also more frequently performs refresh operations (and more frequently \actnack{}'s activate commands) as the refresh period reduces. However, \mech{} alleviates the performance overhead of refresh operations by periodically refreshing only one lock region \omcr{3}{at a time} while allowing ACT commands targeting other lock regions to proceed \omcr{2}{in parallel}.} 

\ignorerev{
 \begin{figure}[!h]
     \centering
     \includegraphics[width=\linewidth]{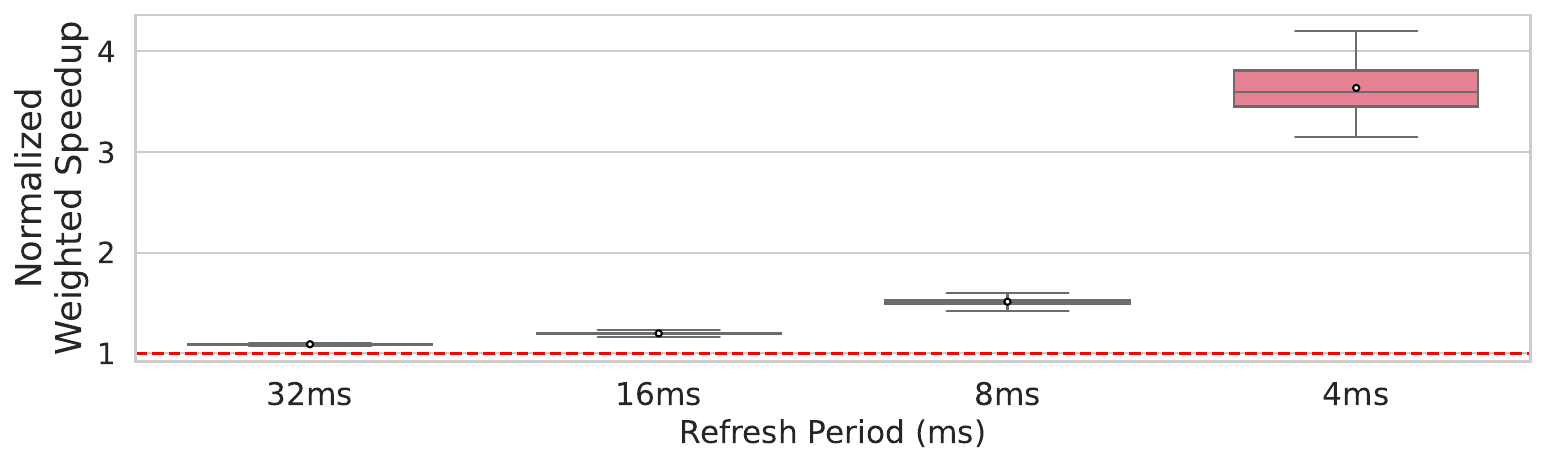}
     \caption{\iqrev{\atbcr{3}{Normalized weighted speedup improvement of 4c-high workloads with SMD-FR
     for different refresh periods}}}
     \label{fig:smd_scaling}
 \end{figure}
 }

\microrevlabel{AQ3}\microreva{We \omcr{3}{also} evaluate \omcr{3}{a} baseline system with per-bank refresh at a \SI{4}{\milli\second} refresh period. Because a bank is busy 70.9\% of the time doing refresh operations, SMD provides 1.7$\times{}$ average speedup across 4c-high workloads over the baseline system with per-bank refresh.}
\atbcrcomment{2}{Add refresh plot}

\iqrev{We conclude that \mech{}'s performance benefits will likely increase for future DRAM chips that require high\microrevlabel{AQ3}\microreva{er refresh rates as 1) the number of rows in a DRAM bank increases to improve DRAM density or 2) the refresh period reduces to improve error tolerance for shrinking DRAM technology or increasing operating temperatures with tighter DRAM-system integration (e.g., high bandwidth memory~\cite{ramalingam2016hbm,jedec2021hbm})}.} 

\subsection{Sensitivity to Number of Cores}


\atbcrcomment{2}{Removed overclaim.}
We evaluate the performance of \mech{}-FR for a \SI{32}{\milli\second} refresh period 
using highly memory-intensive \atbcr{2}{1-, 2-, 4-, and 8-core} workloads. 
At \omcr{3}{a} 32ms refresh period, SMD-FR provides \atbcr{2}{8.0\%, 8.0\%, 8.7\%, and} 8.1\% average
normalized weighted speedup for 
1-, 2-, 4-, and 8-core high memory intensity workloads, respectively. 
\atbcr{2}{We conclude that SMD-FR provides substantial performance improvements
across 1-, 2-, 4-, and 8-core workloads. We attribute
the difference between the provided performance improvement
across different numbers of cores to workload and memory access interleaving differences.}


\subsection{\microreva{\atbcr{2}{SMD-} vs. Memory-Controller-Based Scrubbing}}
\label{subsec:conventional_dram_scrubbing}

\begin{figure}[!h]
    \centering
    \includegraphics[width=\linewidth]{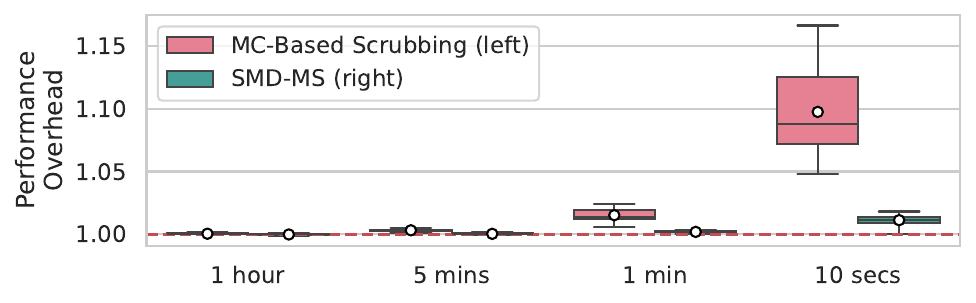}
    \caption{\microreva{\omcr{2}{MC-based} scrubbing \omcr{2}{(DDR4)} vs. \mech{}-based scrubbing}}
    \label{fig:scrubbing_comparison}
\end{figure}

\microrevlabel{AQ6}\microreva{Fig.~\ref{fig:scrubbing_comparison} compares the performance 
\hext{overheads} of \hext{memory-controller-based memory scrubbing} to \mech{}-based memory 
scrubbing (\sms{}) across
\texttt{4c-high} workloads. We show performance overhead normalized to the baseline system.}

\microreva{\hext{We make two observations. First,} both \omcr{2}{MC-based} scrubbing and \sms{} have
small performance overhead\omcr{3}{s} for scrubbing periods \hext{of 5 minutes and
larger} because scrubbing operations are infrequent at such periods.
\hext{Second,} \omcr{2}{MC-based} scrubbing causes 1.5\%/8.8\% average slowdown for \omcr{2}{a} 1
minute/10 second scrubbing period (up to 2.4\%/14.3\%), while \sms{} causes
only \omcr{2}{0.2/1.1\% average} (up to 0.3\%/1.8\%) slowdown. \omcr{2}{MC-based} scrubbing has high overhead at low
scrubbing periods because moving data from DRAM to the MC to perform scrubbing
is inefficient compared to performing scrubbing \omcr{2}{completely inside the} DRAM \omcr{2}{chip} using \sms{}.
Scrubbing at high rates may become necessary for future DRAM chips as their
reliability characteristics continuously worsen.} 
\microreva{We conclude that \sms{} performs memory scrubbing more efficiently than
conventional MC-based scrubbing and it enables scrubbing at high rates with
\hext{small performance overheads}.}

\subsection{Sensitivity to RowHammer Threshold}
\label{subsec:graphene_act_threshold_sweep}

\begin{figure}[!h]
    \begin{yellowb}
        \centering
        \includegraphics[width=\linewidth]{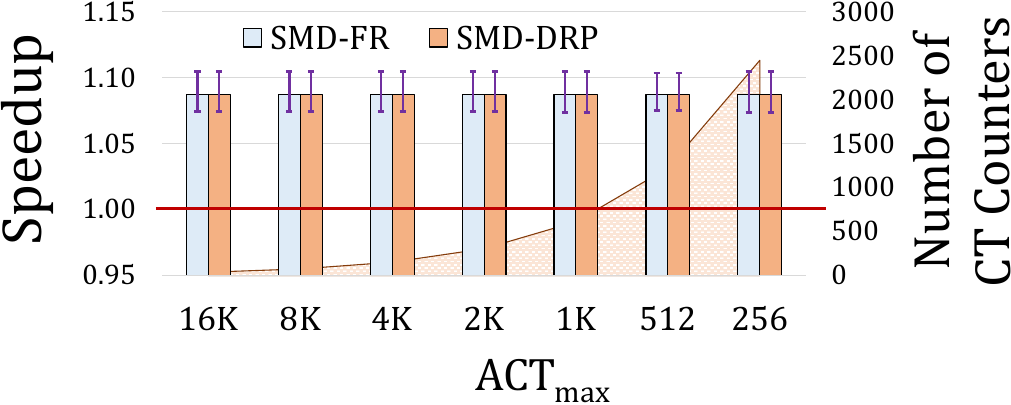}
    \end{yellowb}
    
    \caption{\drp{}'s sensitivity to $ACT_{max}$}
    \label{fig:graphene_act_threshold_sweep}
\end{figure}

We analyze the \drp{}'s sensitivity to the maximum row activation threshold
($ACT_{max}$). Fig.~\ref{fig:graphene_act_threshold_sweep} shows the average
speedup that \fr{} and \drp{} achieve for different $ACT_{max}$ values across
\texttt{4c-high} workloads compared to the DDR4 baseline. When evaluating
\drp{}, we use \fr{} as a DRAM refresh mechanism.

We observe that \drp{} incurs negligible performance overhead on top of \fr{}
even for extremely small $ACT_{max}$ values. This is because \drp{} generates
very few neighbor row refreshes as \texttt{4c-high} is a set of benign workloads
that do not repeatedly activate a single row many times.

Although the performance overhead of \drp{} is negligible, the number of Counter Table (CT) entries required is significantly large for small $ACT_{max}$ values.
For $ACT_{max} = 16K$, \drp{} requires $38$ counters per bank, and the number of counters required increase linearly as $ACT_{max}$ reduces, reaching $2449$ counters at the lowest $ACT_{max} = 256$ that we evaluate.

\subsection{Sensitivity to RowHammer Blast Radius}
\label{subsec:blast_radius_sweep}
    
We analyze the performance overheads of \prp{} and \drp{} when refreshing a
different number of neighbor rows upon detecting a potentially aggressor row.
Kim et al.~\cite{kim2020revisiting} show that, in some DRAM chips, an aggressor
row can cause bit flips also in rows that are at a greater distance than the two
victim rows surrounding the aggressor row. Thus, it may be desirable to configure a RowHammer protection mechanism to refresh more neighbor rows than the two rows that are immediately adjacent to the aggressor row. 

\begin{figure}[!h]
    \begin{yellowb}
        \centering
        \includegraphics[width=\linewidth]{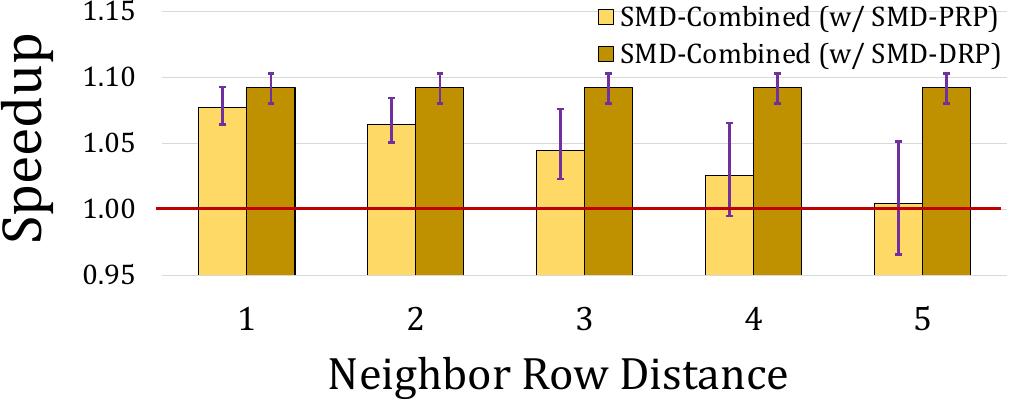}
    \end{yellowb}
    
    \caption{Sensitivity to the number of neighbor rows affected by RowHammer}
    \label{fig:blast_radius_sweep}
\end{figure}

Fig.~\ref{fig:blast_radius_sweep} shows the average speedup that
\mech{}-Combined (separately with \prp{} and \drp{}) achieves for different
number of neighbor rows refreshed across \texttt{4c-high} workloads compared to
the DDR4 baseline. The \emph{Neighbor Row Distance} values on the x-axis
represent the number of rows refreshes on each of the two sides of an aggressor
row (e.g., for neighbor row distance of $2$, \prp{} and \drp{} refresh four
victim rows in total).

We make two key observations from the figure. First, \prp{} incurs large performance overheads as the neighbor row distance increases. This is because, with $P_{mark} = 1\%$, \prp{} perform neighbor row refresh approximately for every 100th \cmdact{} command, and the latency of this refresh operation increases with the increase in the number of victim rows. Second, the performance overhead of \drp{} is negligible even when the neighbor row distance is five. This is because, \drp{} detects aggressor rows with a higher precision than \prp{} using area-expensive counters. As the \texttt{4c-high} workloads do not repeatedly activate any single row, \drp{} counters rarely exceed the maximum activation threshold, and thus trigger neighbor row refresh only a few times.

\subsection{SMD-based Probabilistic\\RowHammer Protection vs. PARA~\cite{kim2014flipping}}
\label{subsec:para_comparison}

\begin{figure}[!h]
    \begin{subfigure}{.48\linewidth}
        \begin{yellowb}
            \centering
            \includegraphics[width=\linewidth]{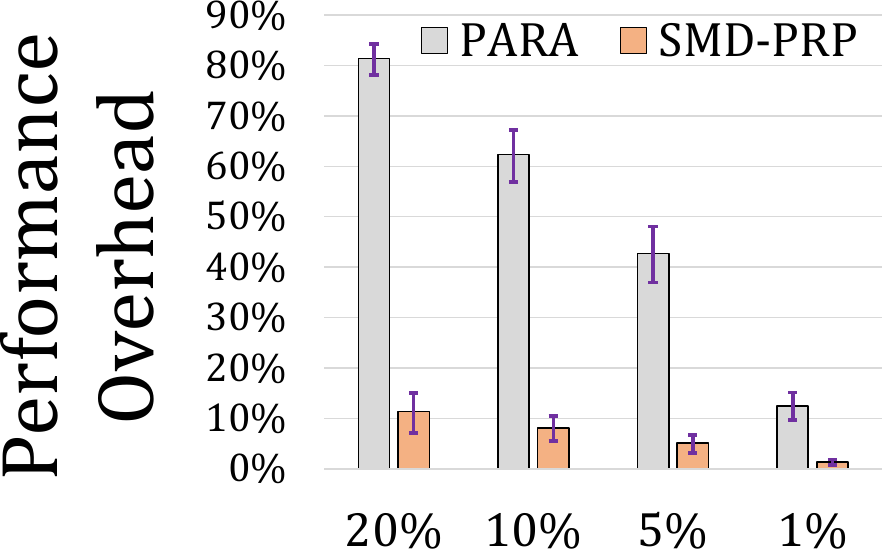}
        \end{yellowb}
    \end{subfigure}\hfill
    \begin{subfigure}{.48\linewidth}
        \begin{yellowb}
            \centering
            \includegraphics[width=\linewidth]{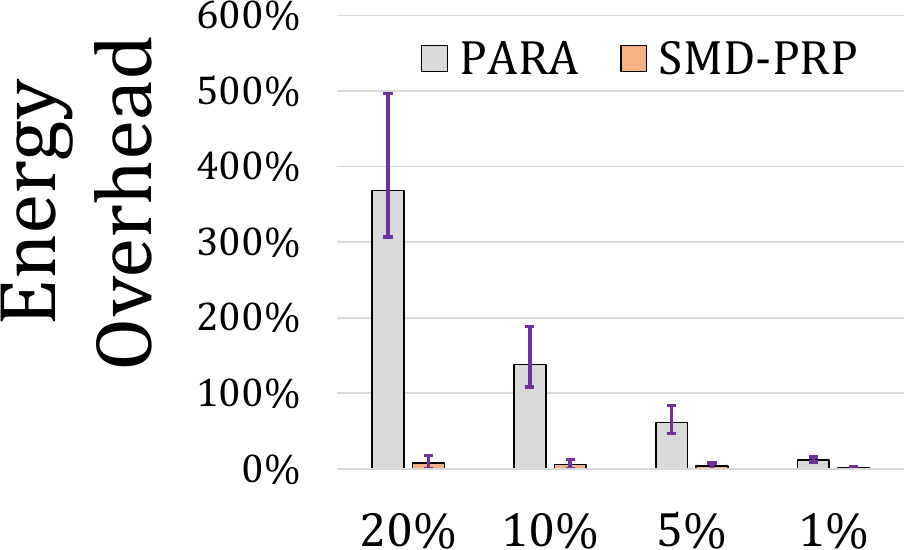}
        \end{yellowb}
    \end{subfigure}
    \caption{PARA vs. \prp{}}
    \label{fig:para_comparison}
\end{figure}

\hext{Fig.~\ref{fig:para_comparison} compares the performance and energy
overheads of PARA implemented in the MC (as proposed by Kim et
al.~\cite{kim2014flipping}) for DDR4 and \prp{} for different neighbor row
activation probabilities (i.e., $P_{mark}$) across \texttt{4c-high} workloads.
\emph{PARA} represents the performance and energy overheads with respect to
conventional DDR4 with no RowHammer protection. Similarly, \prp{} represents the
performance and energy overheads with respect to a \mech{} chip, which uses
\fr{} for periodic refresh, with no RowHammer protection.}

\hext{We make two observations. First, the performance and energy consumption of
MC-based PARA scales poorly with $P_{mark}$. At the default $P_{mark}$ of 1\%,
PARA incurs 12.4\%/11.5\% average performance/DRAM energy overhead. For higher
$P_{mark}$, the overheads of PARA increase dramatically to 81.4\%/368.1\% at
$P_{mark}$ of 20\%. Second, \prp{} is significantly more efficient than PARA. At
the default $P_{mark}$ of 1\%, \prp{} incurs only 1.4\%/0.9\% performance/DRAM
energy overheads and at $P_{mark}$ of 20\% the overheads become only
11.3\%/8.0\%. \prp{} is more efficient than PARA mainly due to enabling access
to non-locked regions in a bank while \prp{} performs neighbor row refreshes on
the locked region.

We conclude that \prp{} is a highly-efficient RowHammer protection that incurs
small performance and DRAM energy overheads even with high neighbor row refresh
probability, which is critical to protect future DRAM chips that may have
extremely high RowHammer vulnerability.}

\subsection{\actnack{} Divergence Across Chips}
\label{subsec:divergence-evaluation}


\addressdivergence{In~\cref{subsubsec:act_nack_divergence}, we explain divergence in \mech{}
maintenance operations can happen when different DRAM chips in the same rank
perform maintenance operation at different times. Such a divergence leads to a
partial row activation when the activated row is in a locked region in some DRAM
chips but not in others. To handle partial row activations, we develop three
policies.}



\begin{figure}[!h]
    \begin{subfigure}{\linewidth}
        \begin{yellowb}
            \centering
            \includegraphics[width=\linewidth]{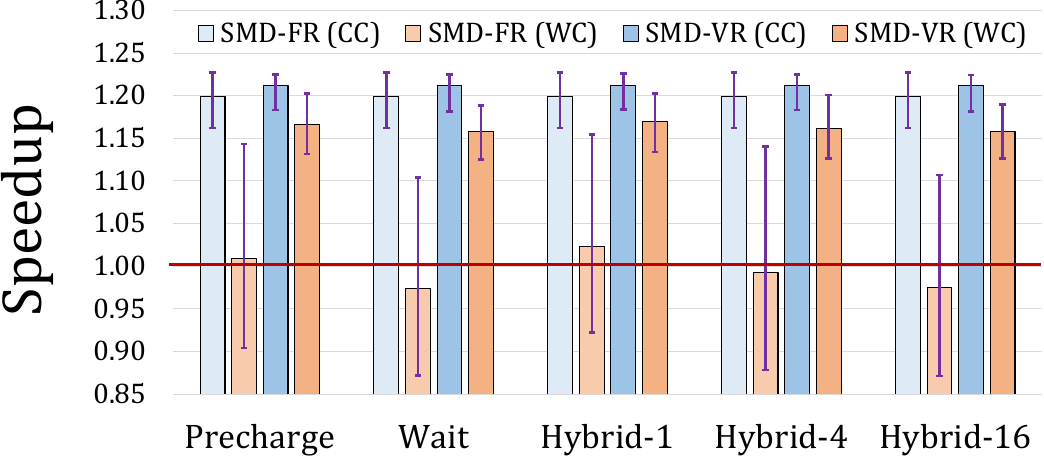}
        \end{yellowb}
    \end{subfigure}

    \begin{subfigure}{\linewidth}
        \begin{yellowb}
            \centering
            \includegraphics[width=\linewidth]{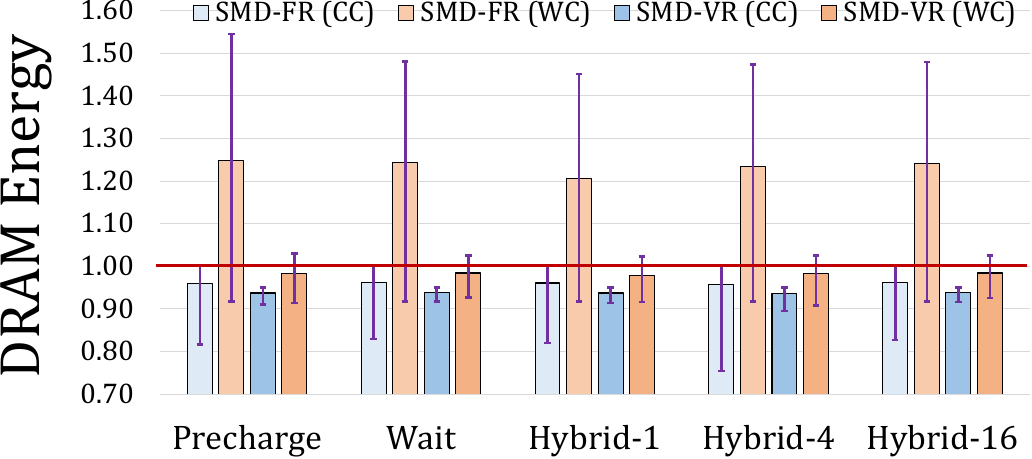}
        \end{yellowb}
    \end{subfigure}

    \caption{Comparison of different policies for handling refresh divergence across DRAM chips.}
    \label{fig:worst_case_ref_distrib}
\end{figure}

\addressdivergence{In Fig.~\ref{fig:worst_case_ref_distrib}, we compare the performance \hext{and
energy savings} of \fr{} and \vr{} when using the three \actnack{} divergence
handling policies \hext{across} \texttt{4c-high} workloads 
for \trefw{} = \SI{16}{\milli\second}. \hext{The plots show
results for the common-case (CC) and worst-case (WC) scenarios with regard to
when maintenance operations happen across different \mech{} chips in the same
rank. In the common-case scenario, the DRAM chips generally refresh the same row
at the same time due to sharing the same DRAM architecture design. However,
refresh operations in some of the DRAM chips may still diverge during operation
depending on the refresh mechanism that is in use. For example, \vr{} refreshes
retention-weak rows, whose locations may differ across the DRAM chips, at a
higher rate compared to other rows, resulting in divergence in refresh
operations across the DRAM chips in a rank. In the worst-case scenario, we
deliberately configure the DRAM chips to refresh different rows at different
times. For this, we 1) delay the first refresh operation in $chip_i$ by $i
\times l{ref}$, where $0 \leq i < Num{chips/rank}$ and $l_{ref}$ is the latency
of a single refresh operation, and 2) set the \emph{Lock Region Counter (LRC)}
of $chip_i$ to $i$.}}

\addressdivergence{We make three key observations. First, the performance \hext{and DRAM energy
consumption} only slightly varies across different \hext{divergence handling}
policies \hext{in both common-case and worst-case scenarios}. This is because,
after a partial row activation happens, a different row that is not in a locked
region in all off of the DRAM chips often does not exist in the memory request
queue. As a result, the \emph{Precharge} and \emph{Hybrid} policies perform
similarly to the \emph{Wait} policy.}

\addressdivergence{Second, \fr{} performs worse \hext{than the DDR4 baseline} in the worst-case
refresh distribution scenario (i.e., average slowdown of 2.6\% with the
\emph{Wait} policy). Certain individual workloads experience even higher
slowdown (up to 12.8\% \hext{with the \emph{Wait} policy}). The reason is that,
when $N$ different DRAM chips lock a region at different times, the total
duration during which the lock region is unavailable becomes $N$ times the
duration when all chips simultaneously refresh the same lock region. This
significantly increases the performance overhead of refresh operations.}

\addressdivergence{Third, \vr{} does not suffer \hext{much} from the divergence problem \hext{and
outperforms the DDR4 baseline even in the worst-case scenario. This is} because
\vr{} mitigates the DRAM refresh overhead by significantly reducing the number
of total refresh operations by exploiting retention-time variation across DRAM
rows. \hext{Thus, the benefits of eliminating many unnecessary refresh
operations surpass the overhead of \actnack{} divergence.}}

\addressdivergence{We conclude that 1) although \fr{} suffers from noticeable slowdown for the
worst-case scenario, which should not occur in a well-designed system, it still
provides comparable performance to conventional DDR4 and 2) \hext{\vr{}
outperforms the baseline and saves DRAM energy even in the worst-case scenario}.}

\ignorerev{
\subsection{\microrevb{Maintenance Operation Improvements\\Supported by \mech{}}}

\omcrcomment{2}{This could be a place to show variable rate refresh and other RH mechanisms’ results}\microrevlabel{BQ1}\microrevb{A large set of prior work proposes various ideas and techniques that mitigate DRAM refresh overhead~\cite{chang2014improving, liu2012raidr,
qureshi2015avatar, nair2014refresh, baek2014refresh, bhati2013coordinated,
cui2014dtail, emma2008rethinking, ghosh2007smart, isen2009eskimo,
jung2015omitting, kim2000dynamic, luo2014characterizing, kim2003block,
liu2012flikker, mukundan2013understanding, nair2013case, patel2005energy,
stuecheli2010elastic, khan2014efficacy, khan2016parbor, khan2017detecting,
venkatesan2006retention, patel2017reaper, riho2014partial, hassan2019crow,
kim2020charge, nguyen2018nonblocking, kwon2021reducing} and improve RowHammer mitigation performance~\cite{lee2018twice,
seyedzadeh2017counter,son2017making,yaglikci2021blockhammer, park2020graphene,
you2019mrloc, seyedzadeh2018cbt, aweke2016anvil, cojocar2019exploiting,
kim2014flipping, frigo2020trrespass, herath2015these, van2018guardion,
brasser2016can, konoth2018zebram, qureshi2022hydra, woo2023scalable, wi2023shadow, kim2023ddr5, bennett2021panopticon}. Many of these ideas and techniques could improve in-DRAM autonomous maintenance operations. At a high-level, any idea or technique that is independent from the organization of multiple DRAM chips into a module or a rank (i.e., any idea that can be leveraged by some technique in a single DRAM chip) can be implemented in an \mech{} chip. Implementing such an idea or technique in an \mech{} chip does \emph{not} require further changes to the DRAM interface.}\atbcrcomment{2}{Should we move this subsection to 5.4. (other use cases for smd)?}

\microrevb{One prior technique mitigates DRAM refresh overhead~\cite{nguyen2018nonblocking} by cleverly leveraging DRAM module organization and rank-level ECC to alleviate the DRAM bandwidth overhead of refresh operations. While \mech{} does \emph{not} preclude this technique from being implemented, implementing it as mainly described in~\cite{nguyen2018nonblocking} would require modifications to the DRAM interface. However, the key idea of~\cite{nguyen2018nonblocking} could still be leveraged inside an \mech{} chip by combining \atbcr{2}{1)~on-die ECC, which already exists (e.g., in DDR5~\cite{jedec2024jesd795c})
and 2)~finer-granularity (\atbcr{2}{DRAM-mat}-level) refresh operations (as described in~\cite{nguyen2018nonblocking}) in a future DRAM design.}} 
}

\section{Related Work}
\label{sec:related_work}

{This is the first work that gives DRAM chips the ability (i.e., the \omcr{2}{\emph{breathing room}})}
to autonomously perform maintenance operations with \omcr{2}{low-cost} simple
changes to \omcr{2}{the} existing \omcr{2}{rigid} DRAM interface.
%
%
{No prior work proposes setting the MC free from managing DRAM maintenance
 operations nor studies the system-level performance and energy impact of
 autonomous \omcr{2}{DRAM} maintenance mechanisms.}
\hht{We briefly discuss relevant prior works.}

\noindent\textbf{Changing the DRAM Interface.}
%
\hht{Several prior works~\cite{udipi2011combining, ham2013disintegrated,
fang2011memory, cooper2012buffer,stuecheli2018open} propose using high-speed serial links and
packe\revcommon{t}-based protocols in DRAM chips.}
\mech{} differs from prior works in two key aspects. First, none of \hht{these}
works describe how to implement maintenance mechanisms completely within DRAM.
\hht{We propose \atbcr{3}{eight} \mech{}-based maintenance mechanisms
(\cref{sec:maintenance_mechanisms})
\atbcr{3}{and rigorously evaluate three (\cref{subsec:smd_refresh}, \cref{subsec:smd_rowhammer_protection}, and 
\cref{subsec:smd_ecc_scrubbing}) of them}.}
Second, prior works significantly \omcr{3}{modify or even overhaul} the existing DRAM interface, which
makes their proposals more difficult to adopt compared to \mech{}, which adds
only a single \actnack{} signal to the existing DRAM interface \hht{and requires
slight modifications in the MC}.

\omcrcomment{3}{Feels terse. Maybe give which mechanisms
we implement in each subsection.}
\atbcrcomment{3}{Running out of space...}
\noindent\textbf{Mitigating DRAM Refresh Overheads.}
\hht{Many} previous works~\cite{chang2014improving, liu2012raidr,
qureshi2015avatar, nair2014refresh, baek2014refresh, bhati2013coordinated,
cui2014dtail, emma2008rethinking, ghosh2007smart, isen2009eskimo,
jung2015omitting, kim2000dynamic, luo2014characterizing, kim2003block,
liu2012flikker, mukundan2013understanding, nair2013case, patel2005energy,
stuecheli2010elastic, khan2014efficacy, khan2016parbor, khan2017detecting,
venkatesan2006retention, patel2017reaper, riho2014partial, hassan2019crow,
kim2020charge, nguyen2018nonblocking, kwon2021reducing} propose techniques to
reduce \omcr{2}{performance and energy overheads of} DRAM refresh. 
\omcr{2}{SMD can \omcr{3}{seamlessly} integrate many of these techniques along with other efficient
maintenance operations (\cref{sec:maintenance_mechanisms}).}

\noindent\textbf{\omcr{2}{Read Disturbance} Protection.}
Many prior works~\mitigatingRowHammerAllCitations{} propose techniques for
RowHammer \omcr{2}{and RowPress~\cite{luo2023rowpress}} protection. 
\hhh{{One can
use \mech{} to implement} \omcr{2}{these} or new RowHammer protection
mechanisms.}

\noindent\textbf{Memory Scrubbing.}
\omcr{2}{Various} prior works report that the overhead of memory scrubbing is small as
low scrubbing rates (e.g., scrubbing period of 24
hours~\cite{siddiqua2017lifetime, rooney2019micron, jedecddr5c}, 45 minutes
per 1~GB scrubbing rate~\cite{schroeder2009dram}, only when
idle~\cite{meza2015revisiting}) \hht{are generally} sufficient. \omcr{2}{However,} the cost of
scrubbing \hht{can} increase for future DRAM chips due to
increasing DRAM bit error rate~\cite{patel2021harp,nair2013archshield,cha2017defect} and increasing DRAM chip density~\cite{qureshi2015avatar}. \sms{} enables
efficient scrubbing \hht{by eliminating off-chip data transfers}.


\noindent
\reva{\textbf{\revlabel{RAC2}Leveraging Subarray-level Parallelism.} Prior works overlap the latency of accessing multiple subarrays \iscanew{by modifying} the DRAM architecture~\cite{kim2012case, zhang2014cream, chang2014improving}. SMD's maintenance-access parallelization builds on the basic design proposed in SALP~\cite{kim2012case} and refresh-access parallelization introduced in~\cite{chang2014improving,zhang2014cream}. A \omcr{3}{DRAM} interface based on these prior works~\cite{kim2012case, zhang2014cream, chang2014improving} \emph{still} needs to \omcr{3}{be modified} for each new maintenance operation that the \omcr{2}{JEDEC standard dictates DRAM} to implement. SMD, in contrast, allows the implementation of new maintenance mechanisms {with a single, simple change to the \omcr{3}{DRAM}} \omcr{2}{interface}.}

\noindent
\atbcrcomment{3}{Will keep if we have space}\hpcarevlabel{Rev.A/C1}\hpcareva{\textbf{Compute Express Link (CXL)~\cite{cxlmanual}.} CXL is a cache-coherent interconnect for computing systems. CXL does \emph{not} define the interface between a memory controller and a DRAM module/chip. Even with CXL, the memory controller chip has to deal with the management complexity of DRAM. SMD can be used in conjunction with CXL to ease management complexity in computing systems.}


\section{Conclusion}
\label{sec:conclusion}

\atbcr{1}{We introduced a new, low-cost framework, \omcr{2}{Self-Managing DRAM (SMD),} for accelerating the adoption of new
in-DRAM maintenance operations.}
\atbcr{1}{With a one-time \omcr{2}{low-cost} modification to the DRAM interface, \mech{} 
enables implementing new} in-DRAM maintenance
operations with no further changes to the DRAM interface, memory controller, or
other system components. Using \mech{}, we implement efficient \omcr{2}{in-DRAM} maintenance
mechanisms for DRAM refresh, RowHammer protection, and memory scrubbing. 
\hhf{We show that these mechanisms enable a higher performance, more energy efficient, and, at the same time, more \omcr{2}{robust} DRAM system.}
We believe \hhf{and hope that} \mech{} \omcr{2}{can} enable practical adoption of
\omcr{2}{future} innovative ideas in DRAM design \omcr{3}{and
inspire better ways of partitioning work between
memory and processor chips.}

\section*{Acknowledgments}

\atbcr{2}{We thank the anonymous reviewers of MICRO 2022, HPCA 2023, ISCA 2023, MICRO 2023, HPCA 2024, ISCA 2024, and MICRO 2024 for the feedback.
We thank the SAFARI Research Group members for {their} valuable \omcr{3}{and constructive} 
feedback \omcr{3}{along with} the stimulating scientific and intellectual environment they provide.
We acknowledge the generous gift funding provided by our industrial partners (especially Google, Huawei, Intel, Microsoft, VMware), which has been instrumental in enabling the research we have been conducting on \atbcr{3}{memory systems}.\atbcrcomment{3}{cite what here?}
This work was also in part supported by the Google Security and Privacy Research Award, \omcr{3}{and} the Microsoft Swiss Joint Research Center.}

\bibliographystyle{IEEEtran}
\bibliography{refs, rh_refs}

\end{document}